\documentclass[12pt]{article}
\usepackage{amsthm}
\usepackage{booktabs}
\usepackage{amsmath}
\usepackage{color}
\usepackage{graphicx}
\usepackage{amsmath, amssymb, amsfonts, amsthm}
\usepackage{graphics}
\usepackage{geometry}
\usepackage{color}
\usepackage{setspace}
\usepackage{epstopdf}
\usepackage{setspace}
\usepackage{caption}
\usepackage{subcaption}
 \usepackage{pdfsync}

 \usepackage[authoryear]{natbib}
  \usepackage{hyperref}
\usepackage[usenames,dvipsnames]{xcolor}
\newcommand{\field}[1]{\mathbb{#1}}
\newcommand{\R}{\field{R}}

\newcommand{\p}{\field{P}}

\newcommand{\E}{\field{E}}
\newcommand{\pp}{\mathcal{P}}
\newcommand{\FF}{\mathcal{F}}

\def\mf{\mathbf}

\def\mk{\mathfrak}

\def\argmax{\mathop{\mbox{argmax}}}
\def\argmin{\mathop{\mbox{argmin}}}
\theoremstyle{Conjecture}
\theoremstyle{example}
\theoremstyle{remark}
\theoremstyle{lemma}
\theoremstyle{definition}
\theoremstyle{proposition}
\theoremstyle{condition}
\newtheorem{theorem}{Theorem}[section]
\newtheorem{algorithm}{Algorithm}[section]
 \newtheorem{corol}[theorem]{Corollary}

\newtheorem{remark}{Remark}[section]
\newtheorem{lemma}{Lemma}[section]
\newtheorem{definition}{Definition}[section]

\newtheorem{proposition}{Proposition}[section]

\def\cod{\stackrel{\cal D}{\longrightarrow}}

\def\lf{\lfloor}
\def\rf{\rfloor}

\begin{document}

\title{Change point analysis of second order characteristics in non-stationary time series}

\author{ {\small Holger Dette}\\
 {\small Ruhr-Universit\"at Bochum }\\
 {\small Fakult\"at f\"ur Mathematik} \\
 {\small44780 Bochum} \\
 {\small Germany }\\
{\small email: holger.dette@ruhr-uni-bochum.de}\\
\and
 {\small Weichi Wu}\\
 {\small University of Toronto }\\
 {\small 100 St. George St.} \\
 {\small Toronto ON M5S 3G3} \\
 {\small Canada }\\
{\small email: weichi.wu@utoronto.ca}\\
\and
{\small Zhou Zhou}\\
 {\small University of Toronto }\\
 {\small 100 St. George St.} \\
 {\small Toronto ON M5S 3G3} \\
 {\small Canada }\\
{\small email: zhou@utstat.toronto.edu}\\
}

 \maketitle
\begin{abstract}

A restrictive assumption in the work on testing for structural breaks in time series
consists in the fact that the model is formulated such that the stochastic process under the null hypothesis of  ``no change-point''
is stationary. This assumption is crucial   to derive (asymptotic)  critical values for the corresponding testing procedures  using an
elegant and powerful mathematical theory, but it might be not very realistic from a practical point of view. For example, if
change point analysis for a particular parameter of the process (such as the variance) is performed, it is not necessary clear
why other parameters (such as the mean or higher order moments) have to stay constant under the hypothesis that there is no change point
in the parameter of interest. \\
This paper develops change point analysis under less restrictive assumptions and
deals with the problem of detecting change points in the marginal variance and correlation structures
of a non-stationary time series. A CUSUM approach is proposed, which is used to test the ``classical'' hypothesis of the
form $H_0: \theta_1=\theta_2$  vs.  $H_1: \theta_1 \not =\theta_2$, where
$\theta_1$ and $\theta_2$ denote second order parameters (such as the variance or
the lag $k$-correlation)  of the process before and after a change point.
The asymptotic distribution of the CUSUM test statistic is derived under the null hypothesis. This distribution depends in a complicated
way on the dependency structure of the nonlinear non-stationary time series and a bootstrap approach is developed to generate
critical values. The results are then extended to test the hypothesis of a {\it  non relevant change point},   i.e.
 $H_0: | \theta_1-\theta_2 | \leq \delta$, which reflects the fact that
  inference should not be changed, if the difference
between  the parameters before and after the change-point is small. \\
In contrast to previous work, our approach does neither require the mean to be constant
nor - in the case of testing  for lag $k$-correlation - that the mean, variance and fourth order joint cumulants are constant under the null hypothesis.
In particular, we allow that the variance has a change point at a different location than the auto-covariance.
The results are illustrated by means of a simulation study, which shows that the new procedures have nice finite sample properties. The central England monthly temperature series are analyzed and significant change points in the variance and lag $1$-correlation are found in the winter monthly temperature at the late 19th century.
\end{abstract}



\section{Introduction}
\def\theequation{1.\arabic{equation}}
\setcounter{equation}{0}

Change point analysis  is a well studied subject in the statistical literature.  Since the seminal
work  on detecting structural breaks in the   mean of  \cite{page1954}  a powerful methodology has been developed to detect
various types  of change points in time series [see for example
\cite{auehor2013} and \cite{jandhyala2013}  for recent reviews of the literature].   Several authors have  argued that in applications
besides the mean  the detection of
changes in the variance or the correlation structure
of a time series  is of importance as well. Typical examples include the discrimination between stages of high and low asset
volatility or the detection of changes in the parameters of an AR($p$) model in order to obtain superior forecasting procedures.
\cite{wichern1976}  studied the change point problem for the variance in a first order
 autoregressive model. These authors pointed out that  - even if  log-return data exhibits a  stationary behavior in the mean -
 the variability is often not constant and
 as a consequence  any conclusions based on the assumption of homoscedasticity could be misleading.
\cite{abrwei1984} and \cite{bauras1985}  used  a Bayesian
 and an ML approach to find change points  in AR-models.  \cite{incltiao1994} proposed a nonparametric CUSUM-type test for changes in the variance of an independent
identically distributed sequence and \cite{leepark2001} derived corresponding results applicable to linear processes [see also
\cite{changevariance} who used the Schwarz information criterion]. Recently \cite{galpen2007} and
 \cite{aueetal2009}  suggested  nonparametric tests for structural breaks in the variance matrix of a multivariate time series,
 while  \cite{davis2006}  and  \cite{prepucdet2014} proposed methods for detecting multiple breaks in piecewise stationary processes. \\
This list of references is by no means complete but an important feature of  the cited references and  most of the literature
on testing for structural breaks
consists in the fact that the model is formulated such that the stochastic process under the null hypothesis of  ``no change-point''
is stationary. This assumption is crucial   to derive (asymptotic)  critical values for the corresponding testing procedures  using an
elegant and powerful mathematical theory such as strong approximations or invariance principles. On the other hand this assumption drastically
restricts the applicability of the  methodology. For example,  \cite{incltiao1994}  and  \cite{aueetal2009}  assume for the construction of a testing procedure
for the hypothesis
\begin{eqnarray} \label{nullvar}
H_0:\  \sigma^2_i= \sigma^2_j ~\ \text{ for all}\  i, j=1,\ldots , n  ~\text{ versus }\  H_1:\  \sigma^2_i \neq \sigma^2_j\ \text{for some}\ i\neq j.
\end{eqnarray}
of a constant variance of a time series
 that the mean of the sequence under consideration does not change in time (as the variance under the null hypothesis).
A similar assumption was made by
 \cite{wiekradeh2012}  in the context of testing for a constant  correlation, where the authors suggested
a CUSUM-type statistic for a change in the correlation of a stationary time series
if  at  the same the means and variances do not change.
However, from a practical point of view,   assumptions of this type are very restrictive and there might be many situations  where one is interested in a change of the variance (or the correlation)  even if the mean
(or the means and  the variances) change gradually in time. In this case
the classical approach is not applicable any more. Recently, \cite{zhou2013}  investigated such a
problem,  in the context of testing for a constant mean, and demonstrated that the classical CUSUM approach
  yields to severe biased testing results if the  assumption of (weak) stationarity (under the null
hypothesis) is not satisfied.

The situation gets even more complicated if one is interested in more sophisticated hypotheses such as
 {\it precise hypotheses} [see \cite{bergdela1987}].
Here  (in the simplest case)  one  assumes
  the existence of a change point  $k \in \{1,\ldots,n \}$
such that
\begin{align}\label{eq137a}
v_1=\sigma_{1}^2=\ldots =\sigma_{k}^2~\not = ~  v_2=\sigma_{k+1}^2=\ldots  =\sigma_{n}^2~,
\end{align}
and is interested in hypotheses   of the form
\begin{align}\label{relvar}
H_0: \Delta:=|v_2-v_1|\leq \delta\ \ \text{versus}\ \ H_1: \Delta:=|v_2-v_1|>\delta
\end{align}
for  some pre-specified constant $\delta>0$. Throughout this paper we call hypotheses of the form
\eqref{nullvar} {\it ``classical''} in order to distinguish these from the precise hypotheses of the form \eqref{relvar}. Although hypotheses of the form \eqref{relvar} have been discussed in other fields [see \cite{chowliu1992} and \cite{mcbride1999}]
the problem of testing precise hypotheses has only recently been considered  by
\cite{dettwied2014} in the context of change point analysis.  These authors point  out that in many cases  a modification
of the  statistical analysis might not be necessary if a change point has been identified but the
difference between the parameters
before and after the change-point is rather small.
In particular, inference might be robust under ``small'' changes of the parameters  and
changing decisions (such as trading strategies or modifying a manufacturing process) might be very
expensive and should therefore only be performed if changes would have serious consequences.
 Testing hypothesis  of the form \eqref{relvar} to detect a structural break
  also avoids the consistency problem mentioned in \cite{berkson1938}, that is: any test will detect negligible
 changes in the parameter if the sample size is sufficiently large.
 \cite{dettwied2014} call the hypotheses of the form \eqref{relvar} hypotheses of a  {\it non  relevant} (null hypothesis)  and
 {\it relevant change point} (alternative),
 and according to their argumentation only  relevant change points  should be detected, because one has to distinguish scientific from statistical significance.
   \\
 Although the formulation of the testing problem in the form \eqref{relvar}
 is appealing, the construction of corresponding tests faces several   mathematical challenges. In particular, one has to deal with the problem of non-stationarity (even under the null hypothesis of a {\it non relevant change point}).
For example,
 \cite{dettwied2014}    developed a CUSUM-type test for    the hypotheses in \eqref{relvar}, which is only  applicable
 under the assumption that the time series before and after the
change point is strictly stationary. From a practical point this assumption
seems to be very strong and not very realistic.

The present paper is devoted to the construction of change point tests for the second-order characteristics of a non-stationary time series, in particular  changes in the variance and the lag $k$-correlation. We consider   piecewise locally stationary processes as
discussed by  \cite{zhou2013}, which  are introduced in Section \ref{sec2}. Section \ref{sec3} is devoted to   the ``classical'' change point problem for the variance or lag $k$-correlation   of a piecewise locally stationary process. We propose a CUSUM approach based on nonparametric residuals and prove weak convergence of the corresponding CUSUM statistic. It turns out that the limiting distribution
depends in a complicated  way on the dependence structure of the piecewise locally  stationary process, and for this reason a wild  bootstrap approach is developed and its consistency
proved. The methodology is very general and applicable in many situations where the assumptions of classical tests are not satisfied. For example, in the problem of testing the ``classical'' hypothesis of a change in the lag $k$-correlation we do neither assume that the mean, variance or higher order joint cumulants of the nonstationary sequence are constant nor that the change in the variance and the lag $k$-correlation occur at the same location. Furthermore, we discover in this paper that the stochastic errors produced in the nonparametric estimation of the mean and variance function are asymptotically negligible in the second-order CUSUM statistic. The result is of particular interest and  highly non-trivial because the order of the latter nonparametric errors are larger than the $1/\sqrt{n}$ convergence rate of the CUSUM test.

Section \ref{sec4} is devoted to the problem of testing the hypothesis of a non relevant change in the variance or lag
$k$-correlation. We use the CUSUM approach proposed in \cite{dettwied2014} to obtain a test for the hypothesis \eqref{relvar} and its analogue in the case of lag $k$-correlations. Asymptotic normality of a corresponding $\mathcal{L}_2$-type statistic is established and   a wild bootstrap method is developed, which addresses the particular structure of the hypotheses in relevant change point analysis.
 To  our best knowledge  resampling procedures  for this type of change point analysis in non-stationary nonparametric problems have not
    been considered in the literature so far.
 The finite sample properties of the new procedures are investigated by means of a simulation study in Section \ref{sec5}. In Section \ref{sec:data}, we analyze the central England monthly temperature series and illustrate the usefulness of the proposed methodology in identifying second order change points in climate data.  Finally, all proofs and technical details are deferred to an appendix and an online supplement, respectively.

\section{Piecewise locally stationary processes} \label{sec2}
\def\theequation{2.\arabic{equation}}
\setcounter{equation}{0}

We start  introducing some notations, which we frequently use  throughout this paper.
For a (real valued) random variable $X$ and  $p\geq 1$ we denote  by $\|X\|_p=(\E|X|^p)^{1/p}$  the $\mathcal{L}_p$ norm of $X$.
The symbol   $\cod $  means weak convergence of real valued random variables (convergence in distribution).
 For any interval  $\mathcal{I} \subset \R $ and nonnegative integer $q$ define $ \mathcal{C}^{q}( \mathcal{I})$
 as the set of $q$  times continuously differentiable
functions $f:  \mathcal{I} \to \R $ and $\mathcal{C} (\mathcal{I})=\mathcal{C}^0(\mathcal{I})$.
Let $\{\varepsilon_i\}_{i \in \mathbb{Z}}$ denote a sequence of independent identically distributed (i.i.d.) random
variables and  denote by $\FF_i=\sigma(...,\varepsilon_0,...,\varepsilon_{i-1},\varepsilon_i)$ the sigma field generated by
$\{ \varepsilon_j | j \leq i  \}$. We define the sigma field
$\FF_i^{(j)}=\sigma(...,\varepsilon_{j-1},\varepsilon'_{j},\varepsilon_{j+1}...,\varepsilon_i)$, where $\{\varepsilon'_i\}_{i \in \mathbb{Z}}$ is an independent copy of
$\{\varepsilon_i\}_{i \in \mathbb{Z}}$, and $\FF_i^*=\FF_i^{(0)}$ for short.
In the following discussion we will also make frequent use of  the projection operator $\pp_j(\cdot)=\E(\cdot|\FF_j)-\E(\cdot|\FF_{j-1})$.

In this paper, we consider the model
\begin{align}\label{mod1}
Y_i=\mu(t_i)+e_i, \qquad i=1,\dots,n,
 \end{align}
 where (for the sake of simplicity) $t_i=i/n$ ($i=1,\ldots ,n$) and $\mu$ is a smooth function.
 Note  that formally $\{Y_i\}_{i=1}^n$ is a triangular array of random variables but  we do not reflect this fact in our notation.
 Change point problems for this  model have found considerable attention in the recent literature,
 where most of the work refers to problems of detecting a gradual  change of the mean in the  situation of zero mean  and independent identically distributed
 (i.i.d.)  errors (even assumed to be Gaussian in some cases)
 [see \cite{Mueller1992} for an early reference and    \cite{Mallik2011} and \cite{Mallik2013}  for more recent references]. Recently
 \cite{vogtdett2015} proposed  a generalized CUSUM approach to detect  gradual changes in model \eqref{mod1} using a different concept of local stationarity [see \cite{vogt2012}].
  \\
 In the present  paper we consider non-stationary processes of the form
 \eqref{mod1}
 and  are interested in identifying abrupt changes in
 the second  order properties such as the variance or the correlation at a given lag.  More precisely we consider
 an error process $\{e_i\}_{i=1}^n$  in   \eqref{mod1}, which is
piecewise locally stationary (PLS)  with $r$ breaks for some $r \in \mathbb{N}$. Formally, we use the following definition for a $PLS$ process
and  the concept of ``\emph{physical dependence measure for PLS}'', which is
given in \cite{zhou2013}.
\begin{definition} \label{def1} ~~\\
(1) The sequence $\{e_i\}_{i=1}^n$ is called PLS with $r$ break points  if there exist constants $0=b_0<b_1<...<b_r<b_{r+1}=1$ and nonlinear filters $G_0, G_1,..., G_r$, such that
\begin{align*}
e_i=G_j(t_i,\FF_i),  \mbox{ if } b_j<t_i\leq b_{j+1},
\end{align*}
where  $\FF_i=\sigma(...,\varepsilon_0,...,\varepsilon_{i-1}, \varepsilon_i)$, and $\{\varepsilon_i\}_{i \in \mathbb{Z}}$ is a sequence of  i.i.d. random variables.

(2) Assume that $\max_{1\leq i\leq n}\|e_i\|_p<\infty$ for some  $p \ge 1 $.
 Then for $k>0$, define the $k$th physical dependence measure in $\mathcal{L}_p$-norm as
\begin{align*}
\delta_p(k)=\max_{0\leq i\leq r}\sup_{b_i< t\leq b_{i+1}}\|G_i(t,\FF_k)-G_i(t,\FF_k^*)\|_p,
\end{align*}
where   $\delta_p(k)=0$ if $k<0$.

\end{definition}
For the asymptotic analysis presented later  in this paper we list the following conditions:
\begin{description}
\item (A1) The process $\{e_i\}^n_{i=1}$ is PLS and {\it piecewise stochastic Lipschitz continuous}. This means that there exists a constant
$C > 0$, such that for all $i\in\{0,\dots,r\}$  and all $t,s\in [b_i,b_{i+1}]$ the condition  \begin{align*}
\|G_i(t,\FF_0)-G_i(s,\FF_0)\|_\iota/(t-s)\leq C (t-s)
  \end{align*}
  holds, where     $\iota\geq 8$ and $C$ denotes a positive constant.
   In addition, $\E[e_i]=0$ for all $1\leq i\leq n$, and we assume the existence of a
   strictly positive  \emph{variance function} $\sigma^2: [0,1] \to \mathbb{R}^+$, such that $\sigma^2_i := \sigma^2(t_i)= {\rm Var}(e_i) \ (i=1,\dots,n).$
\item (A2) The second derivative $\ddot{\mu}$ of the function $\mu$ in  model \eqref{mod1}  exists and is Lipschitz continuous on the interval $[0,1]$.
\item (A3) $\max_{0\leq i\leq r}\sup_{t\in(b_i,b_{i+1}]}\|G_i(t,\FF_0)\|_\iota<\infty$ for some $\iota\geq 8$. 
\item (A4) $\delta_\iota(k)=O(\chi^k)$ for some $\chi\in (0,1)$ and some $\iota\geq 8$.
\end{description}
\begin{remark} {\rm ~~\\
a)  We emphasize  that  the bound of $\max_{1 \leq i \leq n}\|e_i\|_p$ in Definition \ref{def1} does not depend on $n$. This assumption is made in order to
 simplify the assumptions and the  proofs in the subsequent discussion. It is also possible to develop corresponding results for
an $n$-dependent bound with an additional complication
in the technical arguments of  the  proofs and in the   assumptions. \\
b)
Note that
the process  $\{e_i^2\}^n_{i=1}$  of squared errors is also PLS. Simple calculations show  that $\{e_i^2\}^n_{i=n}$ satisfies the assumptions (A1), (A3), (A4) with $\iota\geq 4$.
}
\end{remark}

\section{Tests for changes in the second order structure } \label{sec3}
\def\theequation{3.\arabic{equation}}
\setcounter{equation}{0}

Suppose that we observe data $\{Y_i\}_{i=1}^n$ according to model \eqref{mod1}, where the process $\{e_i\}_{i=1}^n$ is PLS  and $\mu$ is
an unknown deterministic trend.
We are interested in testing nonparametrically the ``classical'' hypothesis of a  change point in the variance or
 the lag $k$-correlation.
The important difference to previous
work on this subject [see for example \cite{incltiao1994} or \cite{aueetal2009}] is that in  general the process is NOT assumed to be stationary under the
 null hypothesis of no  change point.  This means - for example - that the approach proposed here can be used to test the hypotheses \eqref{nullvar}, where the mean is not constant. The price for this type of flexibility is that critical values of the asymptotic distribution of the CUSUM statistic are not directly available.
For the solution of this problem we will develop  a bootstrap CUSUM-type  test for the ``classical'' hypotheses of a change point in the variance or lag $k$-correlation,   which is based on residuals from a local linear fit.  For the definition of the local
linear estimator we assume throughout this paper that the corresponding kernel function, say $K$,
is  symmetric  with support $[-1,1]$ satisfying  $\int K(x)dx=1$,
 and define for $b>0$ the function  $K_{b}(\cdot)=K(\frac{\cdot}{b})$. The moments of $K$ and $K^2$ are denoted by
  $\mu_l=\int_{\mathbb{R}}x^lK(x)dx$  and   $\phi_l=\int_{\mathbb{R}}x^lK^2(x)dx$, respectively ($l=0,1,\ldots$).
  We also assume that  $K \in \mathcal{C}^{(2)}([-1,1])$.

  \subsection{Change point tests for  the  variance} \label{sec31}
 Our first goal is to investigate the stability of the variances  $\sigma^2_i = \sigma^2(t_i)  = \mbox{Var} (Y_i) \ (i=1,\dots,n)$
  testing nonparametrically the ``classical'' hypotheses \eqref{nullvar}.
For this purpose  we consider the CUSUM statistic
\begin{align}\label{hatTn}
\hat{T}_n=\max_{1\leq i\leq n}\Big|\hat{S}_i-\frac{i }{n}\hat{S}_{n}\Big|,
\end{align}
where $\hat{S}_i=\sum_{j= 1}^i\hat{e}_j^2$ denotes the sum  of squared nonparametric residuals
$\hat{e}_i= Y_i-\hat{\mu}_{b_n}(t_i)$, and $\hat{\mu}_{b_n}({\cdot})$ is the local linear estimator of
the function  $\mu$ with bandwidth $b_n$, that is
 \begin{align}\label{defhatmu}
(\hat{\mu}_{b_n}(t), \hat{\dot{\mu}}_{b_n}(t))=\argmin_{b_0,b_1} \sum^n_{i=1}\big(Y_i-b_0-b_1(t_i-t)\big)^2K_{b_n}(t_i-t)
\end{align}
[see \cite{fangij1996}].
Weak convergence of the statistic $\hat T_n/\sqrt{n}$   follows   under the additional assumption
\begin{description}\item (A5) The long run variance function
 \begin{equation} \label{sigma1}
\kappa_1^2(t)=\sum_{k=-\infty}^\infty \text{cov}(G^2_i(t,\FF_k),G^2_i(t,\FF_0)) \qquad \text{if $t\in(b_i,b_{i+1}]$},
\end{equation}
exists, $\kappa_1^2(0):=\lim _{t\downarrow0} \kappa^2_1(t)$ exists and    $\inf_{t\in[0,1]}\kappa_1^2(t)>0$.
\end{description} The   following result provides the asymptotic distribution of $\hat T_n$. Its proof is complicated and  therefore deferred to  Section \ref{prooftheorem1}
in the Appendix.

\begin{theorem}\label{theorem1}
If assumptions (A1)-(A5) are satisfied and   $nb^6_n\rightarrow 0$, $nb^3_n\rightarrow \infty$, then
under the null hypothesis   of no change in the variance we have
\begin{align}\label{eqthm1}
\frac{1}{\sqrt{n}}\hat T_n \cod  \mathcal{K}_1  := \sup_{t\in(0,1)}| U_1(t)-tU_1(1)|,
\end{align}
where  $\{U_1(t)\}_{t \in [0,1]}$  is a zero mean Gaussian process with covariance function
\begin{equation}\label{cov1}
\gamma(t,s)=\int_0^{\min(t,s)}\kappa_1^2(r)dr.
\end{equation}
\end{theorem}

\medskip

It follows from the proof of Theorem \ref{theorem1} that the statistic $\frac {1}{\sqrt{n}}\hat T_n$ has the same limit distribution as the statistic which is obtained if the nonparametric residuals $\hat e_i$ are replaced by the ``true'' errors $e_i$ from model \eqref{mod1}. This observation is remarkable and highly non-trivial because the error from the nonparametric estimation is larger than $1/\sqrt{n}$.

As a consequence of Theorem \ref{theorem1}, we obtain - in principle - an asymptotic level $\alpha$ test for the hypothesis \eqref{nullvar} by rejecting $H_0$, whenever
$
\frac {1}{\sqrt{n}} \hat T_n > q_{1- \alpha},
$
where $q_{1 - \alpha}$ is the $(1- \alpha)$-quantile of the distribution of the random variable $\mathcal{K}_1$ in \eqref{eqthm1}.
However, under non-stationarity (more precisely under the PLS assumption),
the function $\kappa^2_1$ defined in \eqref{sigma1} and, as a consequence, the covariance structure of the Gaussian process  $ \{  U_1(t)-tU_1(1)\}_{t\in[0,1]} $
involves the complex dependency structure of the data generating process. Therefore
it is very difficult to estimate  the  critical value $q_{1-\alpha}$ of the
asymptotic distribution of the  CUSUM test statistic directly. As an alternative,  a data-driven critical value will be derived in the following discussion  using  a wild bootstrap method to mimic the distributional properties of the Gaussian process $U_1(\cdot)$.
Following   \cite{zhou2013} we
define for  a fixed window size, say $m$,   the quantity
 \begin{align}\label{new.83}
\hat \Phi_{i,m}=\frac{1}{\sqrt{m(n-m+1)}}\sum_{j=1}^i\Big(\hat S_{j,m}-\frac{m}{n}\hat S_{n}\Big)R_j, \qquad i=1,...,n-m+1,
\end{align}
where $\hat S_{j,m}=\sum_{r=j}^{j+m-1}\hat e_r^2$ and $\{R_i\}_{i\in \mathbb{Z}}$ is a sequence of i.i.d
standard normal distributed random variables, which is  independent of $\{ \varepsilon_i \}_{i \in \mathbb{Z}}$.

\begin{theorem}\label{theorem2}Suppose that the conditions of Theorem \ref{theorem1}   are satisfied. In addition, assume that $m\rightarrow \infty$, $m/n\rightarrow 0$,
$mb_n^4{\log^2n}\rightarrow 0$ and $\frac{m\log^2n}{nb_n^{3/2}}\rightarrow 0$.
Then
\begin{align*}
M_n= \max_{m+1\leq i\leq n-m+1}\Big|{\hat \Phi}_{i,m}-\frac{i}{n-m+1}{\hat \Phi}_{n-m+1,m}\Big|\cod  \mathcal{K}_1
\end{align*}
conditional on $\FF_n$,
where the random variable $\mathcal{K}_1$ is defined in \eqref{eqthm1}.
\end{theorem}

 Theorem \ref{theorem2}  provides an asymptotic level $\alpha$ test for the hypothesis of a constant variance in model \eqref{mod1}, where the critical values are obtained by resampling. The details are summarized in the following algorithm. Some illustrations of this method are given in Section  \ref{sec5} and \ref{sec:data}.
\begin{algorithm} \label{algvar}
~~\\{\rm
[1] Calculate the statistic $\hat{T}_n$ defined in (\ref{hatTn}).

[2] Generate $B$ conditionally $i.i.d$ copies $\{\hat{\Phi}^{(r)}_{i,m}\}_{i=1}^{n-m+1}$ $(r=1,\dots,B)$ of the random variables
$\{\hat{\Phi}_{i,m}\}_{i=1}^{n-m+1}$  defined in \eqref{new.83} and calculate
 $$M_r=\max_{m+1\leq i\leq n-m+1}\Big|\hat{\Phi}_{i,m}^{(r)}-\frac{i}{n-m+1}\hat{\Phi}^{(r)}_{n-m+1,m}\Big|.$$
[3] Let $M_{(1)}\leq M_{(2)}\leq ... \leq M_{(B)}$ denote the order statistics of   $M_1,\dots,M_B$. The null hypothesis
of constant variance is rejected   at level $\alpha$, whenever \begin{equation} \label{testvar}
    \hat{T}_n/\sqrt{n}>M_{\lf B(1-\alpha)\rf}.
    \end{equation}
 The $p$-value of this test  is given   by $1-\frac{B^*}{B}$, where $B^*=\max\{r: M_{(r)}\leq \hat{T}_n/\sqrt{n}\}$.
 }
\end{algorithm}

\begin{remark}\label{consis_test}  {\rm
it follows by similar arguments as given in the proof of Theorem 2, Proposition 3 of \cite{zhou2013}, and Lemma \ref{mu} and   Lemma \ref{boundhat}
 in the appendix of this paper  that the bootstrap test \eqref{testvar} is consistent. In fact the  test  is able to detect  local alternatives
  of the form $\sigma^2(\cdot) = \sigma^2+n^{-1/2}f(\cdot)$, where $f(\cdot)$ is a nonconstant piecewise Lipschitz continuous function.
  }
\end{remark}

\subsection{Changes in the correlation} \label{sec22}

In this section we consider the problem of testing whether there are changes in the correlation $\rho_{i,k}:=corr(Y_i,Y_{i+k})$ for some pre-specified lag $k$. Namely, we are interested
in testing the hypotheses
\begin{align}\label{eq83}
H_0:\  \rho_{i,k}=\rho_{j,k}\ \text{ for all}\  i, j=1,\ldots , n  ~  \text{ versus}\  H_1:\ \rho_{i,k}\neq \rho_{j,k}\ \text{for some}\ i\neq j.
 \end{align}
  A test for the classical hypothesis $H_0: \rho_1 = \rho_2$ in stationary processes can be derived by similar arguments as given
  in \cite{wiekradeh2012} under the additional assumption that the mean and variance  are not changing. However,
 statistical  inference regarding changes the correlation structure  in a general local stationary framework (including non constant mean or variance)
 requires estimates of the covariances and variances.
For this purpose we consider two different estimators. First, let $\hat{\sigma}^2(t_i)=\hat{\sigma}^2_{c_n,b_n}(t_i) $ denote the local linear estimates of $\sigma^2_i =\sigma^2 (t_i) =\mbox{Var}(Y_i)$, which is defined as
\begin{align}\label{defhatmu1}
(\hat{\sigma}^2_{c_n,b_n}(t),\widehat{\dot{\sigma^2}}_{c_n,b_n}(t))=\argmin_{b_0,b_1}\sum_{i=1}^n(\hat{e}^2_{i}-b_0-b_1(t_i-t))K_{c_n}(t_i-t),
\end{align}
where   $\hat{e}_{i}=Y_i-\hat{\mu}_{b_n}(t_i)$ denote the residuals obtained from a fit of the local linear estimate with bandwidth $b_n$.
This estimate is appropriate if there is no structural break in the variance.

If this situation cannot be excluded, a more
refined
estimate for the function $\sigma^2$  is required. To be precise,
 we  also allow the variance to have a
 structural break at a point, say  $\tilde{t}_v$, which does not necessarily coincide with the location of the change point in the lag $k$-correlation. We assume that $\ddot \sigma^2$ is Lipschitz continuous
on the intervals  $(0,\tilde{t}_v)$ and $(\tilde{t}_v,1)$     and that there exists a constant $\zeta>0$, such that $\tilde{t}_v\in[\zeta,1-\zeta]$.
We define    an  estimator, say $t_n^* $, of the change point $\tilde{t}_v$ in the variance by
\begin{align} \label{tstar}
 t^*_n=\argmax_{\lf {n\zeta}\rf\leq i\leq n-\lf n\zeta\rf+1}|\mathcal{M} (i)|,
\end{align} where
\begin{align} \label{mi}
\mathcal{M}(i)= \frac {1}{L}\Big(\sum_{j=i-L+1}^i\hat e_j^2-\sum_{j=i}^{i+L-1}\hat e_j^2\Big)
\end{align}
and
$L \in \mathbb{N} $ is a regularization parameter, which  increases with $n$. Note that the maximum in \eqref{tstar} is not taken over the full range $1 \leq i \leq n$ as recommended in \cite{andrews1993}  [see also \cite{qu2008}]. Finally, the second estimator of the variance function
is defined by
 \begin{align}  \label{varest1}
  \hat{\sigma}_n^{2*} (s)=\tilde\sigma^2_1(s)I(s\leq  t^*_n)+\tilde\sigma^2_2(s)I(s>  t^*_n)~,
\end{align}
where
$\tilde\sigma^2_1$ and $\tilde\sigma^2_2$ are the  local linear estimator of the variance function
 from the samples $Y_{1}, \dots, Y_{\lfloor n  t_n^* \rfloor}$ and $Y_{\lfloor n  t_n^* \rfloor +1},\dots, Y_n$, respectively.
 The following result shows that $t_n^*$ is a consistent estimate of $\tilde{t}_v$.
 A proof can be found in Section \ref{sec61}.
 \begin{lemma}\label{tstar0}
  Assume that $nb^6_n\rightarrow 0$, $nb^3_n\rightarrow \infty$ and  that
Assumptions (A1) - (A4) are satisfied with $\iota>8$. Suppose that the variance function is
twice differentiable on the intervals $(0,\tilde{t}_v)$ and $(\tilde{t}_v,1)$, such that the second derivative $\ddot{\sigma^2}$
is Lipschitz continuous  (here $\tilde{t}_v$ is the location of the change point of the variance,
which is defined as $\tilde{t}_v=1$ if there exists no jump). Let $L=\lf n^{\alpha}\rf$ for $4/\iota<\alpha<1/2$,
then  the estimator  $t_n^*$ defined in \eqref{tstar} satisfies
$t^*_n-\tilde{t}_v=O_p(n^{-(1-\alpha)})$ if $\tilde{t}_v<1$.
 \end{lemma}
  \begin{remark}{\rm Observe that the lower bound for the parameter $\alpha$  in Lemma \ref{tstar0} converges to $0$ as   $\iota\rightarrow \infty$. Consequently, if $0<\tilde{t}_v<1$, the  rate of convergence of the estimator $t_n^*$ is arbitrarily close  to the optimal rate $n^{-1}$ if Assumptions (A1) and (A4) hold for any $\iota > 0$.  }
 \end{remark}

 The lag $k$-correlation at the point $t_i$ is estimated by a local average of the quantities of the form
\begin{align}\label{70aa} \hat{W}_i^k=\frac{\hat{e}_{i}\hat{e}_{i+k}} { \hat{\sigma}^{2}(t_i)},\
 \end{align}
 where $ \hat{\sigma}^{2}$ is either the estimate $\hat{\sigma}^2_{c_n,b_n} $ defined in \eqref{defhatmu1} (if a change point in the variance can be excluded)
 or the estimate  $ \hat{\sigma}_n^{2*}$ defined in \eqref{varest1}. {For convenience, we set $\hat{e}_i=0$, whenever $i>n$, in the following discussion}.
The corresponding partial sum is denoted by
 $\hat{S}_i^W=\sum_{j=1}^i\hat{W}_j^{k}$,
and we consider the CUSUM statistic
\begin{align}\label{deftnc}
\hat{T}^c_n=\max_{1\leq i\leq n}\Big|\hat{S}_i^W-\frac{i}{n}\hat{S}_{n}^W\Big|.
\end{align}
 In order to specify the necessary assumptions for the asymptotic theory,
recall the definition of the random variable  $\hat W^k_i$ in \eqref{70aa} and define an unobservable analogue by
\begin{align}\label{70a}
   W_i^k=\frac {e_ie_{i+k}}{\sigma(t_i)\sigma(t_{i+k})}.
\end{align}
It is  easy to see that $W_i^k$   is $\FF_{i+k}$ measurable and that the process $(W_i^k)^{n-k}_{i=0}$ is PLS.
Define $q$ as the number of break points,  $0=c_0<c_1<...<c_{q}=1$ as the corresponding locations of the breaks and $H_0,H_1,\dots,H_q$
as the corresponding nonlinear filters, that is $W_i^k = H_j(t_i, \FF_{i+k})$ if $c_j < t_i \leq c_{j+1}$.
 For the asymptotic analysis of the CUSUM-test we require the following additional assumption:
 \begin{description}\item (A6) The long run variance function
 \begin{align*}
\kappa_2^2(t)=\sum_{k=-\infty}^\infty \text{cov}(H_i(t,\FF_k),H_i(t,\FF_0))\qquad \text{if $t\in(c_i,c_{i+1}]$},
\end{align*}
exists, the limit $\kappa_2^2(0)=\lim _{t\downarrow0} \kappa^2_2(t)$ exists and $\inf_{t\in[0,1]}\kappa^2_2(t)>0$.
\end{description}

\begin{theorem}\label{theorem5}Assume that {$b_n\rightarrow 0$, $c_n/b_n\rightarrow 0$, $c_nb_n^{-2}\rightarrow \infty$, $nc_n^4\rightarrow 0$, $nb_n^{6}c_n^{-1/2}\rightarrow 0$, $nb_n^{4}c_n^{1/2}\rightarrow \infty$} and suppose that Assumptions (A1) - (A4) and  (A6) are satisfied with $\iota\geq 8$. Assume that $ \hat{\sigma}^{2}$ in   \eqref{70aa}  is either the estimate $ \hat{\sigma}^2_{c_n,b_n} $ defined in \eqref{defhatmu1},
if  there is no change point,  or the estimate  $ \hat{\sigma}_n^{2*}$ defined in \eqref{varest1} if there
exists one change point, say $\tilde{t}_v$,  in the variance function. In the latter case
 let $\sigma^2$ be  strictly positive and  twice differentiable on the intervals $(0,\tilde{t}_v)$ and $(\tilde{t}_v,1)$, such that the second derivative $\ddot{\sigma}^2$
is Lipschitz continuous.
Then under the null hypothesis of no change point in the lag $k$-correlation we have
\begin{align*}
\frac{1}{\sqrt{n}}\hat T_n^c\cod  {\cal K}_2 :=  \sup_{t\in (0,1)}|U_2(t)-tU_2(1)|,
\end{align*} where $\{U_2(t)\mid t \in [0,1]\}$ is a centered Gaussian process   with covariance kernel \begin{equation} \label{cov2}
\gamma(t,s)=\int_0^{\min(t,s)}\kappa_2^2(r)dr.
\end{equation}
\end{theorem}

\medskip

In order to develop a consistent bootstrap test recall the notation \eqref{eqthm1}, define  $S_{j,m}^W =\sum_{r=j}^{j+m-1} \hat W^k_r$ and $\hat \Phi^W_{i,m}$ as in \eqref{new.83}  where   $\hat S_{n}$ and  $\hat S_{j,m}$
are replaced by $\hat S_{n}^W$ and  $\hat S_{j,m}^W$, respectively.

\begin{theorem}\label{theorem5boot}
Assume that the conditions of Theorem \ref{theorem5} are satisfied. If $m\rightarrow \infty$, $m/n\rightarrow 0$, { $\sqrt{m}\big(c_n^2+(\frac{1}{\sqrt{nc_n}}+b_n^2+\frac{1}{\sqrt{nb_n}})c_n^{-1/4}\big)\log n\rightarrow 0$}, then
\begin{align} \label{bootcor}
M_n^W=
\max_{m+1\leq i\leq n-m+1}\Big|{\hat \Phi}_{i,m}^W-\frac{i}{n-m+1}{\hat \Phi}_{n-m+1,m}^W \Big|\cod
{\cal K}_2
\end{align}
 conditional on  $\FF_n$, where the random variable ${\cal K}_2$ is defined in Theorem \ref{theorem5}.
\end{theorem}

\bigskip

Theorem \ref{theorem5} and \ref{theorem5boot}
yield  the consistency of  the following bootstrap  test for the hypothesis of a constant correlation with asymptotically correct type I error rate.
The null hypothesis \eqref{eq83}   is rejected whenever
\begin{equation} \label{testcorr}
\hat{T}^c_n/\sqrt{n}>M_{\lf B(1-\alpha)\rf}^W.
\end{equation}
Here $M_{\lf B(1-\alpha)\rf}^W$ is the $(1-\alpha)$-quantile of  bootstrap sample of the
distribution of the  statistic $M_n^W$ defined in \eqref{bootcor}, which is generated in the same way as
described in Algorithm  \ref{algvar}.

\begin{remark}\label{consistest1}  ~~ \\{\rm
(1)
Assume that under the alternative hypothesis the variance has at most one structural break at the point $\tilde{t}_v$  and that
the second derivative $\ddot \sigma^2 $ of the variance function  is Lipschitz continuous on the intervals $(0,\tilde{t}_v)$ and $(\tilde{t}_v,1)$.
Then it can be shown by similar arguments as indicated in Remark \ref{consis_test}  that the  bootstrap
test \eqref{testcorr} is able to detect  local alternatives converging to the null hypothesis at a rate $n^{-1/2}$. \\
(2) In principle one could always work with the estimator $ \hat\sigma^{2*}_n$ defined  in  \eqref{varest1}. However, a numerical study indicates that  in cases
where there is in fact no change point in the variance the estimator $ \hat \sigma^{2}_{c_n,b_n} $
defined  in   \eqref{defhatmu1} has a better finite sample performance.
Therefore we strictly recommend its use, if a structural change in the variance can be excluded.
}
\end{remark}

\begin{remark}\label{rmk3.2} ~~ {\rm
Although the method described so far refers to the problem of detecting
a change point in a particular lag $k$-correlation, it can easily be extended to the problem of testing for a
change in one (or more) correlations  simultaneously. We illustrate extensions of this type exemplarily
for the problem  of testing the hypotheses
\begin{align}
&H_0:\  \rho_{i,k}=\rho_{j,k}=\rho_k\ \text{ for all}\  i, j=1,\ldots , n, k=1,\dots,q  ~ \label{mulcorr0}  \\
 & H_1: \text{  there  exists  } 1\leq k\leq q \text{ and } i\neq  j  \text{  such that }  \ \rho_{i,k}\neq \rho_{j,k}\   \label{mulcorr1}.
\end{align}
For this purpose recall the notations \eqref{70a} and \eqref{70aa}
and define the vectors  $\mathbf W_i=(W_i^1,\ldots,W_i^q)^T$ and $\hat{\mathbf W_i}=(\hat W_i^1,\ldots,\hat W_i^q)^T$.
For some norm $\| \cdot \|$ on $\R^q$ we consider the statistic
$$
\hat{\mathbf T}_n^c=\max_{1\leq j\leq n-q} \Big \|\sum_{i=1}^j\hat{\mathbf W}_i- \frac {1}{n}\sum_{i=1}^{n-q}\hat{\mathbf W}_i \Big \|.
$$
 Observe that $\{\mathbf W_i\}_{i \in \mathbb{N}}$ is a PLS process, with
 nonlinear filter function, say $\tilde {\mathbf W}$, and (unknown) $\ell$ break points $0=d_0<d_1<...,<d_\ell<d_{\ell+1}=1$.
Assume that  the conditions of Theorem \ref{theorem5}  are satisfied, where assumption  (A6) is now replaced by the condition
 \begin{description}
 \item (A$6^*$) The long run variance function
 \begin{align*}
\tilde \kappa_2^2(t)=\sum_{k=-\infty}^\infty \text{Cov} \ (\tilde{\mathbf W_i}(t,\FF_k),\tilde{\mathbf W}_i(t,\FF_0))  \in \R^{q\times q}  \qquad \text{ if $t\in(d_i,d_{i+1}]$} ,
\end{align*}
exists. 
Let $\lambda (t)=  \lambda_{\min} (\tilde \kappa^2_2 (t))$ denote the smallest eigenvalue of the matric $\tilde \kappa^2(t)$, then $\lim_{t \downarrow 0} \lambda (t)$ exists and $\inf_{t \in [0,1]} \lambda(t)>0$.
\end{description}

Under these assumptions it  can be shown by similar arguments as given in the proof of Theorem  \ref{theorem5}   that
\begin{align*}
\frac{1}{\sqrt{n}}\hat {\mathbf T}_n^c\cod     \sup_{t\in (0,1)} \|\mathbf U_2(t)-t\mathbf U_2(1) \|,
\end{align*}
where $\{\mathbf U_2(t)|t \in [0,1]\}$ is a $q$-dimensional centered Gaussian process  covariance kernel
$
\gamma(t,s)=\int_0^{\min(t,s)}\tilde \kappa_2^2(r)dr   \in \R^{q\times q} .
$ \\
A similar result can also be derived for the new bootstrap procedure. To be precise
define the vectors
 \begin{align}
{\bf \hat  \Phi}_{i,m}=\frac{1}{\sqrt{m(n-m-2q+1)}}\sum_{j=q+1}^i\Big(\hat {\mathbf S}_{j,m}-\frac{m}{n}\hat{\mathbf S}_{n-q}\Big)R_j, \qquad i=q+1,...,n-m-q+1,
\end{align}
where $\hat {\mathbf{S}}_{j,m}=\sum_{r=j}^{j+m-1}\hat {\mathbf W}_r $, $\hat {\mathbf{S}}_{n-q}=\sum_{r=1}^{n-q}\hat {\mathbf W}_r $  and $\{R_i\}_{i\in \mathbb{Z}}$ is a sequence of i.i.d
standard normal distributed random variables, which is   independent of $\{ \varepsilon_i \}_{i \in \mathbb{Z}}$. If the conditions of Theorem \ref{theorem5boot} hold
[where assumption  (A6) is again replaced by (A$6^*$)], then  we have (conditional on $\FF_n$)
\begin{align*}
M_n= \max_{q+1\leq i\leq n-m-q+1}\Big \|{  {\bf \hat  \Phi}}_{i,m}-\frac{i}{n-m-2q+1}{ {\bf \hat  \Phi}}_{n-m-q+1,m}\Big \| ~\cod  \sup_{t\in(0,1)}\| \mathbf U_2(t)-t\mathbf U_2(1) \|.
\end{align*}
These results show  that the bootstrap test \eqref{testcorr} can easily be extended to discriminate between the hypotheses \eqref{mulcorr0} and \eqref{mulcorr1}.}
\end{remark}

\section{Relevant changes of second order characteristics} \label{sec4}
\def\theequation{4.\arabic{equation}}
\setcounter{equation}{0}

After a change point has been detected and localized a modification of the statistical analysis  is necessary, which addresses
the different features of the data generating process before and after the change point. \cite{dettwied2014} pointed out that
in many cases such a modification might not be necessary if the difference between the parameters
before and after the change point is rather small. On the one hand, inference might be robust with respect to small changes
of the variance or correlation structure.
On the other hand, changing decisions (such as trading strategies or modifying a manufacturing process) might be very expensive and only be performed if changes would have serious consequences.
 For these reasons \cite{dettwied2014}  proposed to investigate  the hypothesis \eqref{relvar} of a {\it non  relevant change point}, which will be discussed in this section  for the variance (Section \ref{sec41}) and lag $k$-correlation (Section \ref{sec42})
 in a general non-stationary context (more precisely under the assumption  of PLS).

\subsection{Relevant changes in the variance} \label{sec41}
First we investigate the problem of testing for a non relevant change  in the variance of a time series. Recall
the definition of model \eqref{mod1} and that  $\sigma^2_i = \sigma^2(t_i)= \mbox{Var}(Y_i)$ is the variance of the response at the
point $t_i$.  Throughout this section we  assume the existence  of some  fixed but unknown point $\tilde{t}_v \in [0,1]$ such that
the variance function in Assumption (A1)  is constant on the  intervals $(0,\tilde{t}_v)$ and $(\tilde{t}_v,1)$
and that the variances $\sigma_i^2$ satisfy
\eqref{eq137a} {with $k= \lfloor n\tilde{t}_v \rfloor$}.
 For some pre-specified $\delta>0$, we are interested in testing the hypothesis \eqref{relvar}
 of a non relevant change in the variance. Problems of this type  have recently been discussed in \cite{dettwied2014} under the assumption that the process before and after the change point is stationary and that additionally  the mean of
the process is constant  (even if there is a change of small order in the variance).
 From a practical point of view, assumptions of this type
are not very satisfactory and in this section we will propose a procedure for detecting relevant changes which does not require such strong assumptions.

It turns out that a test for the hypothesis \eqref{relvar} needs   an estimator of the change point, and
we could  use the estimator $t_n^*$ defined in \eqref{tstar} for this purpose. This estimator requires the choice of the regularization parameter
$L$.  However, in the present context such a complicated estimator is in fact not necessary, because the variance before and after the change point is assumed to be constant.
Therefore we introduce the alternative   estimator
\begin{align} \label{estvar}
\tilde t_n=\argmax_{1\leq m\leq n}   \Big (\hat{S}_m-\frac{m}{n}\hat{S}_n\Big )^2,
\end{align}
 where $\hat{S}_m=\sum_{i=1}^m\hat{e}_i^2$ denotes the $m$th partial sum of the squared residuals $\hat{e}_{i}=Y_i-\hat{\mu}_{b_n}(t_i)$
 obtained from a local linear fit.
 Estimators, maximizing  the CUSUM-type statistics have been widely studied in the situation of stationary processes
 [see \cite{jandhyala2013}],  and it turns out that the statistic $\tilde t_n$ has  better finite sample  properties than the estimator  $t_n^*$ defined in \eqref{tstar}, if the variance
function before and after the break point $\tilde{t}_v$ is in fact constant (which is our basic assumption throughout this section). \\
 Let ${\Delta}=v_2-v_1$ denote the  ``true'' difference
before and after the change point.
 Our   first result establishes the asymptotic properties of the estimator $\tilde t_n$ under the PLS assumption and is proved in
Section \ref{sec62}.

\begin{lemma}\label{lemmavar} Assume that $b_n\rightarrow 0$, $nb_n^{3}\rightarrow \infty$, $nb_n^{6}\rightarrow 0$ and  that
Assumptions (A1) - (A5) are satisfied. Suppose that the variance function is strictly positive and constant on the intervals  $(0,\tilde{t}_v)$ and $(\tilde{t}_v,1)$ (here $\tilde{t}_v$ is the location of the change point,
which is defined as $\tilde{t}_v=1$ if there is no jump). The estimate
$\tilde t_n$ defined in \eqref{estvar} has the following properties:
\begin{itemize}
\item[(i)] If $\Delta=0$, then $\tilde t_n$ converges weakly to a $[0,1]$-valued random variable.
\item[(ii)] If $\Delta \not =  0$, then $|\tilde {t}_n-\tilde{t}_v|=O_p(n^{-\alpha}) $ for some $\alpha >1/2$.
\end{itemize}
\end{lemma}
We now use the statistic  \eqref{estvar}  to define estimates of the variance before and after the change point, that is
$$\tilde{\Delta}_{n,1}={1\over \lf n\tilde t_n \rf}
\sum_{j=1}^{\lf n\tilde t_n \rf}\hat{e}_j^2 , \quad
 \tilde{\Delta}_{n,2}= {1 \over n-\lf n\tilde t_n \rf} \sum_{j=\lf n\tilde t_n\rf+1}^{n}\hat{e}_j^2 , $$
and denote by
 $\tilde{\Delta}_n=\tilde{\Delta}_{n,2}-\tilde{\Delta}_{n,1}$   an estimator of the difference $\Delta$.
Using similar arguments as given in  the proof of Lemma \ref{lemmadelta} in Section \ref{sec62}
it follows  that $\tilde{\Delta}_n-{\Delta}=O_p(\frac{\log n}{\sqrt{n}})$.
In order to construct a test for the hypothesis \eqref{relvar} we now consider the statistic
\begin{align}
\check{T}_n=\frac{3}{\tilde t^2_n(1-\tilde t_n)^2}\int_0^1 \hat{U}_n^2(s) ds,
\label{new.58}
\end{align}
where the process $\hat{U}_n$ is the CUSUM process defined by
$$
\hat{U}_n(s)=\frac{1}{n}\sum_{j=1}^{\lf ns\rf}{\hat e^2_j}-\frac{s}{n}\sum_{j=1}^{ n }\hat e^2_j.\\
$$
The following result establishes the asymptotic properties of the statistic $\check{T}_n$.
The proof is omitted because it follows by  similar    but easier arguments as given in the proof
 of Theorem \ref{thm6} below, where we investigate the problem of testing for non relevant changes in the lag $k$-correlation.

\begin{theorem}\label{newthm6}
Suppose that the conditions of Theorem \ref{theorem1}  hold.
\begin{itemize}
\item[(i)] If $\Delta \neq 0$, then
\begin{align} \label{limrelvar}
\sqrt{n}(\check{T}_n - \Delta^2)   \cod {\cal Z}_1 (\Delta) := \frac{  6 }{\tilde{t}_v^2(1-\tilde{t}_v)^2}\int_0^1(U_1(s)-sU_1(1))[s\tilde{t}_v-s\wedge \tilde{t}_v]  {   |\Delta |}   ds,
\end{align}
where $\{U_1(s)\}_{s\in [0,1]}$ denotes the Gaussian process defined in Theorem \ref{theorem1}.
\item[(ii)] If $\Delta = 0$, then $\check{T}_n =O_P(1/n)$.
\end{itemize}
\end{theorem}

It follows from Theorem \ref{newthm6} that an asymptotic level $\alpha$ test for the hypothesis (\ref{relvar}) of a non relevant change in the variance is obtained by rejecting the null hypothesis, whenever
\begin{equation} \label{testrelvar}
 \check T_n   > \delta^2 + \frac {v_{1-\alpha}}{\sqrt{n}},
\end{equation}
where $v_{1-\alpha}$ denotes
    the $(1-\alpha)$-quantile of the distribution of the random variable $ {\cal Z}_1 {( \delta ) } $ defined in \eqref{limrelvar}. We note that this distribution is a
    centered normal distribution with a variance depending on the data generating process in a complicated way, in particular on the long run variance defined in \eqref{sigma1}.    In order to circumvent this problem we will develop a resampling procedure to obtain critical values, where we have to address the particular structure of
    the hypothesis \eqref{relvar} of a non relevant change point in the variance.      To be precise, define
\begin{align}
\hat{V}_j={\hat{e}_j^2} -\hat{ {\Delta}}_n I(j\geq \lf n\tilde t_n \rf).\label{new.61}
\end{align}
Let $\{R_j\}_{j\in \mathbb{Z}}$ denote a sequence of i.i.d.\ standard normal distributed random variables, which is independent of
$\{\FF_i\}_{i\in\mathbb{Z}}$, define $\hat{S}_{j,m}^V=\sum_{r=j}^{j+m-1}\hat{V}_r$, $\hat{S}_{n}^V=\sum_{r=1}^{n}\hat{V}_r$
 and consider the random variables
\begin{align}
\label{new.63}
\hat{\Phi}_{i,m}^V=\frac{1}{\sqrt{m(n-m+1)}}\sum_{j=1}^{n-m+1}\Big(\hat{S}_{j,m}^V-\frac{m}{n}\hat{S}_n^V\Big)R_j. \end{align}
The following result provides a bootstrap approximation for the distribution of the random variable $\mathcal{Z}_1   (1)$. The proof follows by similar but easier arguments as given in Section \ref{proofs}, where a corresponding statement is proved for change point tests in the correlation structure.

\begin{theorem} \label{bootcon}
Assume that the conditions of Theorem \ref{newthm6} are satisfied. In addition, assume that $m\rightarrow \infty$, $m/n\rightarrow 0$,
$mb_n^4{ \log^2n}\rightarrow 0$ and $\frac{m\log^2n}{nb_n^{3/2}}\rightarrow 0$, then
$$
M_n=
 \frac{1}{n}  \frac{ 6}{\tilde t^2_n(1-\tilde t_n)^2}
 \sum_{m+1\leq i\leq n-m+1}\Big(\hat{\Phi}^V_{i,m}-\frac{i}{n-m+1}\hat{\Phi}^V_{n-m+1,m}\Big)\Big(\frac{i\tilde t_n}{n}-\frac{i}{n}\wedge\tilde t_n \Big)
 \Rightarrow {\cal Z}_1 (1)
$$
conditionally on $\mathcal{F}_n$, where $ {\cal Z}_1(1)$  denotes the random variable defined in \eqref{limrelvar}.
\end{theorem}

\bigskip

We summarize the bootstrap   test  for the hypothesis \eqref{relvar}  of a  non relevant change   in the   variance structure
in the following algorithm.
\begin{algorithm}\label{algorithmcorrel} ~~\\ {\rm
[1] Calculate the statistic  $\check{T}_n$ defined in (\ref{new.58}).

[2] Generate $B$  conditionally $i.i.d$ copies $\{\hat{\Phi}^{(r)}_{i,m}\}_{i=1}^{n-m+1}$ ($r=1,2,...,B$)
of the sequence $\{\hat{\Phi}^V_{i,m}\}_{i=1}^{n-m+1}$ defined in \eqref{new.63}
 and calculate
    $$
    M_r^V=\frac{1}{n} \frac{ 6  }{\tilde t_n^2(1-\tilde t_n)^2}  \sum_{m+1\leq i\leq n-m+1}\Big(\hat{\Phi}_{i,m}^{(r)V}-\frac{i}{n-m+1}\hat{\Phi}^{(r)V}_{n-m+1,m}\Big)\Big(\frac{i\tilde t_n}{n}-\frac{i}{n}\wedge\tilde t_n \Big) .
    $$
[3] Let $M_{(1)}^V\leq M_{(2)}^V\leq ... \leq M_{(B)}^V$ denote the order statistics of $M_1^V, \ldots , M_B^V$. Reject the null hypothesis \eqref{relvar} of a non
relevant change in the variance  at level $\alpha$ if
    \begin{equation} \label{bootrel}
    \check{T}_n> \delta^2 + M_{(\lf B(1-\alpha)\rf)}^V \delta/\sqrt{n}.
    \end{equation}
 The $p$-value of this test is given by $1-\frac{B^*}{B}$,  where  $B^*=\max\{r: \delta^2+ \frac{M^V_{(r)}{\delta}}{\sqrt{n}}\leq \check{T}_n\}$.
 }
\end{algorithm}

\begin{remark}\label{discusrelevant}   { \rm
It is of interest to investigate the power of the tests \eqref{testrelvar} and \eqref{bootrel}. For this purpose note that   it follows for $\Delta \not = 0$ from
 \eqref{limrelvar}
 \begin{align} \nonumber
\beta_n (\delta,\Delta) &= \p\Big(\check T_n>\delta^2+\frac{v_{1-\alpha}\delta}{\sqrt{n}}\Big)
=\p\Big(\sqrt{n}\frac{\check T_n-\Delta^2}{ {   |\Delta |}   }>\sqrt{n}\frac{\delta^2-\Delta^2}{  {   |\Delta |}  }+\frac{v_{1-\alpha}\delta}{  {   |\Delta |}  }\Big) \\
&\approx 1 - \Psi \Bigl( \sqrt{n}\frac{\delta^2-\Delta^2}{ {   |\Delta |}  } +  \frac{v_{1-\alpha}\delta}{ {   |\Delta |}  } \Bigr) ,
\label{approx}
\end{align}
where $\Psi$ is the distribution function of the random variable $ {\cal Z}_1  (1) $  (in fact a centered normal distribution).
Therefore,  under the  alternative of a relevant change    $\Delta^2 > \delta^2 $, we have  $ \beta_n(\delta,\Delta) \to 1 $ as $n\to \infty$, which
provides the consistency of the  test \eqref{testrelvar}. On the other hand under the null hypothesis $0< \Delta^2 \le  \delta^2 $ we
have
$$
\lim_{n\to \infty } \beta_n (\delta,\Delta)  =  \left\{ \begin{array}{ll}
0 &  \mbox{ if } 0 < \Delta^2  <  \delta^2 \\
\alpha & \mbox{ if }  \Delta^2  =  \delta^2
\end{array}
\right.
$$
If $\Delta=0$, then $\check{T}_n =O_P(1/n)$ and  $\lim_{n\to \infty } \beta_n (\delta,\Delta)  = 0$,
which means that the test \eqref{testrelvar} has in fact asymptotic   level $\alpha$.  \\
We can also use \eqref{approx}  to investigate the power as a function of the parameter $\delta$
in the hypothesis \eqref{relvar}. For example, we can see that for
sufficiently large  $n$  the power
$\beta_n (\delta,\Delta) $ is approximately $1$  if $\delta\rightarrow 0$, and $\beta_n (\delta,\Delta) $ is approximately $0$
if $\delta\rightarrow \infty$. Moreover,
it is  easy to see that all   statements mentioned in this remark  hold also for the  bootstrap test defined by \eqref{bootrel}.}
\end{remark}

\subsection{Relevant changes in correlation} \label{sec42}

Consider model \eqref{mod1}, denote by $\rho_{i,k}=\mbox{corr}(Y_i,Y_{i+k})$  the correlation at lag $k$ and suppose that
for some unknown $t\in (0,1)$
\begin{align}\label{eq137}
\rho_1=\rho_{1,k}= ... =\rho_{\lf nt\rf,k},\quad \quad \rho_2=\rho_{\lf nt\rf+1,k}= ... =\rho_{n-k,k}.
\end{align}
In this section we   are interested in the problem of testing the hypothesis of a non relevant change in the correlation at lag $k$, that is
\begin{align}\label{hypo3}
H_0:  |\rho_1-\rho_2|\leq \delta\ \text{versus} \  H_1: |\rho_1-\rho_2|> \delta
\end{align}
for some pre-specified   {$\delta>0$}.
\cite{dettwied2014} provided a method for testing the hypothesis \eqref{hypo3}   under
this and the additional  assumption that the process before and after the change point exhibits a stationary behaviour. However, in general local stationary framework the construction of a test is more difficult and will be explained in the following paragraphs.

We denote by  ${\Delta}=\rho_2-\rho_1$  the (unknown) difference before and after the change point
and assume throughout  this section that under the null hypothesis of
a non  relevant change in the correlation  the variance function $\sigma^2$  has either no jumps or has a jump at a  point, say
$\tilde{t}_v$, which does not necessarily coincide with the change  point $t$  in the  correlation structure.
In order to estimate the correlation consistently before and after the change point we
recall the definition of the
 variance estimator
 \eqref{varest1}, which addresses the problem that the variance function before and after the change point is not constant.
We define the CUSUM process
\begin{align} \label{Un}
\hat{U}^{[k]}_n(s)=\frac{1}{n}\sum_{j=1}^{\lf ns\rf}\frac{\hat{e}_j\hat{e}_{j+k}}{  \hat{\sigma}^{2*}_n(t_j)}-\frac{s}{n}\sum_{j=1}^{n-k}\frac{\hat{e}_j\hat{e}_{j+k}}{  \hat{\sigma}^{2*}_n(t_j)},
\end{align}
where $\hat e_i = Y_i - \hat \mu_{b_n}(t_i)$ denotes the nonparametric residuals from the local linear fit
and we use the convention that $\hat{e}_i=0$ for $i\geq n$.
The estimator for the change point of the correlation structure is finally defined by
\begin{equation} \label{changeestcorr}
\hat{t}_n=  \argmax_{1\leq m\leq n}\big (\hat U_n^{[k]}(m/n) \big)^2.
\end{equation}
Note that the statistic $\hat t_n$ depends on the estimator $\hat t_n^* $ for  the change point in the variance, which is defined in
\eqref{tstar}. The first result of this section establishes consistency of this estimate.

\begin{lemma}\label{lemvar}
Suppose that the conditions of Lemma  \ref{tstar0} are satisfied, and that the   conditions for the bandwidths $b_n$ and $c_n$ of Theorem \ref{theorem5} hold.   
The estimate $\hat t_n$
of the change point in the correlation structure defined by \eqref{changeestcorr}
satisfies
\begin{align}
& \hat{t}_n  \cod T_{\max},\ { \text{if}\  | \Delta |=0},\label{new.116} \\
|\hat{t}_n-t| &  =O_p(n^{-\alpha}),\ { \text{if}\ | \Delta |>0},\label{new.115}
\end{align}
for some $\alpha > 1/2$, where $T_{\max}$ is a $[0,1]$-valued random variable.
\end{lemma}
The test for the hypothesis of a non  relevant change will be based on the statistic   \begin{align}\label{76}
\hat{T}^r_n=\frac{3}{\hat{t}_n^2(1-\hat{t}_n)^2}\int_{0}^{1} (\hat{U}^{[k]}_n(s))^2 ds,
\end{align}
where the the process $\hat U_n^{[k]}$ is defined in \eqref{Un}.
The following theorem shows that ${ \hat{T}^r_n}$ is a consistent estimator of $\Delta^2$ and also provides
its  asymptotic  distribution.
\begin{theorem}\label{thm6}
Assume  {that the   conditions for the bandwidths $b_n$ and $c_n$ of Theorem \ref{theorem5} hold},  and  that Assumptions (A1) - (A4) are satisfied with {$\iota\geq 16$}. Suppose further that the variance function is strictly positive, twice differentiable on the intervals  $(0,\tilde{t}_v)$ and $(\tilde{t}_v,1)$, such that the   second derivative $\ddot{\sigma^2}$
is Lipschitz continuous (here $\tilde{t}_v$ is the location of the change point, which is defined as $\tilde{t}_v=1$ if there is no jump). {If there exists a break, then we assume additionally that $\tilde{t}_v\in[\zeta,1-\zeta]$  for some constant $\zeta>0$.}

i) If $\Delta \neq 0$, then
 {\begin{align}\label{revision2-163}
\sqrt{n}(\hat {T}^r_n-\Delta^2)\cod \mathcal{Z}_2 (\Delta):= \frac{6}{{t}^2(1-{t})^2}\int_0^1[U_2(s)-sU_2(1)][st-s\wedge t]{ |\Delta | } ds,
\end{align}}

where the Gaussian process $\{U_2(s)\}_{s \in [0,1]}$ is defined in Theorem \ref{theorem5}.

 { ii) If $\Delta = 0$, then $\hat{T}^r_n =O_P(1/n)$.}
\end{theorem}

A careful inspection of the proof of Theorem \ref{thm6} shows that the statement \eqref{revision2-163} remains correct for any estimator of the change point in the correlation structure, which satisfies \eqref{new.116} and \eqref{new.115}. Moreover,
Theorem \ref{thm6}  yields an asymptotic level $\alpha$ test for the hypothesis \eqref{hypo3}
of a non  relevant change in the correlation structure   by rejecting $H_0$, whenever
$
{ \hat T^r_n} > {\delta}^2 +{ \frac {\bar{v}_{1-\alpha}}{\sqrt{n}}},
 $
where $\bar v_{1-\alpha}$ denotes
    the $(1-\alpha)$-quantile of the distribution of the random variable {$ {\cal Z}_2 (\delta) $} defined in \eqref{revision2-163}. We note
    again  that this distribution is a centered normal distribution with a variance depending on the data generating process in a complicated way.
In order to provide a consistent bootstrap approximation of the distribution  of the random variable ${\cal Z}_2 {  (1)} $, recall the definition
of the  estimator  $\hat t_n$ of the change point in the correlation structure in \eqref{changeestcorr}. We consider the statistics
$
\hat \Delta_{n,1}={1\over \lf n\hat t_n \rf} \sum_{j=1}^{\lf n\hat t_n \rf}
\frac {\hat e_j \hat e_{j+k}}{\hat \sigma^{2*}_n(t_j)} $,
$ \hat \Delta_{n,2}={1\over n-\lf n\hat t_n  \rf} \sum_{j=\lf n\hat t_n \rf+1}^{n-k}
\frac {\hat e_j \hat e_{j+k}}{\hat \sigma^{2*}_n(t_j)}$ and define
\begin{equation}  \label{deltadach}
\hat {\Delta}_n=\hat{\Delta}_{n,2}-\hat{\Delta}_{n,1}
\end{equation}
as  an estimator of the difference $\Delta=\rho_2-\rho_1$.
The next lemma provides consistency of $\hat {\Delta}_n$ and is proved in Section \ref{sec62}.

\begin{lemma}\label{lemmadelta}
Suppose that the conditions of Theorem \ref{thm6} and  Assumption
 (A6) are satisfied,
  then
$
\hat{{\Delta}}_n-{\Delta}=O_p\big(\tfrac{\log n}{\sqrt{n}}\big).
$
\end{lemma}

 Define \begin{align}\label{new.new135}
\hat{A}_j={  \frac{\hat{e}_j\hat{e}_{j+k}}{\hat{\sigma}^{2*}_n(t_j)}}-\hat{{\Delta}}_n I(j\geq \lf n\hat t_n \rf),
\end{align}
where the variance estimator is given by \eqref{varest1},
and let  $\{R_j\}_{j\in \mathbb{Z}}$ be a sequence of  i.i.d.   standard normal distributed random variables,
 which is independent of $\{\FF_i\}_{i\in\mathbb{Z}}$. We introduce the partial sums
 $\hat{S}_{j,m}^A=\sum_{r=j}^{j+m-1}\hat{A}_r,$ $ \hat{S}_{n}^A=\sum_{r=1}^{n}\hat{A}_r$ and define
\begin{align}
\label{107}\hat{\Phi}^A_{i,m}=\frac{1}{\sqrt{m(n-m+1)}}\sum_{j=1}^{n-m+1}\Big( \hat{S}_{j,m}^A-\frac{m}{n}\hat{S}_n^A \Big)R_j ,
\end{align}
then the following result is proved in Section \ref{sec63}.

\begin{theorem}
\label{thm7}
Suppose the conditions of Theorem \ref{thm6} hold and that $m\rightarrow \infty$, $m/n\rightarrow 0$, { $\sqrt{m}\big(c_n^2+(\frac{1}{\sqrt{nc_n}}+b_n^2+\frac{1}{\sqrt{nb_n}})c_n^{-1/4}\big)\log n\rightarrow 0$}, then
\begin{equation}\label{bootcorrel}
M_n^r=
\frac{1}{n} \frac{6 }{\hat t_n^2(1-\hat t_n )^2}
 \sum_{m+1\leq i\leq n-m+1}\Big(\hat{\Phi}^A_{i,m}-\frac{i}{n-m+1}\hat{\Phi}^A_{n-m+1,m}\Big)\Big(\frac{i\hat t_n}{n}-\frac{i}{n}\wedge\hat t_n\Big)
\cod  { {\cal Z}_2    {  (1)}}
\end{equation}
conditional on $\FF_n$, where the random variable ${\cal Z}_2(1)$ is defined in Theorem \ref{thm6}.
\end{theorem}

 Theorem \ref{thm7} provides a consistent asymptotic level $\alpha$
  bootstrap test for the hypothesis of a non relevant change in the correlation structure.
The hypothesis \eqref{hypo3} is rejected, whenever
\begin{equation} \label{testrelcorboot}
\hat{T}^r_n>M_{(\lf B(1-\alpha)\rf)}{ \delta}/\sqrt{n}+\delta^2.
\end{equation}
Here $M_{\lf B(1-\alpha)\rf}^r$ is the $(1-\alpha)$-quantile of  bootstrap sample of the
distribution of the statistic $M_n^r$ defined in \eqref{bootcorrel}, which is generated in the same way as
described in Algorithm  \ref{algorithmcorrel}.
The consistency and the properties of the power function of the bootstrap test follow by similar  arguments as given in
of Remark \ref{discusrelevant} for the test of non relevant change in variance. The details are omitted for the sake of brevity.

\begin{remark}
{\rm
The proposed method  can easily be generalized to address the problem of testing for a
relevant change in several correlations simultaneously. Exemplarily we illustrate such a generalization  in the situation, where one is interested
in detecting  a relevant change in any of  the lag $1$- to lag $q$-correlations.
 Consider model \eqref{mod1} and suppose that
 there exist  time points $t_k\in (0,1)$   such that
\begin{align*}
\rho_1(k)=\rho_{1,k}= ... =\rho_{\lf nt_k\rf,k},\quad \quad \rho_2(k)=\rho_{\lf nt_k\rf+1,k}= ... =\rho_{n-k,k}
\end{align*}
($1\leq k\leq q$).
We are interested in testing the hypotheses
\begin{align*}
&H_0:\  |\rho_{1}(k)-\rho_{2}(k)|\leq \delta_k \text{ for all}\  k=1,\dots,q  ~ \\
& H_1: \text{there exists a lag }  k\in \{1,\ldots , q\} \text{ such that } \ |\rho_{1}(k)-\rho_2(k)|>\delta_k,
\end{align*}
where $\delta_1,\ldots , \delta_q$ are given thresholds.
For each lag $k$,  let $\hat{t}_{n,k}$ denote the estimator for $t_k$ defined in \eqref{changeestcorr}
and define
\begin{align}\label{76}
\hat{T}^r_{n,k}=\frac{3}{\hat{t}_{n,k}^2(1-\hat{t}_{n,k})^2}\int_{0}^{1} \big(\hat{U}^{[k]}_n(s)\big)^2 ds,
\end{align}
where $\hat{U}^{[k]}_n$ is given by \eqref{Un}. Recall the definition of  the $q$-dimensional process $\mathbf U_2(t)=(U_{2,1},\ldots,U_{2,q})^T$ in Remark \ref{rmk3.2},
and suppose that the conditions of Theorem {\ref{thm6}} and (A6$^*$) hold. Then it can
be shown
\begin{align*}
\max_{1\leq i\leq q} |\sqrt{n}(\hat { T}^r_{n,i}-\Delta_i^2)|\cod \mathcal Z^{*}=\max_{1\leq i\leq q}\left| \frac{6}{{t_i}^2(1-{t_i})^2}\int_0^1[U_{2,i}(s)-sU_{2,i}(1)][st_i-s\wedge t_i]{  |\Delta_i | } ds\right|,
\end{align*}
where $\Delta_k =\rho_{1}(k)-\rho_{2}(k)$ ($1\leq k\leq q$). Similarly,  bootstrap methodology can be developed considering
consistent estimates, say $\hat{{\Delta}}_{k,n}$, of $\Delta_k$. Let
\begin{align}
\hat{A}_j^{[k]}={ \frac{\hat{e}_j\hat{e}_{j+k}}{\hat{\sigma}_n^{2*}(t_j)}}-\hat{{\Delta}}_{k,n} I(j\geq \lf n\hat t_k \rf) ,
\end{align}
and consider the vector $\hat{\mathbf A}_j=(\hat{A}_1^{[1]},\ldots,\hat{A}_q^{[k]})^T$. Further, define
$\hat{\mathbf S}_{j,m}^A=\sum_{r=j}^{j+m-1}\hat{\mathbf A}_r, $ $ \hat{\mathbf S}_{n-q}^A=\sum_{r=1}^{n-q}\hat{\mathbf A}_r$ and
\begin{align}
{ \bf \hat\Phi}^A_{i,m}=\frac{1}{\sqrt{m(n-2q-m+1)}}\sum_{j=q+1}^{n-m+1}\Big( \hat{\mathbf S}_{j,m}^A-\frac{m}{n}\hat{\mathbf S}_{n-q}^A \Big)R_j ~
\end{align}
($ i=q+1,...,n-m-q+1$), where $\{R_j\}_{j\in \mathbb{Z}}$ is a sequence of i.i.d.\ standard normal distributed random variables, which is independent of
$\{\FF_i\}_{i\in\mathbb{Z}}$. Let ${ \bf \hat\Phi}^{A^{[k]}}_{i,m}$ be the $k_{th}$ entry of ${\bf \hat \Phi}^A_{i,m}$  $(1\leq k\leq q)$ and define
\begin{equation}
M_{n,k}^r=
\frac{1}{n} \frac{6 }{{\hat t_{n,k}}^2(1-{\hat t_{n,k}} )^2}
 \sum_{m+1\leq i\leq n-m+1}\Big({\bf \hat \Phi}^{A^{[k]}}_{i,m}-\frac{i}{n-m+1}{\bf \hat \Phi}^{A^{[k]}}_{i,m}\Big)\Big(\frac{i{ }}{n}\hat t_{n,k} -\frac{i}{n}\wedge{\hat t_{n,k}}\Big).
\end{equation}
Then (conditional on $\FF_n$)
$\max_{1\leq k\leq q}|\Delta_kM_{n,k}^r| \cod \mathcal Z^*$,
 which provides the consistency of a corresponding bootstrap test.
}
\end{remark}

\section{Finite sample properties}   \label{sec5}
\def\theequation{5.\arabic{equation}}
\setcounter{equation}{0}

In this section we investigate the finite sample properties of the proposed tests by means of a simulation study.
In all examples considered   we used the function
 $\mu(t)=8(-(t-0.5)^2+0.25)$ as mean  function and a sequence of independent
 identically  normal distributed random variables   $\{\varepsilon_j\}_{j\in \mathbb{Z}}$ in the definition
 of the errors  $e_i=G_j(i/n,\FF_i)$ in model \eqref{mod1}, where $\FF_ i = \sigma (\ldots, \varepsilon_0, \ldots, \varepsilon_i) $.
 The dependency structures differ by different choices for the nonlinear filter $G_j$.
 The sample size is  $n=500$ and all results are based on $2000$ simulation runs. In each run,
 the critical values are generated by $B=2000$ bootstrap replications.

\begin{figure}[t]
\centering
  \includegraphics[width=14cm,height=12cm]{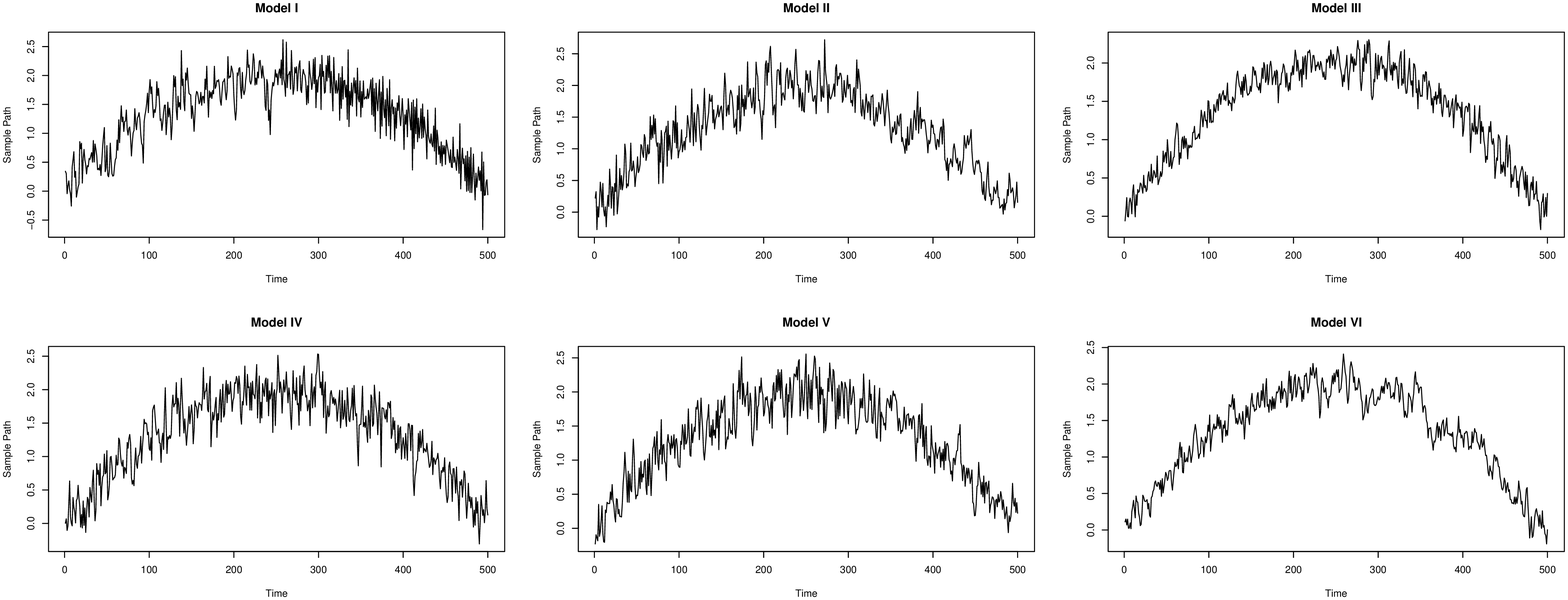}
\caption{\it Typical sample paths of the processes corresponding to model I-VI.}
  \label{fig:mini:subfig:fige01}
\end{figure}

\subsection{Change point tests for the variance}

In this section we investigate the finite sample properties of the bootstrap tests for the ``classical'' hypothesis \eqref{nullvar}  of a constant variance and for the null
hypothesis \eqref{relvar} of a non relevant change in the variance. It turned out that for change point analysis of   the variance
the {\it Minimal Volatility (MV) method} for the selection
of the bandwidth
provided slightly better  results than the commonly used cross validation.   The  MV-method has been advocated in \cite{polromwol1999}. To be  precise consider  a sufficiently wide interval $(a,b)\subset (0,1)$ and define  the points $d_i:=a+(i-1)(b-a)/l\in(a,b) $, $i=1,2,\ldots ,l$
as potential bandwidths. For each $d_i$, we calculate the statistic $\hat T_n(i)$, which is defined as  the test statistic  $\hat T_n$
 in \eqref{hatTn}  with bandwidth $b_n=d_i$. We then calculate for $i=4,\ldots ,l-3$ the standard errors
 $SD(i):=sd(\hat T_n(j),i-3\leq j\leq i+3)$  of different subsamples of these statistics. Finally, the   bandwidth $b_n$ in the interval  $(a,b)$
 is chosen  as  $b_n=d_{i^*} $,  where ${i^*} =\argmin_{4\leq i\leq l -3} SD(i)$. \\
 In the following we discuss three models for the  filters of the innovations in model \eqref{mod1}.

\begin{description}
\item (I) $G(t,\FF_i)=H(t,\FF_i)/4$, where
$H(t,\FF_i)=0.5H(t,\FF_{i-1})+\varepsilon_i$ for $t\leq 0.5$, and $H(t,\FF_i)=-0.5H(t,\FF_{i-1})+\varepsilon_i$ for $t>0.5$.
\item (II) $G(t,\FF_i)=H(t,\FF_i)\sqrt{1-a^2(t)}/4$, where
$H(t,\FF_i)=\sum_{j=0}^\infty a^{j}(t)\varepsilon_{i-j}$ and $a(t)=1/4+t/2$.
\item (III) $G(t,\FF_i)=H(t,\FF_i)\sqrt{1-a^2(t)}/8$ for $t\leq 0.5$, and $G(t,\FF_i)=H(t,\FF_i)\sqrt{2(1-b^2(t))}/8$ for $t>0.5$.
Here $H(t,\FF_i)=\sum_{j=0}^\infty a^{j} (t)\varepsilon_{i-j}$ for $t\leq 0.5$, and $H(t,\FF_i)=\sum_{j=0}^\infty b^{j}(t) \varepsilon_{i-j}$ for $t>0.5$,
$a(t)=1/4+t/2$ and $b(t)=0.5-(t-0.5)^2$.
 \end{description}
 Model (I) is a piecewise stationary process. The correlation has a structural break at the point  $t=0.5$. Model (II) is a locally stationary process. The MA coefficient is smoothly time-varying. By our construction, the variance of model (I) and model (II) remains constant, which means that both models correspond to the ``classical'' null hypothesis in \eqref{nullvar}. Model (III) is a piecewise locally stationary process. The correlation of model (III) has a jump at the point $t=0.5$. The MA coefficients before and after the point
 $t=0.5$ are smoothly varying. The variance of model (III) also has a jump at $t=0.5$ and remains constant before and after the jump, respectively. This model corresponds to the null hypothesis in \eqref{relvar} of a non relevant change point in the variance. Typical trajectories of the processes corresponding to model (I) - (III)
 are displayed in the upper part of Figure \ref{fig:mini:subfig:fige01}.

\begin{table}[htbp]
  \centering
  \caption{\it Simulated Type I error of the tests \eqref{testvar} and \eqref{bootrel} for a change in  the  variance for various bandwidths and the  bandwidth
  calculated by the MV-method (last line). Left and middle column: test for the hypothesis \eqref{nullvar} (model (I) and (II)). Right column: test for the hypothesis
  \eqref{relvar} (model (III)). }
    \begin{tabular}{rrrrrrr}
    \toprule
    model & \multicolumn{2}{c}{I} & \multicolumn{2}{c}{II} & \multicolumn{2}{c}{III} \\
    \midrule
    $b_n/\alpha$ & 5\%   & 10\%  & 5\%   & 10\%  & 5\%   & 10\% \\
    0.025 & 19.15 & 31.55 & 30.45 & 44.9  & 2.45  & 4.95 \\
    0.05  & 8.8   & 16.9  & 15.2  & 25.9  & 5.4   & 8.7 \\
    0.075 & 5.7   & 13.1  & 10.25 & 18.75 & 7.55  & 11.9 \\
    0.1   & 6.05  & 12.8  & 8.85  & 16.3  & 8.15  & 12.7 \\
    0.125 & 4.75  & 10.95 & 7.95  & 15.2  & 7.4   & 12.15 \\
    0.15  & 5.55  & 11.9  & 7.3   & 14.5  & 8.75  & 14.05 \\
    0.175 & 4.1   & 10.2  & 6.5   & 14.5  & 8     & 13.05 \\
    0.2   & 4.6   & 10.7  & 4.9   & 11.15 & 7.55  & 11.8 \\
    0.225 & 3.55  & 8.85  & 4.75  & 10.3  & 7.3   & 12.85 \\
    0.25  & 3     & 8     & 5.15  & 10.8  & 8.35  & 14.75 \\
    0.275 & 3.05  & 9.3   & 4.75  & 10.7  & 7.8   & 14.15 \\
    0.3   & 4.15  & 9.55  & 5.6   & 10.35 & 16.35 & 26.65 \\
    MV    & 4.7   & 11.4  & 6.75  & 13.9  & 6.75  & 12.3 \\
    \bottomrule
    \end{tabular}%
  \label{tab:addlabel}%
\end{table}%

For model  (I) and (II)  we investigate approximation of the nominal level of the bootstrap test  \eqref{testvar}
 for the classical hypothesis \eqref{nullvar} of a change
in the variance  over the interval $[0,1]$.   The corresponding rejection probabilities are displayed in the left and middle column of Table
\ref{tab:addlabel}. We display the  simulated type I error  using different bandwidths in the interval $ (0.025,0.3)$ and
the  bandwidth calculated by the MV-method (last line). We observe that the results are rather stable with respect to the choice of $b_n$. Only
if $b_n \le 0.1$ the level is overestimated. In particular the bandwidth calculated by the MV-method  yields good results for both models.

  For model (III), we are interested in testing  the hypothesis \eqref{relvar} of non relevant changes in variance, where  the threshold is given by $\delta=1/64$.
  The corresponding rejection probabilities are shown in the right column of Table \ref{tab:addlabel} for the case that  $\Delta=\delta=1/64$. Note that this
  choice corresponds to the boundary of the null hypothesis, and according to Remark \ref{discusrelevant}   the nominal level is smaller in the interior of the null hypothesis,
  i.e. $\Delta < \delta$ (these results are not displayed for the sake of brevity).
  Interestingly - compared to the problem of testing the ``classical'' hypothesis \eqref{nullvar} - the method is more sensitive with respect to the
  choice of the bandwidth. However, the MV-method yields a rather accurate approximation of the nominal level.

\subsection{Change point tests for a lag $k$-correlation}

 We now investigate the same properties of the tests for  changes in the lag $1$-correlation. For this purpose we consider the following models.
 \begin{description}
\item (IV)    $G(t,\FF_i)=H(t,\FF_i)\sqrt{1- (t-0.5)^2}/4$, where $H(t,\FF_i)=0.3H(t,\FF_i)+\varepsilon_i$.
\item (V)  $G(t,\FF_i)=H(t,\FF_i)\sqrt{c(t)}/4$ for $t\leq 0.5$, and $G(t,\FF_i)=H(t,\FF_i)\sqrt{d(t)}/4$ for $t>0.5$, where
$c(t)=1- (t-0.5)^2$, $d(t)=1-\frac{1}{2}\sin t$ and  $H(t,\FF_i)=0.3H(t,\FF_i)+\varepsilon_i$.
\item (VI)  $G(t,\FF_i)=H(t,\FF_i)\sqrt{1- (t-0.5)^2}/8$, where $H(t,\FF_i)=0.5H(t,\FF_{i-1})+\varepsilon_i$
for $t\leq 0.5$, and $H(t,\FF_i)=0.7H(t,\FF_{i-1})+\varepsilon_i$ for $t>0.5$.
\end{description}

Model (IV) is a locally stationary processes. The variance of the process is time-varying, but the correlation remains constant. Model (V)
 and model (VI) are piecewise locally stationary processes, where the variance  has a change  point. Before and after the jump, the variance varies smoothly.
 The correlation of model  (IV) and (V) is constant, while the correlation of model (VI) has a break at $t=0.5$.
 Typical trajectories corresponding to these processes are depicted in the lower part of Figure \ref{fig:mini:subfig:fige01}

Note that the change point analysis by the tests proposed in Section \ref{sec41} and \ref{sec42}  requires the choice of two bandwidths
 in the local
linear estimates of the mean and variance.
We  use a generalized cross validation method introduced by \cite{zhouwu2010}
to select the bandwidth for estimating the mean function. Then we apply this cross validation procedure
 again to select the bandwidth for estimating the variance function. The parameters $L$ and $\zeta$ in the estimator
 \eqref{tstar} are chosen as $L= \lfloor n^{1/3}\rfloor $ and $\zeta = 0.016$, respectively.

 The corresponding rejection probabilities
 are displayed in Table \ref{tab:addlabel2} for various bandwidths $b_n$ in the interval $ (0.025,0.3)$. At each fixed $b_n$, the bandwidth for variance $c_n$ is calculated by cross validation.
  \ The results  for the bandwidths
 calculated by cross validation are displayed in the last row of the table. The left and middle
 column correspond to the ``classical''  hypothesis
 whether the correlation remains constant (model (IV) and (V)). We observe a very accurate approximation of the nominal level if the
 bandwidth is chosen such that  $b_n < 0.1$. If $b_n \in (0.1,0.3)$ the nominal level is slightly underestimated.
 For both models the generalized cross validation proposed by   \cite{zhouwu2010}
 yields a rather accurate approximation of the nominal level,

 The right column of  Table \ref{tab:addlabel2}  shows the simulated type I error
 of the test  \eqref{hypo3} for a non  relevant change in correlation with  $\delta=0.2$.
  Again the case
  $\delta=\Delta=0.2$ is displayed  in model (VI) corresponding to the boundary of the null hypothesis, and the simulated rejection probabilities are smaller if $ | \Delta| < \delta$.
  Compared to the test for a non relevant change in the variance (see  Table \ref{tab:addlabel}) the
  test for a non relevant change in the correlation is rather stable with respect to the choice
  of the bandwidth. Also the proposed cross validation methodology  performs reasonably well, which is reported in the last line
  of  Table \ref{tab:addlabel2}.


\begin{table}[htbp]
  \centering
  \caption{\it Simulated Type I error of the tests for a change in the lag $1$-correlation  for various bandwidths and the  bandwidth
  calculated by generalized cross validation (last line). Left and middle column: test for the hypothesis \eqref{eq83} (model (IV) and (V)).
  Right column: test for the hypothesis  \eqref{hypo3} (model (VI)). }
    \begin{tabular}{rrrrrrr}
    \toprule
    model & \multicolumn{2}{c}{IV} & \multicolumn{2}{c}{V} & \multicolumn{2}{c}{VI} \\
    \midrule
    $b_n/\alpha$ & \multicolumn{1}{c}{5\%} & 10\%  & 5\%   & 10\%  & 5\%   & 10\% \\
    0.025 & 4.7   & 10.4  & 4.9   & 10.3  & 6     & 8.95 \\
    0.05  & 3.75  & 8.5   & 4.5   & 9.1   & 6     & 9.25 \\
    0.075 & 4.15  & 9.7   & 4.4   & 9.6   & 7.05  & 9.9 \\
    0.1   & 3.4   & 8.1   & 3.4   & 8.1   & 6.05  & 9.55 \\
    0.125 & 3.85  & 9.1   & 4.05  & 8.25  & 6.4   & 9.35 \\
    0.15  & 3.35  & 8.35  & 3.2   & 8.25  & 5.3   & 7.7 \\
    0.175 & 2.85  & 8.15  & 3.5   & 8.3   & 4.9   & 6.75 \\
    0.2   & 2.6   & 7.6   & 3.95  & 8.9   & 4.5   & 6.95 \\
    0.225 & 2.9   & 7.75  & 3.1   & 8.05  & 4     & 6 \\
    0.25  & 3.1   & 8     & 3.45  & 8.25  & 5.15  & 7.6 \\
    0.275 & 2.8   & 8.75  & 3.5   & 8.9   & 6.3   & 8.9 \\
    0.3   & 3.05  & 8.55  & 3.15  & 8.35  & 5.9   & 8.8 \\
    CV    & 5     & 10.15 & 4.75  & 9.65  & 6.75  & 10.45 \\
    \bottomrule
    \end{tabular}%
  \label{tab:addlabel2}%
\end{table}%

\begin{figure}[htbp]
\centering
  \includegraphics[width=16cm,height=8cm]{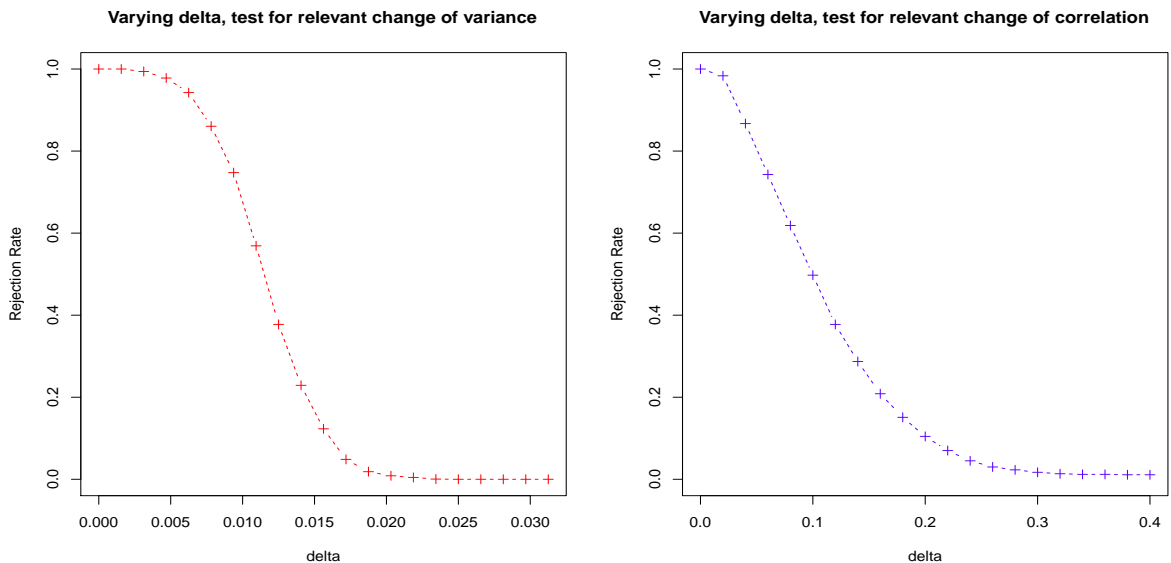}
\caption{\it Simulated rejection probabilities of the test       \eqref{hypo3}  for the hypothesis of  a non relevant change in the variance (left panel) and lag $1$-correlation (right panel) 
as a function of the threshold $\delta \in[0,2\Delta]$ in the hypothesis \eqref{relvar}.}
  \label{fig:mini:subfig:fige03}
\end{figure}

Finally,  we display in Figure  \ref{fig:mini:subfig:fige03} the simulated rejection probabilities
of the tests for the hypotheses  \eqref{relvar} and  \eqref{hypo3}  for  a non relevant change in the variance or correlation,
respectively, as a function of the parameter   $\delta \in[0,2\Delta] $.
 The significance level is chosen as  $0.1$. As expected the probability of rejection decreases with $\delta$ (see also the discussion in Remark \ref{discusrelevant}).  

\subsection{Power properties}
In this section we investigate the power of the  tests proposed in this paper considering the following four scenarios.

\begin{description}
\item (I')  $G(t,\FF_i)=(\mf 1(t\leq 0.5)+\mf 1(t>0.5)\sqrt{1+\lambda})H(t,\FF_i)/4$, where $\lambda > -1$,
 $H(t,\FF_i)=0.5H(t,\FF_{i-1})+\varepsilon_i$ for $t\leq 0.5$, and $H(t,\FF_i)=-0.5H(t,\FF_{i-1})+\varepsilon_i$ for $t>0.5$.
\item (II') $G(t,\FF_i)=H(t,\FF_i)\big[
\mf 1 (t\leq 0.5) \sqrt{1-a^2(t)} + \mf 1 (t>0.5) \sqrt{(2+\lambda)(1-b^2(t))}\big]/8  $, where
 $H(t,\FF_i)=\sum_{j=0}^\infty a^j(t)\varepsilon_{i-j}$ for $t\leq 0.5$, and $H(t,\FF_i)=\sum_{j=0}^\infty b^j(t) \varepsilon_{i-j}$ for $t>0.5$
and the function $a$ and $b$ are defined by  $a(t)=1/4+t/2$ and $b(t)=0.5-(t-0.5)^2$, respectively.
 \item (III')  $G(t,\FF_i)=\sqrt{  1-(t-0.5)^2  }/4 H(t,\FF_i) $, where $H(t,\FF_i)=0.3H(t,\FF_i)+\varepsilon$ for $t\leq0.5$, and $H(t,\FF_i)=(0.3-\lambda)H(t,\FF_i)+\varepsilon$ for $t>0.5$.
\item (IV')  $G(t,\FF_i)=H(t,\FF_i)\sqrt{  1-(t-0.5)^2  }/8$, where
 $H(t,\FF_i)=(0.5-\lambda)H(t,\FF_{i-1})+\varepsilon_i$ for $t\leq 0.5$, and $H(t,\FF_i)=0.7H(t,\FF_{i-1})+\varepsilon_i$ for $t>0.5$,
\end{description}

\begin{figure}[htbp]
\centering
   \includegraphics[width=16cm,height=16cm]{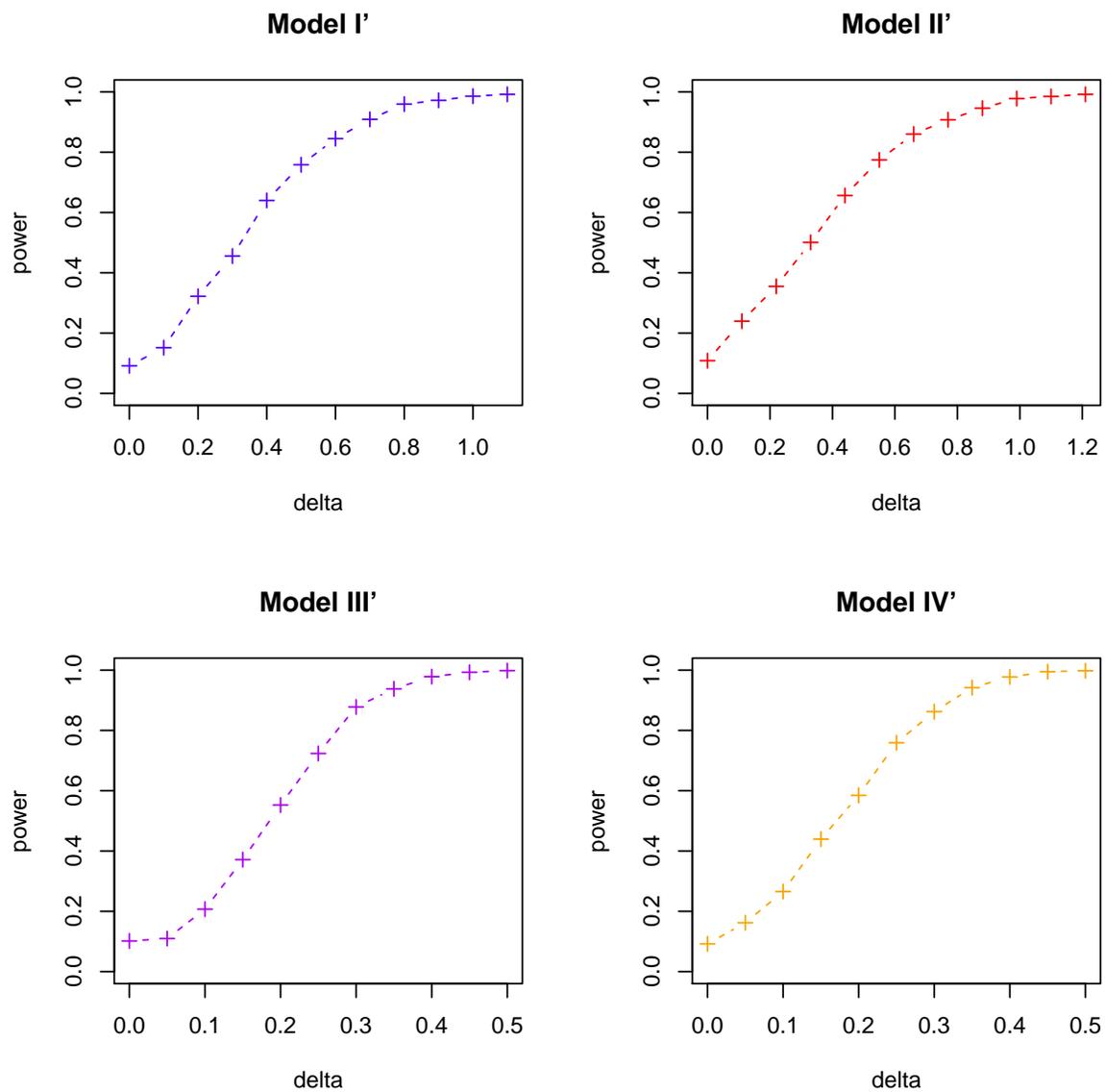}
\caption{\it Simulated power of tests for a change in the variance or lag $1$-correlation. Left upper panel: test for  a constant variance defined in \eqref{testvar} (model  (I')).
Right  upper panel: test for  a non relevant change in the variance defined in \eqref{bootrel} (model (II')).
Left lower panel: test for a constant lag $1$-correlation defined in \eqref{testcorr} (model  (III')).
Right  lower  panel: test for a  non relevant change in the lag $1$-correlation defined in \eqref{testrelcorboot}  (model (IV')).}
  \label{fig:mini:subfig:fige02}
\end{figure}

Model (I') is used to study the power of the test \eqref{testvar} where the  case $\lambda=0$ corresponds to the null hypothesis of a constant variance.
The power properties of the test \eqref{bootrel} for a non relevant change in the variance is investigated
in model (II'). Here we test the hypotheses $H_0:\Delta\leq 1/64$ versus   $H_1: \Delta >1/64$, where the case $-2<\lambda\leq0$ corresponds to the null hypothesis. Similarly, the power of the test for a constant lag $1$-correlation \eqref{testcorr} is studied in model (III') (again the
case  $\lambda=0$ corresponds to the null hypothesis of a constant correlation) and the corresponding hypotheses  $H_0:\Delta\leq 0.2$ versus
$H_1: \Delta>0.2$ of a non relevant change in the lag $1$-correlation are  investigated
in model (IV') (here the case  $-0.4\leq\lambda\leq0$  corresponds to the null hypothesis).
The rejection probabilities for various values of $\lambda$ are displayed in
   Figure \ref{fig:mini:subfig:fige02}. We observe that the  proposed methodology can detect
(relevant) changes in the variance or correlation  with reasonable size.

\section{Data Analysis}\label{sec:data}
\begin{figure}[htbp]
\centering
  \includegraphics[width=16cm,height=8cm]{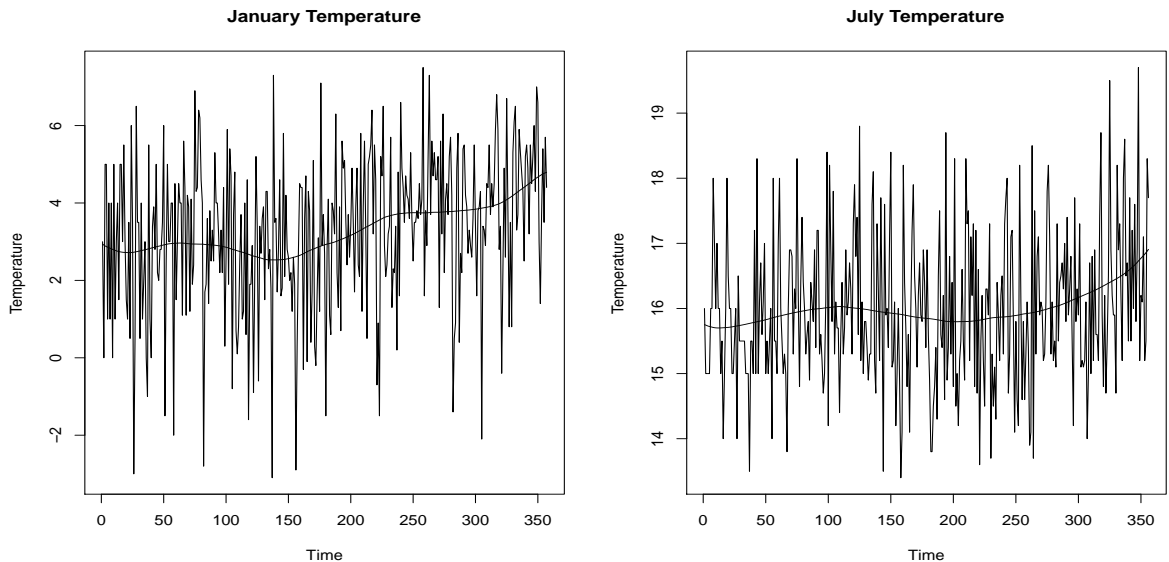}
 \caption{\it Temperature in $^{\circ}$C of UK from 1659--2015 in January (left panel) and  July (right panel). The lines are the fitted trends of the means by local linear  regression.}
  \label{fig:mini:subfig:fige04}
\end{figure}

\def\theequation{6.\arabic{equation}}
\setcounter{equation}{0}
Various scientific data show that since the late 1800s, the mean global temperature starts to increase significantly. For example, during 1906--2005, the Earth's average surface temperature rose by $0.74\pm0.18 $ $^{\circ}$C, with the rate of warming also increasing with time. On the other hand, however, there are much fewer studies on possible changes in the second order characteristics (especially the correlations) of temperature time series, both globally and regionally.
In this section, we are interested in identifying possible changes in the second order structures of regional temperature in the recent three centuries. To this end, we analyse the Hadley Centre Central England Temperature (HadCET) data from 1659--2015. These data can be downloaded from http://www.metoffice.gov.uk/hadobs/hadcet/. We present the analysis results of monthly temperature series in January and July as representatives of the winter and summer monthly temperature patterns in central England. The time series are shown in Figure \ref{fig:mini:subfig:fige04}. There are apparent increasing trends in both time series.

We first test the constancy in the variance and lag $1$-correlation for the January data. In our analysis  the critical values are generated by $8000$ bootstrap replications. The bandwidths were chosen as described in Section \ref{sec5}. The results are summarized in Table \ref{tab:dataana}.  The test \eqref{testvar} rejects  the null hypothesis of no change point in the variance at $5  \%$ level.  We then use the statistic \eqref{estvar} to estimate the change point  and obtain $\tilde t_n=226$, which corresponds to the year of $1884$. Next we apply the test \eqref{testvar} again to the periods before and after the identified change point and conclude that there are no further structural breaks   in the  variance during the two periods. The estimates of the variance before and after the year $1884$  are given by
 $4.05$ and $2.85$, respectively. Next we apply \eqref{testcorr}  to testing the constancy in the lag $1$-correlation, where we use the statistic \eqref{tstar} to estimate the change points in the variance with $\zeta=0.14$ and $L=38$. We identify $t^*_n=242$, which corresponds to the year of $1900$. The result is close to the one which is obtained by the estimator \eqref{estvar}. The null hypothesis of no change points in the lag $1$-correlation is rejected at $5 \%$ level [see Table \ref{tab:dataana}]. Next we use the statistic \eqref{changeestcorr} to identify the location of the change point of the first order correlation and obtain $\hat t_n=213$, which corresponds to the year of $1871$. Again we investigate the existence of further changes in the    lag $1$-correlation before and after the  year $1871$  and conclude that there are no further structural breaks in the lag $1$-correlation during the two periods. The estimates of the lag $1$-correlation before and after the break point
  are equal to $-0.108$ and $0.231$, respectively.

\begin{figure}[htbp]
\centering
  \includegraphics[width=16cm,height=8cm]{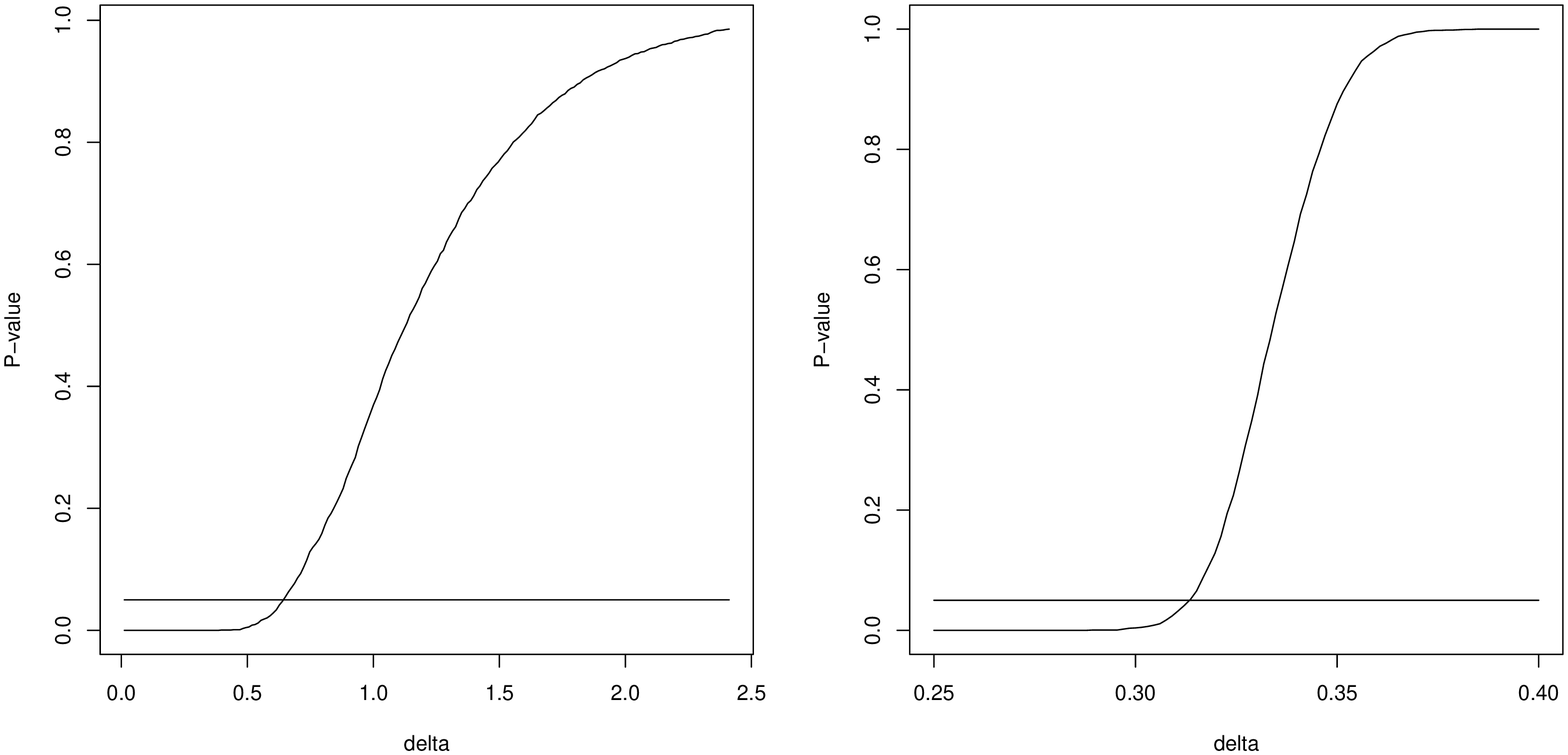}
 \caption{\it  p-values of the bootstrap test for a relevant change   in the  variance (left panel) and lag $1$-correlation
 (right panel)    for different values of the threshold $\delta$. The horizontal line marks the significance
level $0.05$.}
  \label{fig:mini:subfig:fige05}
\end{figure}

The   results from Section \ref{sec4} enable us to perform tests for relevant changes   in the variance and lag $1$-correlation for the January data. Figure \ref{fig:mini:subfig:fige05} displays the $p$-values of the tests for a relevant change in the variance and log $1$-correlation for  different values of the threshold $\delta$. At the 5\% significance level, we conclude that there exists a relevant change with size $\delta=0.645$ in the  variance  and a relevant change with size $\delta= 0.313$ in the lag $1$-correlation.

For comparison, we also analyse the July data. For the variance,  we choose the bandwidths  $b_n=0.205$ and $m=33$. The test statistic \eqref{hatTn} is 1.90, together with the simulated $90\%$ critical value 1.83 and $95\%$ critical value 2.04. Hence we cannot reject the null hypothesis of no change point in the variance at the $5\%$ significance level. For the correlation, the bandwidths are chosen as $b_n=0.26$, $c_n=0.06$ and $m=25$. The test statistic  \eqref{deftnc} is 0.84, and the simulated $90\%$ critical value is 1.11, and the $95\%$ critical value is 1.25. Hence, again we cannot reject the null hypothesis of no change points in lag $1$-correlation at the $10\%$ significance level.

In conclusion, our data analysis suggests that besides the mean trend, there exists strong evidence  indicating structural changes in the second order structures of monthly temperatures in central England. Further, the latter changes are inhomogeneous among different seasons, in the sense that changes in the variance and correlation are more significant in the winter than in the summer. This implies that winter temperatures in central England have become more unstable and more difficult to predict since the late 19th century.  Finally, we also locate the change point in the second order structure of the HadCET temperate data through all three ways we proposed in our paper.
The three change points, $1871$, $1884$ and $1900$, are quite close. Our findings suggest that the time of change in the second order structure of our data coincides with that of the mean global temperature identified in various previous studies.

\begin{table}[htbp]
  \centering
  \caption{{\it Tests for the existence of a change point in the variance and lag $1$-correlation in the HadCET data.
   $v^*_\alpha$ denotes the critical values obtained by the bootstrap procedure.
   ``Whole'' represents the whole period, ``Before'' and ``After'' represent the period before and after the detected change point, respectively.}}
    \begin{tabular}{|c|c|c|c|c|c|c|}
    \hline
          & \multicolumn{3}{|c|}{Variance} & \multicolumn{3}{|c|}{lag $1$-Correlation} \\
    \hline
       &Whole\  \ \ & Before\ \ \  & After\ \ \  & Whole\ \ \  & Before\ \ \  & After \ \ \ \\
    \hline
    Test Stat. & 5.29**  & 2.82  & 3.34  & 1.53**  & 0.64  & 0.71 \\
    $v^*_{90\%}$   & 4.56  & 4.66  & 5.1   & 1.31  & 0.76  & 0.95 \\
    $v^*_{95\%}$ & 5.11  & 5.22  & 5.67  & 1.45  & 0.85  & 1.07 \\
    $b_n$     & 0.155 & 0.26  & 0.26  & 0.23  & 0.19  & 0.21 \\
    $m$     & 40    & 30    & 18    & 19    & 32    & 11 \\
    $c_n$     & --   & --    & --    & 0.05    & 0.06    & 0.059 \\
    \hline
    \end{tabular}%
  \label{tab:dataana}%
\end{table}%

\ \\
{\bf Acknowledgements.}
The work of H. Dette has been supported in part by the
Collaborative Research Center ``Statistical modeling of nonlinear
dynamic processes'' (SFB 823, Teilprojekt A1, C1) of the German Research Foundation (DFG). Z. Zhou's research has been supported in part by NSERC of Canada.
The authors would like to thank
Martina Stein who typed this manuscript with considerable technical
expertise.

\bigskip
 \bibliographystyle{apalike}




\section{Proofs of main results}  \label{proofs}
\def\theequation{7.\arabic{equation}}
\setcounter{equation}{0}

In this section we provide proofs
of the main results, where some of the technical details are deferred to an online supplement.
Throughout this section  the symbol $\Rightarrow$ denotes weak convergence of a
stochastic process in $\mathcal{C}(0,1)$ with the uniform topology.

\subsection{Proof of Theorem   \ref{theorem1},  \ref{theorem2}, Lemma \ref{tstar0}, Theorem   \ref{theorem5} and  \ref{theorem5boot} } \label{sec61}

\subsubsection{Proof of Theorem \ref{theorem1}} \label{prooftheorem1}

In order to study the asymptotic properties of the statistic $\hat T_n$ we introduce
the random variable
\begin{align*}
T_n=\max_{1 \leq i\leq n}\Big|S_i-\frac{i}{n}S_{n}\Big|,
\end{align*}
where $S_i$ is the $i$th partial sum of the PLS $\{e_i^2\}_{i=1}^n$, that is
$S_i=\sum_{j=1}^ie_j^2$.
It   follows from the results of \cite{zhou2013}
that  $\frac{1}{\sqrt{n}} T_n \cod \sup_{t \in (0,1)} | U_1(t)-t U_1(1)|$,
 where $\{U_1(t)\}_{t \in [0,1]}$ is a centered Gaussian process with covariance kernel \eqref{cov1}.

  We will show below  that for   the maximum deviation between $\hat{S}_i$ and $S_i$ satisfies
\begin{equation}
\label{maxdevvar}
\max_{1 \leq i\leq n}|S_i-\hat{S}_i|=  O_p(\sqrt{nb_n} + nb_n^3 + b_n^{-1}).
\end{equation}
 From  \eqref{maxdevvar} we obtain
 $ T_n-\hat{T}_n=  O_p(\sqrt{nb_n} + nb_n^3 + b_n^{-1})=o(\sqrt{n})$,
where the last estimate follows from our  choice of the bandwidth $b_n$. This
 yields the assertion of Theorem \ref{theorem1}.

\medskip
For a proof of  the remaining estimate \eqref{maxdevvar} we use the decomposition
 \begin{align} \label{decomp}
 {S}_i-\hat S_i =A_{n,i}+B_{n,i},
\end{align}
where the quantities $A_{n,i}$ and $B_{n,i}$ are defined by
\begin{align*}
A_{n,i}=2\sum_{j=1}^ie_j\big( \hat{\mu}_{b_n} (t_j)-\mu(t_j)\big), \quad B_{n,i}=\sum_{j=1}^i\big(\mu(t_j)-\hat{\mu}_{b_n}(t_j)\big)^2.
\end{align*}
Observing the estimate (\ref{revision2-34}) in Section \ref{proofstechnical1} of the technical appendix we have  $0\leq B_{n,i}\leq B_{n,n}=O_p(b_n^{-1}+nb_n^4)$ for all $1\leq i\leq n$, which implies
\begin{align}\label{Bni}
\max_{1\leq i\leq n}B_{n,i}=O_p(b_n^{-1}+nb_n^4).
\end{align}
By Lemma \ref{mu} (which is proved in Section \ref{proofstechnical1})
 it follows that
\begin{align*}
&\max_{{ \lf nb_n\rf}\leq i\leq n-\lf nb_n\rf}\Big|A_{n,i}-2\sum_{j=\lf nb_n\rf+1}^i a_{n,j} -2\sum_{j=1}^{\lf nb_n\rf}e_j( \hat{\mu}_{b_n}(t_j)-\mu(t_j))\Big|=O_p(n\chi_n), \\
& \max_{n-\lf nb_n\rf+1\leq i\leq n}\Big|A_{n,i}-2\sum_{j=\lf nb_n\rf}^{n-\lf nb_n\rf} a_{n,j} -2\sum_{j=1}^{\lf nb_n\rf-1}e_j(\hat{\mu}_{b_n}(t_j)-\mu(t_j)) \\
& \qquad \qquad \qquad \qquad \qquad \qquad  - 2\sum_{j=n-\lf nb_n\rf+1}^{{ i}}e_j(\hat{\mu}_{b_n}(t_j)-\mu(t_j))\Big|=O_p(n\chi_n),\notag
\end{align*}
where  $\chi_n=b_n^3+\frac{b_n}{n}$, and
\begin{align}\label{ajn}
a_{n,j} = \frac {e_j}{nb_n} \sum^n_{s=1} K_{b_n} \big( \frac {s-j}{n}\big) e_s \qquad (j=1,\dots,n).
\end{align}
A further application of the estimate (\ref{revision2-34}) in Section \ref{proofstechnical1} and the Cauchy-Schwarz inequality gives
\begin{eqnarray*}
\Big \|\max_{1\leq j\leq \lf nb_n\rf}\Big|\sum_{i=1}^j e_j\big(\mu(t_j)-\hat{\mu}_{b_n}(t_j)\big)\Big| \Big \|_2 &\leq& \sum_{i=1}^{\lf nb_n\rf}\|e_j\|_4\|\mu(t_j)-\hat{\mu}_{b_n}(t_j)\|_4 \\
&=& O(\sqrt{nb_n}+nb_n^3), \\
\Big \|\max_{n-\lf nb_n\rf+1\leq j\leq n}\Big|\sum_{i={ n-\lf nb_n\rf+1}}^je_j\big(\mu(t_j)-\hat{\mu}_{b_n}(t_j)\big)\Big| \Big\|_2 &=&O(\sqrt{nb_n}+nb_n^3).
\end{eqnarray*}
This implies that
\begin{equation} \label{Ani}
\max_{1 \leq i \leq n} |A_{n,i}| \leq \max_{\lfloor nb_n\rfloor  \leq i \leq n - \lfloor nb_n \rfloor} |\bar A_{n,i}| + O_p (\sqrt{nb_n} + nb^3_n),
\end{equation}
where $\bar A_{n,i} = 2 \sum^i_{j=\lfloor nb_n \rfloor} a_{n,j}$ and $a_{n,j}$ is defined in \eqref {ajn}.

In the following we derive an estimate for the first term on the right-hand side of \eqref{Ani}. For this purpose we consider  the random variables  $\tilde{e}_{s,m}=\E(e_s|\varepsilon_{s},...,\varepsilon_{s-m})$ and note that the sequence $(\tilde e_{s,m})^n_{s=1}$ is $m$-dependent.  Now  define $a^{(m)}_{n,j}=e_j\sum_{s=1}^nK_{b_n}\big(\frac{s-j}{n}\big)\tilde{e}_{s,m}/(nb_n)$ and
  \begin{align*}
\bar{A}^{(m)}_{n,i}=2\sum_{j=\lf nb_n\rf}^ia^{(m)}_{n,j},
\end{align*}
then a similar argument as given in the proof of Theorem 1 of \cite{zhou2014}    shows that
\begin{align*}
\max_{1\leq j\leq n}\Big\|\sum_{s=1}^nK_{b_n}\big(\frac{s-j}{n}\big)(\tilde{e}_{s,m}-e_s)\Big\|_4\leq C\sqrt{nb_n}m\chi^m
\end{align*}
for some constant $\chi \in (0,1)$.
  By the  Cauchy-Schwartz inequality it follows that
     \begin{eqnarray} \label{eq43}\nonumber
\Big\|\max_{\lf nb_n\rf \leq i\leq n-\lf nb_n\rf}|\bar{A}_{n,i}-\bar{A}^{(m)}_{n,i}|\Big\|_2&\leq&\Big\| \frac{2}{nb_n}\sum_{j=\lf nb_n\rf}^{n-\lf nb_n\rf}|e_j| \sum_{s=1}^nK_{b_n}\big(\frac{s-j}{n}\big)(\tilde{e}_{s,m}-e_s)|\Big\|_2 \\ &=&O(\sqrt{n}m\chi^mb_n^{-1/2}).
\end{eqnarray}
With the notations $\tilde{a}^{(m)}_{n,j}=\tilde{e}_{j,m}\sum_{s=1}^nK_{b_n}(\frac{s-j}{n})\tilde{e}_{s,m}/(nb_n)$ and
$\tilde{A}^{(m)}_{n,i}=2\sum_{j=\lf nb_n\rf}^i\tilde{a}^{(m)}_{n,j}$
 it is easy to see that \begin{align} \label{615}
 \Big \|\max_{\lf nb_n\rf\leq i\leq n-\lf nb_n\rf}|\tilde{A}^{(m)}_{n,i}-\bar{A}^{(m)}_{n,i}|\Big \|_2\leq \frac{2}{nb_n}\sum_{j=1}^n \|e_j-\tilde{e}_{j,m} \|_4\Big\|\sum_{s=1}^nK_{b_n}(\frac{s-j}{n})\tilde{e}_{s,m}\Big\|_4.
\end{align}
Now an elementary calculation via Burkholder's inequality shows
\begin{align*}
 \max_{1 \leq j \leq n} \Big\|\frac {1}{nb_n}\sum_{s=1}^nK_{b_n}\Big(\frac{s-j}{n}\Big)\tilde{e}_{s,m} \Big\|_4=O\Big(\frac{1}{\sqrt{nb_n}}\Big),
\end{align*}
and by a similar argument as given in the proof of Theorem 1 of \cite{zhou2014}   we have   for some constant $\chi \in (0,1)$ the estimate
$\max_{1\leq j\leq n}\|\tilde{e}_{j,m}-e_j\|_4=O(\chi^m)$.
 This  gives for the left-hand side of \eqref{615} \begin{align*}
\Big\|\max_{\lf nb_n\rf\leq i\leq n-\lf nb_n\rf}|\bar{A}^{(m)}_{n,i}-\tilde{A}^{(m)}_{n,i}|\Big\|_2=O(\sqrt{n/b_n}\chi^m),
\end{align*}
and an application of \eqref{eq43} yields
\begin{align}\label{1.26}
\Big\|\max_{\lf nb_n\rf\leq i\leq n-\lf nb_n\rf}|\bar{A}_{n,i}-\tilde{A}^{(m)}_{n,i}|\Big\|_2=O(\sqrt{n/b_n}m\chi^m).
\end{align}
A tedious but straightforward calculation shows  that  $\pp_{j-l}(\tilde{e}_{j,m}\tilde{e}_{i,m})=0$ for $l>2m$.
For example,  if  $i\geq j-m$, then  by definition,
$\tilde{e}_{j,m}\tilde{e}_{i,m}$ is $\sigma (\varepsilon_{j-2m},\varepsilon_{j-2m+1},...,\varepsilon_{i})$ measurable.
Consequently,  $\E(\tilde{e}_{j,m}\tilde{e}_{i,m}|\FF_{j-l})=\E(\tilde{e}_{j,m}\tilde{e}_{i,m}|\FF_{j-l-1})=\E(\tilde{e}_{j,m}\tilde{e}_{i,m})$ if
$l>2m$, which gives $\pp_{j-l}(\tilde{e}_{j,m}\tilde{e}_{i,m})=0$. The other cases  $i\leq j-l-1$ and
 $ j-l\leq i\leq j-m-1$ are treated similarly, and details are omitted for the sake of brevity.
 Observing $\pp_{j-l}(\tilde{e}_{j,m}\tilde{e}_{i,m})=0$ for $l>2m$ we obtain
\begin{align}\label{tilde{A}}
\Big \| \max_{\lf nb_n\rf\leq i\leq n-\lf nb_n\rf} |\tilde{A}^{(m)}_{n,i}-\E\tilde{A}^{(m)}_{n,i}|\Big \|_2 \leq {2} \sum_{l=0}^{2m}\Big\| \max_{\lf nb_n\rf\leq i\leq n-\lf nb_n\rf} |\sum_{j=\lf nb_n\rf}^i\pp_{j-l}\tilde{a}^{(m)}_{n,j}|\Big\|_{2}.
\end{align}
Similar arguments as given in the proof of Theorem 1 in \cite{wu2005}
show
\begin{align*}
&\|\pp_{j-l}\tilde{a}^{(m)}_{n,j}\|_{2} \leq \frac{M}{n}\Big \| \tilde{e}_{j,m} \sum_{s=1}^n\tilde{e}_{s,m}  K_{b_n}\Big(\frac{s-j}{n}\Big)-\tilde{e}^{(j-l)}_{j,m} \sum_{s=1}^n\tilde{e}^{(j-l)}_{s,m}  K_{b_n}\Big(\frac{s-j}{n}\Big) \Big\|_{2},
\end{align*}
and by the triangle inequality it follows that
\begin{align*}
&\|\pp_{j-l}\tilde{a}^{(m)}_{n,j}\|_{2} \leq M(Z_{1,j} +Z_{2,j}),
\end{align*}
where the terms $Z_{1,j}$ and $Z_{2,j}$ are defined by
\begin{align*}
Z_{1,j}&= \frac{1}{nb_n}\Big \| \tilde{e}_{j,m} \sum_{s=1}^n K_{b_n}\Big(\frac{s-j}{n}\Big)\big[\tilde{e}_{s,m}^{(j-l)}-\tilde{e}_{s,m}\big]   \Big\|_{2}, \\
Z_{2,j}&= \frac{1}{nb_n} \Big\| \big[\tilde{e}^{(j-l)}_{j,m}-\tilde{e}_{j,m}\big] \sum_{s=1}^nK_{b_n}\Big(\frac{s-j}{n}\Big)\tilde{e}^{(j-l)}_{s,m}  \Big\|_{2},
\end{align*}
$\tilde{e}_{s,m}^{(j)}=\E(e_s^{(j)}|\varepsilon_{s-m}, \ldots ,\varepsilon_j', \ldots  ,\varepsilon_s)$  for $s-m\leq j\leq s$,   $e_s^{(j)}=G_l(t_s,\FF_s^{(j)})$  for $b_l<t_s\leq b_{l+1}$  and we use the convention $\tilde{e}_{s,m}^{(j)}=\tilde{e}_{s,m}$
 for $j<s-m$ or $j>s$.
 Elementary calculations show  that for  $l\geq 0$
\begin{align*}
\Big\|\sum_{s=1}^nK_{b_n}(\frac{s-j}{n})\tilde{e}^{(j-l)}_{s,m}\Big\|_{4}=O (\sqrt{nb_n}) , \qquad 1 \leq j \leq n,
\end{align*}
 while by definition $\|\tilde{e}_{j,m}^{(j-l)}-\tilde{e}_{j,m}^{}\|_4=0$ for  $l> m$. On the other hand, if $1\leq j\leq n$, $0\leq l\leq m$, we have by Assumption (A4)
 \begin{align*}
\|\tilde{e}_{j,m}^{(j-l)}-\tilde{e}_{j,m}^{}\|_4=\big\|\E(e_j-e_j^{(j-l)}|\varepsilon_{j-m},...,\varepsilon_{j-l}, \varepsilon_{j-l}',\varepsilon_j)\big\|_4\leq M\chi^l,
\end{align*}
which gives  $Z_{2,j}=O( \frac{\chi^l}{\sqrt{nb_n}})$. Observing that $\tilde{e}_{s,m}^{(j-l)}-\tilde{e}_{s,m}=0$ if $s\geq j-l+m+1$ or $s\leq j-l-1$,
  it is easy to see that $Z_{1,j}=O (\frac{m}{nb_n})$.  {It now follows from  Doob's inequality
  \begin{align*}
 \Big\| \max_{\lf nb_n\rf\leq i\leq n-\lf nb_n\rf}|\sum_{j=\lf nb_n\rf}^i\pp_{j-l}\tilde{a}^{(m)}_{n,j}|\Big\|_2=O\Big(\sqrt{n}\Big(\frac{\chi^l}{\sqrt{nb_n}}+\frac{m}{nb_n}\Big)\Big),
  \end{align*}}
and we obtain  from (\ref{tilde{A}}) that
\begin{align} \label{equa68}
\Big\|\max_{\lf nb_n\rf\leq i\leq n-\lf nb_n\rf}|\tilde{A}^{(m)}_{n,i}-\E\tilde{A}^{(m)}_{n,i}|\Big\|_{2}=O\Big(\frac{m^2}{n^{1/2}b_n}+(b_n)^{-1/2}\Big).
\end{align}
Finally, similar arguments as given in the proof of Lemma 5 in \cite{zhouwu2010} show
$$
\max_{\lf nb_n\rf\leq i\leq n-\lf nb_n\rf} \E[\tilde{A}_{i,m}]=O\Big(\sum_{i=1}^n\sum_{j=1}^n\chi^{|i-j|}/(nb_n)\Big)=O(b_n^{-1}).
$$
Observing \eqref{Ani}, \eqref{1.26}
and \eqref{equa68} and taking $m=M \log n$ for a sufficiently large constant $M>0$ yields
$
\max_{1 \leq i \leq n} |A_{n,i}| =  O_p (\sqrt{nb_n} + nb_n^3 +b_n^{-1}).
$
Consequently, the assertion   \eqref{maxdevvar} follows from \eqref{decomp}, \eqref{Bni} and this estimate.

 \subsubsection{Proof of Theorem  \ref{theorem2}}  \label{prooftheorem2}
We recall the definition of $\hat \Phi_{i,n}$ in \eqref{new.83} and define on the interval $[0,1]$   the linear interpolation
\begin{align}\label{new.84}
\hat{\tilde{\Phi}}_{m,n}(t)=\hat \Phi_{\lfloor nt \rfloor,m}+(nt-\lfloor nt \rfloor) (\hat \Phi_{\lfloor nt \rfloor +1,m}-\hat \Phi_{\lfloor nt \rfloor,m}).
\end{align}
The assertion follows if the weak convergence
$$
\{\hat{\tilde{\Phi}}_{m,n}(t)\}_{t \in [0,1]} \Rightarrow \{ U_1(t) \}_{t \in [0,1]}
$$
  conditional on $\FF_n$ can be established.
 For a proof of this statement define
 ${\Phi}_{i,m}$ and  $\tilde{\Phi}_{m,n}(t)$   by replacing the nonparametric residuals $\{\hat{e}_i\}_{i=1}^n$  by the (non-observable) errors  $\{e_i\}_{i=1}^n$  in the definition \eqref{new.83} and \eqref{new.84} of
${\hat \Phi}_{i,m}$ and  ${\hat{\tilde \Phi}}_{m,n}(t)$, respectively. Note that similar arguments as given in the proof of Theorem 3 in \cite{zhou2013} show that \{$\tilde \Phi_{m,n}(t)\}_{t \in [0,1]} \Rightarrow \{U_1(t)\}_{t \in [0,1]}$.  The assertion of Theorem \ref{theorem2} then follows from the estimate
\begin{align} \label{estthm2}
\sup_{t\in [0,1]}\big|\tilde{\Phi}_{m,n}(t)-\hat{\tilde{\Phi}}_{m,n}(t)\big|=O_p\Bigl( \Bigl(\frac{m\log^2n}{nb_n^{3/2}}\Bigr)^{1/2}+\sqrt{m}b_n^2\log n\Bigr).
\end{align}
In order to prove \eqref{estthm2} let  $C$ denote a sufficiently large  constant, which may vary from line to line in the following calculations, and consider the event
 $$A_n=\Big\{\sup_{t\in[0,1]}|\hat{\mu}_{b_n}(t)-\mu(t)|\leq C\Big(\frac{\log n}{\sqrt{nb_n}b_n^{1/4}}+b_n^2\log n\Big)\Big\}.$$
 By Lemma \ref{maxdevvar1} of  Section \ref{proofstechnical1}  in the technical appendix we have that
$\lim_{n\rightarrow \infty}\p(A_n)=1$. This yields for $1\leq j\leq n-m+1$ the estimate \begin{align*}
\E[(S_{j,m}-\hat{S}_{j,m})^2I(A_n)] & \leq \E\Big[\sum_{r=j}^{j+m-1}(e_r-\hat{e}_r)^2\sum_{r=j}^{j+m-1}(2e_r+\hat{e}_r-e_r)^2I(A_n)\Bigr]
\notag\\
& \leq C\Big(\frac{m^2\log^2n}{nb_n^{3/2}}+m^2b_n^4\log^2n\Big).
\end{align*}
Similarly, it follows that
$\E [(S_{n}-\hat{S}_{n})^2\frac{m^2}{n^2}I(A_n) ]\leq C (\frac{m^2\log^2n}{nb_n^{3/2}}+m^2b_n^4\log^2n )$,
which gives
 \begin{align*}
\|({\Phi}_{n-m+1}-\hat{\Phi}_{n-m+1})I(A_n)\|^2_2&= \frac{1}{m(n-m+1)} \sum_{j=1}^{n-m+1}\E\Big[\big(S_{j,m}-\hat{S}_{j,m}-\frac{m}{n}(S_{n}-\hat{S}_{n})\big)^2 I(A_n)\Big]\notag\\
&\leq  C\Big(\frac{m\log^2n}{nb_n^{3/2}}+mb_n^4\log^2n\Big).
\end{align*}
An application of Doob's inequality and Proposition \ref{prop01} in Section \ref{proofstechnical4}   finally yields
\begin{align*}
\max_{1\leq i\leq n-m+1}|\Phi_{i,m}-\hat{\Phi}_{i,m}|=O_p\Big(\Big(\frac{m\log^2n}{nb_n^{3/2}}\Big)^{1/2}+\sqrt{m}b_n^2\log n\Big).
\end{align*}
The estimate \eqref{estthm2} now follows from  this result  and  definition (\ref{new.84}), which completes the proof of Theorem \ref{theorem2}.

\subsubsection{Proof of Lemma \ref{tstar0}}
 Define $\mathcal N(i)=\frac {1}{L}\big(\sum_{j=i-L+1}^i e_j^2-\sum_{j=i}^{i+L-1} e_j^2\big)$
 and recall the definition of $\mathcal M(i)$  in \eqref{mi}. By similar arguments as given in the proof of Lemma \ref{maxdevvar1}   in the technical appendix (note that $\iota>8$) we have
$
 \|\mathcal M(i)-\mathcal N(i)\|_4=b_n^2+\frac{1}{\sqrt{nb_n}}
$,
and Proposition \ref{prop:2.1} yields
 \begin{align}\label{4.13}
 \max_{L\leq i\leq n-L+1}|\mathcal M(i)-\mathcal N(i)|=O_p\Big(n^{1/4}b_n^2+{\frac{1}{n^{1/4}b_n^{1/2}}}\Big).
 \end{align}
 Consider the case that $i\in B:=\{i:|t_i-\tilde{t}_v|>L\}$. Then by our assumption on the  variance function, there exists a large constant $C$, such that
 $
 |  \E\mathcal N(i) | \leq CL/n
$
for $L\leq i\leq n-L+1$, $i\in B$.
 By    Lemma \ref{maxdevvar1} and
 Lemma \ref{hatsigma} in the technical appendix it now follows
 $\|\mathcal N(i)-\E \mathcal N(i)\|_{\iota/2}\leq CL^{-1/2}$
 ($L\leq i\leq n-L+1$, $i\in B$),
 which gives
\begin{align*}
\max_{L\leq i\leq n-L+1,i\in B} |\mathcal{N}(i)|=O_p(L^{-1/2}n^{2/\iota}+L/n).
\end{align*}
Combining this estimate  with \eqref{4.13} yields
\begin{align*}
\max_{L\leq i\leq n-L+1,i\in B} |\mathcal{M}(i)|=O_p\Big(L^{-1/2}n^{2/\iota}+L/n+n^{1/4}b_n^2+{ \frac{1}{n^{1/4}b_n^{1/2}}}\Big).
\end{align*}
Similarly, we can show that
$
 \mathcal{M}(\lf n\tilde{t}_v\rf)=\sigma(t_{v}^+)-\sigma(t_{v}^-)+O_p\big(n^{1/4}b_n^2+{ \frac{1}{n^{1/4}b_n^{1/2}}}+L^{-1/2}+{ L/n}\big).
$
The choice of $L$  implies that
\begin{align*}
\lim_{n\rightarrow \infty}\p\Big(|\mathcal{M}(\lf n\tilde{t}_v\rf)|>\max_{L\leq i\leq n-L+1,i\in B} |\mathcal{M}(i)|\Big)=1,
\end{align*}
which completes the proof of Lemma \ref{tstar0}.\hfill $\Box$\\

\subsubsection{Proof of Theorem \ref{theorem5}}

We restrict ourselves to the case of a variance function with no change point and the corresponding estimator
\eqref{defhatmu1}. The statement for the estimator \eqref{varest1} follows by similar arguments as given in Theorem \ref{thm6},
where we deal with the problem of testing
for  relevant changes in the  correlation.  \\
Recall the definition   \eqref{70a}, define $S_i^W = \sum_{j=1}^i W_j^k$ as the corresponding partial
sum and  consider the CUSUM statistic
$
 T^c_n=\max_{1\leq i\leq n} |S_i^W-\frac{i}{n}S_{n}^W | ,
$
 We will show the estimate
 \begin{equation}\label{corr}
  \max_{1\leq i\leq n}| {\hat{S}_i^W-{S}_i^W}|=O_p({ nc_n^2+nb_n^3c_n^{-1/4}+b_n^{-1}c_n^{-1}}),
  \end{equation}
  which implies
   ${\hat{T}^c_n-{T}^c_n}= O_p({ nc_n^2+nb_n^3c_n^{-1/4}+b_n^{-1}c_n^{-1}})$.
 It follows from \cite{zhou2013} that ${T}^c_n/\sqrt{n}$ converges weakly to the distribution
of the random variable ${\cal K}_2$ defined in Theorem \ref{theorem5}. By our choice of  the bandwidth
$b_n$ we have   ${ nc_n^2+nb_n^3c_n^{-1/4}+b_n^{-1}c_n^{-1}}=o(\sqrt{n})$, and the assertion of Theorem  \ref{theorem5}
follows.

For the sake of simplicity we omit in  the subscripts $c_n,b_n$  in the variance estimator $\hat \sigma_{c_n,b_n}$ and the superscript $k$ in the definition $\hat W^k_i, W^k_i$ the proof of the estimate \eqref{corr}.
With the notation $\tilde{W}_i=\frac{e_ie_{i+k}}{\sigma^2(t_i)}$  we obtain
  \begin{align} \label{78}
\max_{1 \leq j \leq n} \Big| \sum_{i=1}^j \left(W_i-\tilde{W}_i\right)  \Big | \leq \max_{1 \leq j \leq n}\sum_{i=1}^j\frac{|e_ie_{i+k}|\cdot | \sigma(t_i)-\sigma(t_{i+k} )|}{\sigma^2(t_i)\sigma(t_{i+k})} = O_p(1),
\end{align}
where we have used the fact that the variance function is Lipschitz continuous.
 Let $\bar{W}_i=\frac{\hat{e}_i\hat{e}_{i+k}}{{\sigma^2}(t_i)}$  denote the analogue of $\hat W_i$,
 where the estimate $\hat \sigma^2(t_i)$ has been replaced by the ``true'' variance   $\sigma^2(t_i)$.
 By a careful inspection of the proof of Theorem \ref{theorem1}, it can be seen that
 \begin{align}\label{revision2-93}
 \max_{1\leq j\leq n}\Big|{ \sum_{i=1}^j}\left(\bar{W}_i-\tilde{W}_i\right)\Big|=O_p (\sqrt{nb_n} + nb_n^3 + b_n^{-1}).
 \end{align}    Define
\begin{align*}
\Lambda_j:=\sum_{i=1}^j(\hat{W}_i-\bar{W}_i)=\sum_{i=1}^j\frac{\hat{e}_i\hat{e}_{i+k}(-\hat{\sigma}^2 (t_i)+\sigma^2(t_i))}{\hat{\sigma}^2 (t_i)\sigma^2(t_i)},
\end{align*}
then our next goal is to estimate $\max_{1\leq j\leq n}|\Lambda_j|$.
For this purpose we  consider the random variable
\begin{align*}
\bar{\Lambda}_j:=\sum_{i=1}^j\frac{\hat{e}_i\hat{e}_{i+k}(-\hat{\sigma}^2 (t_i)+\sigma^2(t_i))}{\sigma^4(t_i)}
\end{align*}
(here  the estimator in the denominator has been  replaced by the true variance function), and obtain
\begin{align}\label{tildeapprox}
\max_{1\leq j\leq n}|\Lambda_j-\bar{\Lambda}_j|\leq \sum_{i=1}^{n}\frac{|\hat{e}_i\hat{e}_{i+k}|(\hat{\sigma}^2 (t_i)-\sigma^2(t_i))^2}{\hat{\sigma}^2 (t_i)\sigma^4(t_i)}.
\end{align}
For the expectation of the right-hand side it follows
\begin{align}\label{eq94}
\E\Big[\sum_{i=1}^{n}\frac{|\hat{e}_i\hat{e}_{i+k}|(\hat{\sigma}^2(t_i)-\sigma^2(t_i))^2}{\hat{\sigma}^2 (t_i)\sigma^4(t_i)}\Big]\leq C \sum_{i=1}^n\|\hat{e}_i\|_4\|\hat{e}_{i+k}\|_4\|(\hat{\sigma}^2 (t_i)-\sigma^2(t_i))^2\|_2.
\end{align}
By   Lemma  {\ref{maxdevvar1} of  Section \ref{proofstechnical1}  in the technical appendix we have that
\begin{equation}\label{normmu}
\|\hat{\mu}_{b_n}(t)-\mu(t)\|_4=O\Big(b_n^2+\frac{1}{\sqrt{nb_n}}\Big),
\end{equation}
which implies
$\|\hat{e}_i\|_4\leq C$.
On the other hand, Corollary \ref{corosigma1} in Section \ref{proofstechnical1} shows
\begin{align}\label{revision2-100}
\max_{1\leq i\leq n}\|(\hat{\sigma}^2 (t_i)-\sigma^2(t_i))^2\|_2=O\Big(b_n^4+\frac{1}{nb_n}+c_n^4+\frac{1}{nc_n}\Big),
\end{align} and we obtain   from \eqref{tildeapprox}, \eqref{eq94} and Proposition \ref{prop:2.1} in Section \ref{proofstechnical4} the estimate
\begin{align}\label{63}
\max_{1 \leq j \leq n}|\Lambda_j| \leq \max_{1 \leq j \leq n}|\bar\Lambda_j| +\max_{1\leq j\leq n }| \Lambda_j-\bar \Lambda_j|=
\max_{1 \leq j \leq n}|\bar \Lambda_j| + O_p(nb_n^4+b_n^{-1}+nc_n^4+c_n^{-1}).
\end{align}
Now the remaining problem  is to  derive an appropriate estimate for the quantity $\max_{1\leq j\leq n}|\bar\Lambda_j|$.
For this purpose note that $\bar \Lambda_j =   \bar \lambda_{j,1} + \bar \lambda_{j,2}$, where
 \begin{eqnarray*}
\bar{\lambda}_{j,1}&=&\sum_{i=1}^j\frac{(\hat{e}_i\hat{e}_{i+k}-e_ie_{i+k})(\sigma^2(t_i)-\hat{\sigma}^2 (t_i))}{\sigma^4(t_i)},\\
\bar{\lambda}_{j,2}&=&\sum_{i=1}^j\frac{ e_ie_{i+k} (\sigma^2(t_i)-\hat{\sigma}^2 (t_i))}{\sigma^4(t_i)}.
\end{eqnarray*}
By   Lemma \ref{mu}, Corollary \ref{corosigma1}  of Section \ref{proofstechnical1}  and the estimate  (\ref{normmu})  it is easy to see that
\begin{eqnarray}\label{lambda-j1}
\E\Big[\max_{1 \leq j\leq n}|\bar{\lambda}_{j,1}|\Big] &\leq &\sum_{i=1}^{n}\frac{  \|\hat{e}_i\hat{e}_{i+k}-e_ie_{i+k}\|_2}{\sigma^4(t_i)}
\|\sigma^2(t_i)-\hat{\sigma}^2 (t_i)\|_2=O(\underline \pi_n), \\
\label{lambda-j3}
 \E\Big[\max_{1 \leq j\leq \lf nb_n+nc_n\rf}|\bar{\lambda}_{j,2}|\Big]
 &\leq  &\sum_{i=1}^{\lf nb_n+nc_n\rf}\frac{ \|e_ie_{i+k}\|_2 }{\sigma^4(t_i)}
 \|\sigma^2(t_i)-\hat{\sigma}^2 (t_i)\|_2=O(\pi_n),  \\
\label{lambda-j3'}
\max_{n- \lf nb_n+nc_n\rf \leq j\leq n}|\bar{\lambda}_{j,2}|
&\leq & |\bar{\lambda}_{n-\lf nb_n+nc_n\rf-1,2}|+\sum_{i=n-\lf nb_n+nc_n\rf}^{n}\frac{ |e_ie_{i+k}| }{\sigma^4(t_i)}|\sigma^2(t_i)-\hat{\sigma}^2 (t_i)| 
\notag\\
&\leq & \max_{\lf nb_n+nc_n\rf \leq j\leq n-\lf nb_n+nc_n\rf-1}|\bar{\lambda}_{j,2}|+\sum_{i=n-\lf nb_n+nc_n\rf}^{n}\frac{ |e_ie_{i+k}| }{\sigma^4(t_i)}|\sigma^2(t_i)-\hat{\sigma}^2 (t_i)|
\notag\\
&   = & \max_{\lf nb_n+nc_n\rf \leq j\leq n-\lf nb_n+nc_n\rf-1}|\bar{\lambda}_{j,2}|+O_p(\pi_n).
\end{eqnarray}
where the constants $\underline \pi_n$ and $ \pi_n$ are given by
$\underline \pi_n = nb_n^2c_n^2+\sqrt{\frac{n}{c_n}}b_n^2+\sqrt{\frac{n}{b_n}}c_n^2+\frac{1}{\sqrt{b_nc_n}}, $
$ \pi_n =  (nb_n+nc_n)(b_n^2+c_n^2+\frac{1}{\sqrt{nb_n}}+\frac{1}{\sqrt{nc_n}}), $
respectively.

In order to prove a corresponding estimate for  the remaining term $\max_{\lf nb_n+nc_n\rf \leq j\leq n-\lf nb_n+nc_n\rf}|\bar{\lambda}_{j,2}|$
in \eqref{lambda-j3'} 
 we study the asymptotic behavior of the quantity $\hat{\sigma^2}(t)-\sigma^2(t)$.
By Lemma \ref{hatsigma} in Section \ref{proofstechnical1} it easily follows that
\begin{align}\label{54}
\sup_{t\in\mk{T}_n}\Big|\hat{\sigma}^2 (t)-\sigma^2(t)- \frac{\mu_2\ddot{\sigma}^2 (t)c_n^2}{2} -\frac{1}{nc_n}\sum_{i=1}^nK_{c_n}(t_i-t)(\hat{e}^2_i-\E({e}_i^2))\Big|=O\Big(c_n^3+\frac{1}{nc_n}\Big).
\end{align}
We now consider the decomposition
\begin{align*}
 \sum_{i=1}^nK_{c_n}(t_i-t)\big(\hat{e}^2_i-\E({e}_i^2)-(e_i^2-\E (e_i^2))\big) =
\sum_{i=1}^nK_{c_n}(t_i-t)Q_i,
\end{align*}
where $Q_i=Q_{1,i}+Q_{2,i}$, $Q_{1,i}=2e_i[\mu(t_i)-\hat{\mu}(t_i)]$, $Q_{2,i}=[\mu(t_i)-\hat{\mu}(t_i)]^2$.
By Lemma \ref{mu} in Section \ref{proofstechnical1} we obtain
\begin{align*}
\sup_{\lf nb_n\rf\leq i\leq n-\lf nb_n\rf} \Bigl |\hat{\mu}_{b_n}(t_i)-\mu(t_i)-{ \frac{\mu_2\ddot{\mu}(t_i)}{2}b_n^2}-\frac{1}{nb_n}\sum_{j=1}^ne_jK_{b_n}(t_j-t_i)\Bigr |=O(b_n^3+\frac{b_n}{n}).
\end{align*}
The triangle inequality  and Proposition \ref{prop:2.1} in Section \ref{proofstechnical4} imply
\begin{align}\label{57}
\Big\|\sup_{t\in \mk{T''_n}}\Big|\sum_{i=1}^nK_{c_n}(t_i-t)\big[Q_{1,i}-\frac{2e_i}{nb_n}\sum_{j=1}^ne_jK_{b_n}(t_i-t_j)-{ \mu_2\ddot{\mu}(t_i)b_n^2e_i}\big]\Big|\Big\|_4=O(nb_n^3c_n^{3/4}),
\end{align}
where we use the notation  $\mk{T}''_n=[b_n+c_n,1-b_n-c_n]$.
Similar arguments as given in the calculation of   $\max_{\lf nb_n\rf \leq i\leq n-\lf nb_n\rf}|A_{n,i}|$ in the proof of Lemma \ref{maxdevvar} and the summation by parts formula show
\begin{align*}
\Big\|\sup_{t\in\mk{T}''_n}\frac{2}{nb_n}\Big|\sum_{i=1}^nK_{c_n}(t_i-t)e_i\sum_{j=1}^ne_jK_{b_n}(t_i-t_j)\Big|\Big\|_2={O(b_n^{-1})},\\
{\Big\|\sup_{t\in\mk{T}''_n}\Big|\sum_{i=1}^nK_{c_n}(t_i-t)\mu_2\ddot{\mu}(t_i)b_n^2e_i\Big|\Big\|_2={O(n^{1/2}b_n^2)},}
\end{align*}
{and (\ref{57}) gives
$  \big\|\sup_{t\in \mk{T}''_n} \big|\sum_{i=1}^n K_{b_n}(t_i-t) Q_{1,i}\big| \big \|_2=O(nb_n^{3}c_n^{3/4}+b_n^{-1}+n^{1/2}b_n^2)
$}.  On the other hand, note that
\begin{align*}
 \Big \|\sup_{t\in\mk{T}''_n}\Big|\sum_{i=1}^nK_{c_n}(t_i-t)Q_{2,i}\Big|  \Big \|_2& \leq  R_{n,1} + R_{n,2}
\end{align*}
where
\begin{align}
R_{n,1}  &=   \Big  \|\sup_{t\in\mk{T}''_n} \Big  |\sum_{i=1}^nK_{c_n}(t_i-t) \Big (\frac{1}{nb_n}\sum_{j=1}^ne_j K_{b_n}(t_i-t_j) +{ \frac{\mu_2\ddot{\mu}(t_i)}{2}b_n^2}\Big )^2\Big|  \Big \|_2 \notag
\\
R_{n,2}  &=    \Big \| \sup_{t\in\mk{T}''_n}\Big|\sum_{i=1}^nK_{c_n}(t_i-t) \Big (\mu(t_i)-\hat{\mu}(t_i)+\frac{1}{nb_n}\sum_{j=1}^ne_j K_{b_n}(t_i-t_j)+{ \frac{\mu_2\ddot{\mu}(t_i)}{2}b_n^2}\Big  )\notag
\\& \qquad \qquad \times \Big  (\mu(t_i)-\hat{\mu}(t_i)-\frac{1}{nb_n}\sum_{j=1}^ne_j  K_{b_n}(t_i-t_j)-{ \frac{\mu_2\ddot{\mu}(t_i)}{2}b_n^2}\Big  ) \Big |  \Big \|_2 . \notag
\end{align}
Proposition \ref{prop:2.1} in Section \ref{proofstechnical4} and similar calculations as given in the proof of   Lemma \ref{maxdevvar}   show that
\begin{align*}
R_{n,1}=O\Big(nc_nc_n^{-1/2}(\frac{1}{nb_n}+{ b_n^4})\Big)=O(c_n^{1/2}b_n^{-1}+{ nc_n^{1/2}b_n^4}),
\end{align*}
while a further application of Lemma \ref{mu} in Section \ref{proofstechnical1}   yields
\begin{align}\label{61}
R_{n,2}=O \Big(\frac{nb_n^3c_n}{\sqrt{nb_n}}c_n^{-1/2}+{ nb_n^{5}c_n^{1/2}}\Big)=O(\sqrt{n}b_n^{5/2}c_n^{1/2}+{ nb_n^{5}c_n^{1/2}}).
\end{align}
Consequently, combining the arguments in  (\ref{54})-(\ref{61}), it follows that
\begin{align}\label{62}
\Big\|\sup_{t\in \mk{T}''_n}\Big|\hat{\sigma^2}(t)-\sigma^2(t)-{ \frac{\mu_2\ddot{\sigma^2}(t)c_n^2}{2}}-\frac{1}{nc_n}\sum_{i=1}^nK_{b_n}(t_i-t)\big(e^2_i-\E(e_i^2)\big)\Big|\Big\|_2= O(\bar \pi_n),
\end{align}
where {  $\bar \pi_n=c_n^3+\frac{1}{nc_n}+b_n^3c_n^{-1/4}+\frac{1}{nb_nc_n}+\frac{b_n^2}
{\sqrt{n}c_n}+c_n^{-1/2}b_n^{-1}n^{-1}+c_n^{-1/2}b_n^{4}+b_n^{5/2}(nc_n)^{-1/2}+b_n^5c_n^{-1/2}$.}
Recall the definition of $\tilde W_i$, define   $Z_i'=e^2_i-\E e_i^2$,
then it follows from \eqref{62} that
\begin{align}\label{eq113}
&\E\Big( \max_{\lf nb_n+nc_n\rf\le j\le n-\lf nb_n+nc_n\rf}\Big|\bar{\lambda}_{j,2}+\sum_{i=1}^j \frac{{ \tilde{W}_i}(\sum_{j=1}^nK_{c_n}(t_j-t_i)Z_i'+{ \mu_2\ddot{\sigma}^2 (t_i)nc_n^3/2)}}{\sigma^2(t_i)nc_n}\Big|\Big)\leq \notag\\
&\sum_{i=\lf nb_n+nc_n \rf}^{n-\lf nb_n+nc_n\rf}\|e_ie_{i+k}\|_2 \Big \|\frac{\hat{\sigma}^2 (t_i)-\sigma^2(t_i)-\frac{1}{nc_n}\sum_{j=1}^nK_{c_n}(t_j-t_i)Z_i'-{ \mu_2\ddot{\sigma}^2 (t_i)c_n^2/2}}{\sigma^4(t_i)}\Big\|_2 \\\notag & \qquad \qquad \qquad \qquad = O(n\bar \pi_n).
\end{align}
By the Cauchy-Schwarz inequality   we obtain
 $ { \|  \tilde{W}_i-\tilde{W}_{i}^{(m)}\|_4}=O(\chi^{|i-m|}),\ \ \|Z_i'-Z_{i}'^{(m)}\|_4=O(\chi^{|i-m|})$, where
 $Z_i'^{(m)}= (e^{(m)}_i)^2-\E (e^{(m)}_i)^2$,
${\tilde{W}_i^{(m)}}=\frac{e_i^{(m)}e_{i+k}^{(m)}}{\sigma(t_i)^2}$, and
 \begin{align*}
\ e_i^{(m)}=G_j(t_i,\FF^{(m)}_i),\mbox{  if  } \ b_j<t_i\leq b_{j+1}.
\end{align*}
Hence, with similar arguments as given in the proof of Lemma 5 of \cite{zhouwu2010} we get
\begin{align*} \label{eq116}
\max_{\lf nb_n+nc_n\rf\le j\le n-\lf nb_n+nc_n\rf}\E \Big [\sum_{i=\lf nb_n+ nc_n \rf}^j \frac{\tilde W_i\sum_{j=1}^nK_{c_n}(t_j-t_i)Z_j'}{{\sigma^2}(t_i)nc_n}\Big ]=O(c_n^{-1}).
\end{align*}
Then by a similar $m$-dependent approximating technique as given in the proof of Lemma \ref{maxdevvar}  we get
\begin{align*}
\max_{\lf nb_n+nc_n\rf\leq j\leq n-\lf nb_n+nc_n\rf}\Big| \sum_{i=\lf nb_n+nc_n\rf}^j
 \frac{ \sum_{j=1}^n { \tilde {W_i}} K_{c_n}(t_j-t_i)Z_j' - \E [ \tilde {W_i} K_{c_n}(t_j-t_i)Z_j'  ]}{ {\sigma^2}(t_i)nc_n} \Big|  =O_p(c_n^{-1}).
\end{align*}
Similarly, and more easily one obtains
\begin{equation}
\max_{\lf nb_n+nc_n\rf\le j\le n-\lf nb_n+nc_n\rf}\Big|\sum_{i=1}^j e_ie_{i+k}\mu_2\ddot{\sigma^2}(t_i)b_n^2/(2\sigma^4(t_i)) \Big|=O_p(nc_n^2). \label{revi2-6.32}
\end{equation}
Hence, it follows from \eqref{eq113} and \eqref{revi2-6.32} that
 $\max_{\lf nb_n+nc_n\rf\leq j\leq n-\lf nb_n+nc_n\rf}|\bar{\lambda}_{j,2}|= O_p(n\bar \pi_n+nc_n^2)$,
 which implies, observing \eqref{lambda-j1} - \eqref{lambda-j3'}, $$\max_{\lf nb_n+nc_n\rf\leq j\leq n-\lf nb_n+nc_n\rf}|\bar{\Lambda}_j|= O_p(\pi_n+\underline \pi_n+n\bar \pi_n+nc_n^2) .$$ 
 Combining this result with the estimates \eqref{78}, \eqref{revision2-93} and \eqref{63}, and by our choice of the bandwidths, we have that
  $$
 \max_{1 \leq j \leq n} \Big | \sum^j_{i=1} ( \hat W_i - W_i ) \Big | { =}
   O_p(nc_n^2+nb_n^3c_n^{-1/4}+b_n^{-1}c_n^{-1}),
 $$
 which establishes the estimate  \eqref{corr} and completes the proof of Theorem \ref{theorem5}.

\subsubsection{Proof of Theorem \ref{theorem5boot}}
We restrict ourselves to the case of a variance function with no change point and the corresponding estimator
\eqref{defhatmu1}. The statement for the estimator \eqref{varest1} follows by similar arguments as given in Theorem \ref{thm7},
where we deal with the problem of testing
for  relevant changes in the  correlation.  \\
Recall the definition of  $\hat W_i$ and  $W_i$  in \eqref{70aa} and \eqref{70a} and introduce the notation
   $\tilde W_i=\frac{e_ie_{i+k}}{\sigma^2(t_i)}$ (again the superscript is omitted in our notation).
We consider  the corresponding partial sums  $S_{j,m}=\sum_{r=j}^{j+m-1}W_r$,
$\tilde S_{j,m}=\sum_{r=j}^{j+m-1}\tilde W_r$ and  $\hat  S_{j,m}^W=\sum_{r=j}^{j+m-1} \hat  W_r$
and define  $S_{n}=\sum_{r=1}^{n}W_r$, $\tilde S_{n}=\sum_{r=1}^{n}\tilde W_r$, $\hat S_{n}^W=\sum_{r=1}^{n}\hat
W_r$. Similarly, define $(\Phi_{i,m}$, $\tilde{\Phi}_{m,n}(t))$,
 ($\Phi^o_{i,m}$, $\tilde{\Phi}^o_{m,n}(t)$) and
($\hat \Phi^W_{i,m}$, $\hat{\tilde{\Phi}}^W_{m,n}(t)$)   by replacing   ($\hat S_{n}$, $\hat S_{j,m}$) in the definitions
(\ref{new.83}), (\ref{new.84})   of $\hat \Phi_{i,m}$ and $\hat{\tilde{\Phi}}_{m,n}(t)$
  by $(S_n,S_{j,m})$,  ($\tilde S_{n}$, $\tilde S_{j,m}$) and ($\hat S^W_{n}$, $\hat S^W_{j,m}$), respectively.
   It then follows  from the results  \cite{zhou2013} that
$
\{ \tilde{\Phi}_{m,n}(t)\}_{t \in [0,1]}\Rightarrow \{U_2(t)\}_{t \in [0,1]},
$
where $\{U_2(t)\}_{t \in [0,1]}$ is a centered Gaussian process with mean $0$ and covariance kernel
\eqref{cov2}. The assertion of Theorem \ref{theorem5boot}}
is now a consequence of  the estimate
\begin{align} \label{corol3}
\sup_{t\in [0,1]}|\tilde{\Phi}_{m,n}(t)-\hat{\tilde{\Phi}}_{m,n}(t)|=O_p\big(\sqrt{m}\delta_n\big),
\end{align}
where {$\delta_n=\big(c_n^2+(\frac{1}{\sqrt{nc_n}}+b_n^2+\frac{1}{\sqrt{nb_n}})c_n^{-1/4}\big)\log n$}.
For a proof of this estimate we consider the events
\begin{eqnarray*}
A_n&=&\Big\{\sup_{t\in [0,1]}|\hat{\mu}_{b_n}(t)-\mu(t)|\leq C\frac{\log n}{\sqrt{nb_n}b_n^{1/4}}+Cb_n^2\log n\Big\}, \\ B_n&=&\Big\{\sup_{t\in[0,1]}|\hat{\sigma}^2_{}(t)-\sigma^2(t)|\leq C\Big(c_n^2+(\frac{1}{\sqrt{nc_n}}+b_n^2+\frac{1}{\sqrt{nb_n}})c_n^{-1/4}\Big)\log n.\Big\},
 \end{eqnarray*}
 then by   \eqref{revision2-35} and   Corollary \ref{corosigma2} in Section \ref{proofstechnical1}
 it follows that $\lim_{n\rightarrow \infty}\p(A_n\cap B_n)=1$. Using similar arguments as in the proof of the
 estimate \eqref{estthm2} it can be shown that
\begin{align*}
\|({\Phi}^o_{n,m}-{\Phi}_{n,m})I(A_n\cap B_n)\|^2\leq C\Big(\frac{m}{n^2}\Big),\\
\|({\Phi}^o_{n,m}-\hat{\tilde{\Phi}}_{n,m})I(A_n\cap B_n)\|^2\leq Cm\delta_n^2.
\end{align*}
The assertion of Theorem  \ref{theorem5boot} now follows by a similar argument as given in the proof of Theorem \ref{theorem2}. \hfill $\Box$

\subsection{Proof of Lemma   \ref{lemmavar} - \ref{lemmadelta}} \label{sec62}

In order to simplify the notation define
 $G_n(m)=S_m-\frac{m}{n}S_n$,  $\hat{G}_n(m)=\hat{S}_m-\frac{m}{n}\hat{S}_n$, where $S_m=\sum_{i=1}^me_i^2$, $\hat{S}_m=\sum_{i=1}^m\hat{e}_i^2$.   Then it is easy to see  that the estimator  $\tilde t_n$  of the change point in the variance function defined in \eqref{estvar} can be represented as
$
\tilde t_n=\frac{1}{n}\argmax_{1\leq m\leq n}(\hat{G}_n(m))^2.
$
Similarly, we introduce the notation
$$
\check{G}_n(m)=\sum_{j=1}^{m}\frac{\hat{e}_j\hat{e}_{j+k}}{\hat{\sigma}^{2*}(t_j)}-\frac{m}{n}\sum_{j=1}^{ n }\frac{\hat{e}_j\hat{e}_{j+k}}{\hat{\sigma}^{2*}(t_j)}
$$
and obtain the representation
$\hat{t}_n=\frac{1}{n}\argmax_{1\leq m\leq n}(\check{G}_n(m))^2$
for the estimator of the change point in the correlation function defined  in \eqref{changeestcorr}.

\medskip

\subsubsection{Proof of Lemma \ref{lemmavar}}
 Recall that Section \ref{sec41} considers the problem of testing for a non relevant change in the variance
and that  under null hypothesis, we have  $\sigma^2(s)=\sigma^2$ for $s\leq {\tilde{t}_v}$  and $\sigma^2(s)=\sigma^2+\gamma$
for $s>{ \tilde{t}_v}$, where $\gamma$ is an   unknown (without loss of generality) positive constant. A simple calculation shows that
\begin{align}\label{eq170}
f_n(m):=\E G_n(m)=n(m(n)t(n)-m(n)\wedge t(n))\gamma,
\end{align}
where we used the notation $m(n)=m/n$ and $t(n)=  \lfloor n{\tilde{t}_v} \rfloor/n$.
By Proposition 5 of \cite{zhou2013}, on a possibly richer probability space, there exist $i.i.d$ standard normal variables, say $\{V_i\}_{i\in \mathbb{Z}}$, such that
\begin{align}\label{Gaussapprox}
\max_{1\leq i\leq n} \Big|S_i-\E(S_i)-\sum_{j=1}^i\kappa_1(t_j)V_j\Big|=o_p(n^{1/4}\log n),
\end{align}
 {where $\kappa_1$ is defined in assumption (A5).}
By the  arguments given in Section \ref{prooftheorem1}  we have
\begin{align}\label{revision2-169}
\max_{1\leq m\leq n}|G_n(m)-\hat{G}_n(m)|=O_p(\varrho_n),
\end{align}
where $\varrho_n=b_n^{-1}+nb_n^3+\sqrt{nb_n}$.
Now a similar reasoning as given in the proof of Lemma 5 of \cite{zhouwu2010}, Assumption (A3) (A4) and (A5) yield that there exists a constant $C$ such that $\kappa_1^2(s)\leq C$ for all $s \in [0,1]$. 
Then it is easy to see that $\|\Xi_n\|_2^2=O(n)$. By Doob's inequality, we have that
\begin{align}\label{maxgaussian}
{\max_{1\leq j\leq n}|\Xi_j|=O_p(\sqrt{n}),}
\end{align}
and observing (\ref{Gaussapprox}) we obtain
\begin{align*}
\max_{1\leq m\leq n}\big| G^2_n(m) - \hat{G}^2_n(m) \big|=\max_{1\leq m\leq n}|G_n(m)+\hat{G}_n(m)||G_n(m)-\hat{G}_n(m)| = {O_p(n\varrho_n)}.
\end{align*}
Define $\hat{V}_n(m)=\hat{G}^2_n(m)-\hat{G}^2_n(\lf nt \rf)$, note that $\hat{V}_n(\lf nt \rf)=0$ and consider a constant $\alpha \in (\frac {1}{2}, \frac {2}{3})$, such that $n^{1-\alpha}/\varrho_n\rightarrow \infty$.
 Observing the definition \eqref{eq170} and the estimate {(\ref{Gaussapprox}),
it follows that \begin{align}\label{revision2-173}
\max_{1\leq m\leq n}\Big|G_n^2(m)-\Big(f_n(m)+\Xi_m-\frac{m}{n}\Xi_n\Big)^2\Big|=O_p(n^{5/4}\log n).
\end{align}
By (\ref{maxgaussian}), we have $\max_{1\leq m\leq n}(\Xi_m-\frac{m}{n}\Xi_n)^2=O_p(n)$, and together with (\ref{revision2-169}) and (\ref{revision2-173})  this yields }
\begin{align}\label{hat140}
&\max_{ m \in {\cal M}_n } \hat{V}_n(m)=\max_{ m \in {\cal M}_n }     \left[G_n^2(m)-G_n^2(\lf nt\rf)\right]+O_p(n\rho_n) =\max_{ m \in {\cal M}_n }  \Big\{f^2_n(m)-f^2_n(\lf nt\rf) \notag\\
& ~~~~~~~~+2(f_n(m)-f_n(\lf nt \rf))\Xi_m+2f_n(\lf nt \rf)(\Xi_m-\Xi_{\lf nt\rf})\notag\\
&~~~~~~~~-2\frac{m}{n}f_n(m)\Xi_n+2\frac{\lf nt\rf}{n}f_n(\lf nt\rf)\Xi_n\Big\}+O_p(n\varrho_n+n^{5/4}\log n),
\end{align}
where the maxima are taken over the set ${\cal M}_n = \{m ~|~{\lf n\tilde{t}_v\rf-\lf n^{1-\alpha/2}\rf\leq m\leq \lf n\tilde{t}_v\rf-\lf n^{1-\alpha}\rf}|\}$.
Observing the definition of $f_n(m)$ in \eqref{eq170} we have  for some positive constant $C$,
\begin{align}\label{hat141}
\max_{ m \in {\cal M}_n }  [f^2_n(m)-f^2_n(\lf nt\rf)]\leq -Cn^{2-\alpha},
\end{align}
and (\ref{maxgaussian}) implies
\begin{align}\label{hat142}
&&\max_{ m \in {\cal M}_n }  [f_n(m)-f_n(\lf nt\rf)]\Xi_m=O_p(n^{3/2-\alpha/2}\log n), \\
\label{hat143}
&&\max_{ m \in {\cal M}_n }   [\frac{m}{n}f_n(m)-\frac{\lf nt\rf}{n} f_n(\lf nt\rf)]\Xi_n=O_p(n^{3/2-\alpha/2}\log n).
\end{align}
Using the representation
$\Xi_m-\Xi_{\lf nt\rf}=\sum_{i=m+1}^{\lf nt\rf}\sigma(t_i)V_i$
 and
similar arguments as in the derivation of   (\ref{maxgaussian}) yields
\begin{align*}
\max_{ m \in {\cal M}_n }  [\Xi_m-\Xi_{\lf nt\rf}]=O_p(n^{1/2(1-\alpha/2)}\log n).
\end{align*}
Consequently,
\begin{align}\label{hat145}
\max_{ m \in {\cal M}_n }  f_n(\lf nt\rf)[\Xi_m-\Xi_{\lf nt\rf}]=O_p(n^{3/2-\alpha/4}\log n).
\end{align}
By our choice of $\alpha$,  (\ref{hat140}) - (\ref{hat145})
it now follows that
\begin{align}\label{hatthatv}
{ \p\Big(\limsup_{n\rightarrow \infty}\max_{\lf n\tilde{t}_v\rf-\lf n^{1-\alpha/2}\rf\leq m\leq \lf n\tilde{t}_v\rf-\lf n^{1-\alpha}\rf}\hat{V}_n(m)=-\infty\Big)=1}.
\end{align}
On the other hand, similar arguments give the estimates
\begin{eqnarray*}
\max_{1\leq m\leq \lf n\tilde{t}_v\rf-\lf n^{1-\alpha/2}\rf}[f^2_n(m)-f^2_n(\lf nt\rf)]&\leq& -Cn^{2-\alpha/2}, \\
 \max_{1\leq m\leq \lf n\tilde{t}_v\rf-\lf n^{1-\alpha/2}\rf}f_n(\lf nt\rf)[\Xi_m-\Xi_{\lf nt\rf}]&=&O_p(n^{3/2}\log n),\\
   \max_{1\leq m\leq \lf n\tilde{t}_v\rf-\lf n^{1-\alpha/2}\rf}[\frac{m}{n}f_n(m)-\frac{\lf nt\rf}{n}f_n(\lf nt\rf)]\Xi_n &=& O_p(n^{3/2}\log n), \\
   \max_{1\leq m\leq \lf n\tilde{t}_v\rf-\lf n^{1-\alpha/2}\rf}[f_n(m)-f_n(\lf nt\rf)]\Xi_m&=&O_p(n^{3/2}\log n) ,
\end{eqnarray*}
and by our choice of $\alpha$ we obtain
 $\p(\limsup_{n\rightarrow \infty}\max_{1\leq m\leq \lf n\tilde{t}_v\rf-\lf n^{1-\alpha/2}\rf}\hat{V}_n(m)=-\infty)=1$.
 Combined with (\ref{hatthatv}) this gives
 $\p(\limsup_{n\rightarrow \infty}\max_{1\leq m\leq \lf n\tilde{t}_v\rf-\lf n^{1-\alpha}\rf}\hat{V}_n(m)=-\infty)=1$,
 and it can be shown by similar arguments that
$\p(\limsup_{n\rightarrow \infty}\max_{\lf n\tilde{t}_v\rf+\lf n^{1-\alpha}\rf\leq m\leq n }\hat{V}_n(m)=-\infty)=1$.
   Consequently, it follows that
\begin{align*}
\lim_{n\rightarrow \infty}\p(|n\tilde t_n-\lf n\tilde{t}_v\rf|\leq n^{1-\alpha})=1,
\end{align*}
which proves part (ii) of Lemma \ref{lemmavar}.
For the case that the variance has no jump at time $t$, the desired result follows from the fact that $\hat{G}_n(m)/\sqrt{n}$ converges weakly to some Gaussian process $\{U_1(s)-sU_1(1)\}_{s \in [0,1]}$, which implies   $\tilde t_n \cod \tilde{T}=\argmax_{s\in (0,1)}|U_1(s)-sU_1(1)|$, where the Gaussian process $\{U_1(s)\}_{s \in [0,1]}$ is defined in  Theorem \ref{theorem1}.
\medskip

\subsubsection{Proof of Lemma \ref{lemvar}}  By a careful  examination of the proof of Theorem \ref{thm6} (see the next section) it follows that
\begin{align*}
\max_{1\leq m\leq n}|\check{G}_n(m)-\bar{G}_n(m)|=O_p({nc_n^2+nb_n^3c_n^{-1/4}}+n^{1-\alpha}\log n+b_n^{-1}c_n^{-1}),
\end{align*}
where \begin{align*}
\bar{G}_n(m)=\sum_{j=1}^{m}\frac{{e}_j{e}_{j+k}}{{\sigma^2}(t_j)}-\frac{m}{n}\sum_{j=1}^{ n }\frac{{e}_j{e}_{j+k}}{{\sigma^2}(t_j)}.
\end{align*}
Write $\varrho_n'={nc_n^2+nb_n^3c_n^{-1/4}+b_n^{-1}c_n^{-1}}+n^{1-\alpha}\log n$, { where $\alpha$ is defined in the proof of Lemma \ref{lemmavar}}. Let $1/2<\alpha'<2/3$, such that $n^{1-\alpha'}/\varrho'_n\rightarrow \infty$.
Then using similar arguments as given in the proof of Lemma \ref{lemmavar} we can show that
\begin{align*}
\hat{t}_n-t_n=O_p(n^{-\alpha'}),
\end{align*}
if there is a change in the correlation at time $t$. On the other hand,
if there is no change in correlation, we have that $\hat{t}_n \cod  {T}_{\max}$, where $T_{\max}=\argmax_{t\in (0,1)}|U_2(t)-tU_2(1)|$, and the stochastic process  $\{U_2(s)\}_{s \in [0,1]}$ is defined in Theorem \ref{theorem5}.

\medskip

\subsubsection{Proof of Lemma \ref{lemmadelta}}  Recall the definition of \eqref{deltadach} and define
$
\Delta_{n,1}={1\over \lf n t \rf} \sum_{j=1}^{\lf n  t \rf}{W}_j $,  $ \Delta_{n,2}={1\over n-\lf n  t  \rf} \sum_{j=\lf n  t \rf+1}^{n} {W}_j
$ (the superscript $k$ is again omitted).
From  the proof of Theorem \ref{theorem5} we have that
\begin{align*}
\Delta_{n,1}-\E[\Delta_1]= O_p\Big(\frac{1}{\sqrt{n}}\Big),\quad \Delta_{n,2}-\E[\Delta_2]= O_p\Big(\frac{1}{\sqrt{n}}\Big).
\end{align*}
Since $\Delta = \E [\Delta_{2}] - \E [\Delta_{1}]$   we have
$\Delta_n:=\Delta_{n,2}-\Delta_{n,1}={\Delta}+O_p(1/\sqrt{n})$.
In order to prove this estimate we introduce the notation
${\cal A}_n=\{|\hat{t}_n-t|\leq \frac{C\log n}{\sqrt{n}}\}$.  Then by {Lemma \ref{lemvar}, we have that $\lim_{n\rightarrow \infty}\p(\mathcal A_n)=1$}. This yields
\begin{align}
&(\Delta_{n,1}-\hat{\Delta}_{n,1})I({\cal A}_n) \notag =I( {\cal A}_n )   ( A_n +B_n + C_n),
\end{align}
where
\begin{align} \label{eq204}
 A_n =    \sum_{j=1}^{\lf nt \rf}  \frac{  W_j}{\lf nt \rf} -\sum_{j=1}^{\lf n\hat{t}_n\rf}  \frac{  W_j }{ \lf nt \rf}   ~, ~B_n =
 \sum_{j=1}^{\lf n\hat{t}_n\rf}  \Big(\frac{ W_j}{\lf nt \rf}  - \frac{  \hat{W}_j}{ \lf nt \rf}  \Big) ~,~C_n =   \sum_{j=1}^{\lf n\hat{t}_n \rf} \Big( \frac{  \hat{W}_j}{ \lf nt \rf}
- \frac{  \hat{W}_j}{ \lf n\hat{t}_n \rf} \Big).
\end{align}
It is easy to see that
 $I(\mathcal{A}_n) A_n=O_p (\frac{\log n}{\sqrt{n}} )$.
 Using the same arguments as in the proof of Theorem \ref{thm6}, we obtain
$I({\cal A}_n) B_n=o(\sqrt{n}/n)=o(1/\sqrt{n})$
and
\begin{align}
I({\cal A}_n) \cdot C_n & =I({\cal A}_n) \sum_{j=1}^{\lf n\hat{t}\rf}\hat{W}_j\frac{\lf n\hat{t}\rf-\lf nt\rf}{\lf nt\rf\lf n\hat{t}\rf}  \leq C  \cdot  I({\cal A}_n)
\sum_{j=1}^{\lf n\hat{t}\rf}\hat{W}_j\frac{\log n}{n\sqrt{n}} \nonumber  \\
\nonumber & =C \cdot   I({\cal A}_n) \Big(\sum_{j=1}^{\lf n\hat{t}\rf}{W}_j+o_p(\sqrt{n})\Big)\frac{\log n}{n\sqrt{n}}\notag
\\   &= I({\cal A}_n) \cdot \Big(\sum_{j=1}^{\lf nt \rf}{W}_j+\sum_{j=\lf nt \rf}^{\lf n\hat{t} \rf}{W}_jI(t\leq \hat{t})-\sum_{j=\lf n\hat{t} \rf}^{\lf n{t} \rf}{W}_jI(t>\hat{t})+o_p(\sqrt{n})\Big)\frac{\log n}{n\sqrt{n}}  \nonumber  \\
& =O_p\Big (\frac{\log n}{\sqrt{n}} \Big).\label{123}
\end{align}
Combining  (\ref{eq204}) - (\ref{123})  and using   Proposition \ref{prop01} in Section \ref{proofstechnical4} shows
$
\Delta_{n,1}-\hat{\Delta}_{n,1}=O_p\big(\frac{\log n }{\sqrt{n}}\big).
$
Similarly, we have
 $\Delta_{n,2}-\hat{\Delta}_{n,2}=O_p(\frac{\log n }{\sqrt{n}})$,
 and the assertion of the
 lemma follows.

\subsection{Proof of Theorem \ref{thm6} and \ref{thm7}} \label{sec63}

\subsubsection{Proof of Theorem \ref{thm6}}
 \label{proofthm6}
We consider the non-observable analogue
\begin{align*}
{T}^r_n=\frac{3}{{t}^2(1-{t})^2}\int_{0}^{1} {U}^2_n(s) ds.
\end{align*}
  of the statistic ${ \hat{T}^r_n}$ defined in \eqref{76}, where the process
  $U_n$ is given by
  \begin{align*}
{U}_n(s)=\frac{1}{n}\sum_{j=1}^{\lf ns\rf}\frac{{e}_j{e}_{j+k}}{{\sigma}(t_j){\sigma}(t_{j+k})}-\frac{s}{n}
\sum_{j=1}^{ n }\frac{{e}_j{e}_{j+k}}{{\sigma}(t_j){\sigma}(t_{j+k})}.
\end{align*}
It follows from the proof of Theorem \ref{theorem5} that
\begin{align}
\{\sqrt{n}(U_n(s)+(s\wedge t-st){\Delta})\}_{s \in [0,1]}\Rightarrow \{U_2(s)-sU_2(1)\}_{s \in [0,1]},\label{100}
\end{align}
whenever $\Delta \not = 0$.  The continuous mapping theorem, elementary calculations, and the identity
$
{3}\int_0^1[st-s\wedge t]^2ds={{t}^2(1-{t})^2}
$
 imply
$
\sqrt{n}( {T}^r_n-\Delta^2)\cod  \mathcal{Z}_2  (\Delta),
$
where the random variable $\mathcal{Z}_2(\Delta)$ is defined in Theorem \ref{thm6}.
Finally, we   show the estimate
\begin{align} \label{compare}
\sqrt{n}(T^r_n-\hat{T}^r_n)=O_p({n^{1/2}c_n^2+n^{1/2}b_n^3c_n^{-1/4}}+n^{-1/2}\zeta_n+n^{-1/2}b_n^{-1}c_n^{-1}),
\end{align}
where $\zeta_n=n^{1-\alpha}+nc_n\big( n^{-\alpha}c_n^{-5/4}+c_n^2+(\frac{1}{\sqrt{nc_n}}+b_n^2+\frac{1}{\sqrt{nb_n}})c_n^{-1/4}\big)$, $\alpha=1-4/\iota-\nu$ for some $\nu>0$,
which completes the proof.

For a proof of this remaining estimate   we
 consider the case $\Delta>0$, where it follows from Lemma \ref{tstar0} that $|{t}^*_n-\tilde{t}_v|=O_p(n^{-\alpha})$.
The proof of the statement in the case $\Delta=0$ (which corresponds to $\hat{t}_n\cod T_{\max}$)
is easier  and omitted for the sake of brevity.
 For the event 
   $A_n=\{|{t}^*_n-\tilde{t}_v|\leq \frac{1}{n^\alpha}\}$ it follows from Lemma \ref{tstar0} that
   $\lim_{n\rightarrow }\p(A_n)=1$. 
Observe that we have   from Lemma \ref{hatsigma}   
\begin{align*}
\sup_{t\in T_n}\Big|\hat{\sigma}_n^{2*}(t)-\sigma^2(t)-\frac{\mu_2\ddot{\sigma}^2(t)c_n^2}{2}-\frac{1}{nc_n}\sum_{i=1}^n(\hat{e}_i^2-\E{e}_i^2)K_{b_n}(t_i-t)\Big|=O\Big(c_n^3+\frac{c_n}{n}\Big),
\end{align*}
where $T_n=[c_n,\tilde{t}_v-n^{-\alpha}-c_n]\cup[\tilde{t}_v+n^{-\alpha}+c_n,1-c_n]$. Also \eqref{boundedsigma2} still holds.
Now similar calculations as given in the proof of  \eqref{corr} yield
\begin{align*}
 &\max_{1\leq l\leq \lf n(\tilde{t}_v-n^{-\alpha}-c_n)\rf}\Big|\sum_{j=1}^l \Big(\frac{{e}_j{e}_{j+k}}{{\sigma}(t_j){\sigma}(t_{j+k})}-\frac{\hat{e}_j\hat{e}_{j+k}}{\hat{\sigma}_n^{2*}(t_j)}\Big) \Big|I(A_n)= O_p({nc_n^2+nb_n^3c_n^{-1/4}+b_n^{-1}c_n^{-1}}),\\
& \max_{\lf n(\tilde{t}_v-n^{-\alpha}-c_n)\rf\leq l\leq \lf n(\tilde{t}_v+n^{-\alpha}+c_n)\rf}\Big|\sum_{j=\lf n(\tilde{t}_v-n^{-\alpha}-c_n)\rf}^l \Big( \frac{{e}_j{e}_{j+k}}{{\sigma}(t_j){\sigma}(t_{j+k})}-\frac{\hat{e}_j\hat{e}_{j+k}}{\hat{\sigma}_n^{2*}(t_j)}\Big) \Big|I(A_n)= O_p(\zeta_n),
\\ &\max_{\lf n(\tilde{t}_v+n^{-\alpha}+c_n)\rf\leq l\leq \lf ns\rf}\Big|\sum_{j=\lf n(\tilde{t}_v+n^{-\alpha}+c_n)\rf}^l \Big(\frac{{e}_j{e}_{j+k}}{{\sigma}(t_j){\sigma}(t_{j+k})}-\frac{\hat{e}_j\hat{e}_{j+k}}{\hat{\sigma}_n^{2*} (t_j)}\Big) \Big|I(A_n)= O_p({nc_n^2+nb_n^3c_n^{-1/4}+b_n^{-1}c_n^{-1}}),
\end{align*}
where we used  Corollary \ref{corosigmastar} for the second statement.
So we have
 \begin{align*}
\sup_{0\leq s\leq 1}n|U_{n}(s)-\hat{U}_n(s)|I(A_n)=O_p({nc_n^2+nb_n^3c_n^{-1/4}+b_n^{-1}c_n^{-1}+\zeta_n}).
\end{align*}
Using the same arguments
and Proposition \ref{prop01} in Section \ref{proofstechnical4} we obtain
\begin{align*}
\sup_{0\leq s\leq 1}n|U_{n}(s)-\hat{U}_n(s)|=O_p({nc_n^2+nb_n^3c_n^{-1/4}+b_n^{-1}c_n^{-1}+\zeta_n}).
\end{align*}
From   (\ref{100})   it follows that $\int_0^1|U_{n}(s)|ds=O_p(1)$. Consequently, we have
\begin{eqnarray*}
&&n^{1/2}\int_0^1[U^2_{n}(s)-\hat{U}^2_n(s)]ds\leq \sup_{0\leq s\leq 1}n^{1/2}|U_{n}(s)-\hat{U}_n(s)|\int_0^1|U_{n}(s)+\hat{U}_n(s)|ds
  \\
&& \qquad {\leq 2n^{1/2}\sup_{0\leq s\leq 1}|U_{n}(s)-\hat{U}_n(s)|\int_0^1|U_{n}(s)|ds+ {n^{1/2}\sup_{0\leq s\leq 1}|U_{n}(s)-\hat{U}_n(s)|^2}}
\\
&& \qquad =O_p({n^{1/2}c_n^2+n^{1/2}b_n^3c_n^{-1/4}+n^{-1/2}\zeta_n}+n^{-1/2}b_n^{-1}c_n^{-1}),
\end{eqnarray*}
and the remaining estimate  \eqref{compare}  follows.

\subsubsection{Proof of Theorem \ref{thm7}}  \label{proofthm7}

Recall the definition of  $\hat{{\Delta}}_n$, $\hat{A}_j$, $\hat{\Phi}^A_{i,m} $ in \eqref{deltadach}, \eqref{new.new135},
  \eqref{107}
and define
\begin{align*}
A_j &=\frac{e_je_{j+k}}{\sigma(t_j)\sigma(t_{j+k})}-{\Delta} I(j\geq \lf nt\rf), \\
\Phi^A_{i,m}& =\frac{1}{\sqrt{m(n-m+1)}}\sum_{j=1}^{n-m+1}(S_{j,m}^A-\frac{m}{n}S_n^A)R_j,
\end{align*}
where $ S_{j,m}^A=\sum_{r=j}^{j+m-1}A_r, $ $ S_{n}^A=\sum_{r=1}^{n}A_r$. We introduce the  processes
\begin{align} \nonumber
{\Phi}^A_{m,n}(s)=\Phi^A_{\lfloor ns \rfloor,m}+ (ns- \lfloor ns \rfloor) (\Phi^A_{\lfloor ns \rfloor+1,m}-\Phi^A_{{\lf ns\rf},m}),\\ \label{eq163}
\hat{\Phi}^A_{m,n}(s)=\hat{\Phi}^A_{\lfloor ns \rfloor,m}+ (ns-\lfloor ns \rfloor)(\hat{\Phi}^A_{\lfloor ns \rfloor+1,m}-\hat{\Phi}^A_{{\lf ns\rf},m}).
\end{align}
and note that it follows from \cite{zhou2013} that
$
\{ {\Phi}^A_{m,n}(s) \}_{s\in [0,1]} \Rightarrow \{ U_2(s) \}_{s\in [0,1]}
$
conditional on $\FF_n$. The assertion of Theorem \ref{thm7} is therefore a consequence of the estimate
\begin{align} \label{estim}
\sup_{s\in (0,1)}|{\Phi}^A_{m,n}(s)-\hat{\Phi}^A_{m,n}(s)|=O_p (\sqrt m\rho_n ),
\end{align}
where $\rho_n=\big(n^{-\alpha}c_n^{-5/4}+c_n^2+(\frac{1}{\sqrt{nc_n}}+b_n^2+\frac{1}{\sqrt{nb_n}})c_n^{-1/4}\big)\log n.$
{
To show \eqref{estim}, define ${\cal A}_n=\{|\hat{t}_n-t|\leq
\frac{C\log n}{\sqrt{n}}\}$, ${\cal B}_n=\{|\hat{{\Delta}}_n-{\Delta}|\leq
\frac{C\log^2 n}{\sqrt{n}}\}$, ${\cal C}_n=\{\sup_{t\in(0,1)}|\hat{\mu}_{b_n}(t)-\mu(t)|\leq C\frac{\log n}
{\sqrt{nb_n}b_n^{1/4}}+Cb_n^2\log n\}$, {$${\cal D}_{1,n}=\Big\{\sup_{t\in[0,\tilde{t}_v-n^{-\alpha})}|\hat{\sigma^{2*}}_{}(t)-
\sigma^2(t)|\leq C\rho_n\Big\},$$ $${\cal D}_{2,n}=\Big\{\sup_{t\in(\tilde{t}_v+
n^{-\alpha},1]}|\hat{\sigma^{2*}}_{}(t)-\sigma^2(t)|\leq C\rho_n\Big\},$$}
 where $C$ is some sufficiently large  constant, $\alpha$ is defined in Corollary \ref{corosigmastar}.
By  construction  it follows
 $\lim_{n\rightarrow \infty}\p(\mathcal{W}_n)=1$, where
 $\mathcal{W}_n={\cal A}_n\cap {\cal B}_n\cap {\cal C}_n\cap {\cal D}_{1,n}\cap {\cal D}_{2,n}$.}
 Let $\tilde{A}_j=\frac{\hat{e}_j\hat{e}_{j+k}}{\hat{\sigma^{2*}}(t_j)}-{{\Delta}} I(j\geq \lf nt\rf)$.
Similarly to  the definition (\ref{107}) and \eqref{eq163},  we define $\tilde S^A_{i,m}$, $\tilde{\Phi}^A_{i,m}$, $\tilde{\Phi}^A_{m,n}(t)$,
where the random variables $\hat A_i$ are replaced by $\tilde A_i$. From the proof of the estimate \eqref{corol3} it follows
\begin{align}
\sup_{t\in (0,1)}|\tilde{\Phi}^A_{m,n}(t)-{\Phi}^A_{m,n}(t)|=O_p\Big(\sqrt{m}\rho_n+\sqrt mn^{-\alpha/2}\Big).\label{112}
\end{align}
On the other hand, \begin{align*}
&\E[(\tilde{S}^A_{j,m}-\hat{S}^A_{j,m})^2I(\mathcal{W}_n)]\notag\\
=&\E\Big[\sum_{r=j}^{j+m-1}I(\mathcal{W}_n)\big({\Delta}\big(I(r\geq \lf nt\rf)-I(r\geq \lf n\hat{t}_n \rf)\big)\big)+\sum_{r=j}^{j+m-1}I(\mathcal{W}_n)({\Delta}-\hat{{\Delta}})I(r\geq \lf n\hat{t}_n\rf)\Big]^2\notag\\
\leq &C \E\Big[\sum_{r=j}^{j+m-1}I(\mathcal{W}_n)\big({\Delta}\big(I(r\geq \lf nt\rf)-I(r\geq \lf n\hat{t}_n \rf)\big)\big)\Big]^2+\Big(\frac{m\log^2 n}{\sqrt{n}}\Big)^2.
\end{align*}
Note that $\E[\sum_{r=j}^{j+m-1}I(\mathcal{W}_n)({\Delta}(I(r\geq \lf nt\rf)-I(r\geq \lf n\hat{t}_n \rf)))]^2=0$ if $j< \lf nt-{C\log n\sqrt{n} }\rf-m+1$ or  $j\geq \lf nt+{C\log n\sqrt{n} }\rf$, and is bounded by $m^2$ if $j\in [\max(1,\lf nt-{C\log n\sqrt{n} }\rf+m-1),\min(n, \lf nt+{C\log n\sqrt{n} }\rf)]$.
Thus we have
\begin{align}
\frac{1}{(n-m+1)m}\sum_{j=1}^{n-m+1}\E[(\tilde{S}^A_{j,m}-\hat{S}^A_{j,m})^2I(\mathcal{W}_n)]\leq \frac{Cm\log n}{\sqrt{n}}+\frac{Cm\log^4n}{n}, \label{new.146}
\end{align}
and similar arguments lead  to
\begin{align}\label{new.147}
\frac{1}{(n-m+1)m}\sum_{j=1}^{n-m+1}\E\Big[(\tilde{S}^A_{n}-\hat{S}^A_{n})^2I(\mathcal{W}_n)\frac{m^2}{n^2}\Big]\leq { \frac{Cm\log^4n}{n}}.
\end{align}
Now   similar arguments as given in the proof of estimate \eqref{corol3} together with (\ref{new.146}) and (\ref{new.147}), yield
\begin{align*}
\sup_{t\in (0,1)}|\tilde{\Phi}^A_{m,n}(t)-\hat{\Phi}^A_{m,n}(t)|=O_p\Big(\Big(\frac{m\log n}{\sqrt{n}}\Big)^{1/2}\Big),
\end{align*}
and the assertion \eqref{estim}  follows from (\ref{112}).

\section{More technical details}  \label{proofstechnical}
\def\theequation{8.\arabic{equation}}
\setcounter{equation}{0}

\subsection{Uniform bounds for nonparametric estimates} \label{proofstechnical1}

\noindent The following two lemmas provide uniform bounds for the estimate $\hat{\mu}_{b_n}$ in the interior $\mk{T}_n=[b_n, 1-b_n]$ and at the boundary $\mk{T}_n'=(0,b_n]\cup[1-b_n,1)$ of the interval $[0,1]$.
\begin{lemma}\label{mu} If assumptions (A1)-(A3) are satisfied and
    $b_n\rightarrow 0$, $nb_n\rightarrow \infty$,    we have
\begin{align*}
\sup_{t\in\mk{T}_n} \Bigl |\hat{\mu}_{b_n}(t)-\mu(t)-\frac{\mu_2\ddot{\mu}(t)}{2}b_n^2-\frac{1}{nb_n}\sum_{i=1}^ne_iK_{b_n}(t_1-t)\Bigr |=O(b_n^3+\frac{b_n}{n}),
\end{align*}
where $\mk{T}_n=[b_n, 1-b_n]$.
\end{lemma}
{\it Proof.}
With the notations   \begin{align*}
S_{n,l}(t)=\frac{1}{nb_n}\sum_{i=1}^n\big(\frac{t_i-t}{b_n}\big)^l K_{b_n}(t_i-t), \qquad
R_{n,l}(t)=\frac{1}{nb_n}\sum_{i=1}^nY_i\big(\frac{t_i-t}{b_n}\big)^lK_{b_n}(t_i-t),
\end{align*}
($l=0,1,...$) we obtain the representation \begin{align}\label{11}
   \begin{bmatrix}
       \hat{\mu}_{b_n}(t)\\
        b_n\hat{\dot{\mu}}_{b_n}(t)
     \end{bmatrix}=
      \begin{bmatrix}
       S_{n,0}(t) & S_{n,1}(t)\\
       S_{n,1}(t) & S_{n,2}(t)
     \end{bmatrix}^{-1} \begin{bmatrix}
       R_{n,0}(t)\\
       R_{n,1}(t)
     \end{bmatrix}=:S_n^{-1}(t)R_n(t),
\end{align}
for the local linear estimate $\tilde \mu_{b_n}$, where the last identity defines the $2 \times 2$ matrix $S_n(t)$ and the vector $R_n(t)$ in an obvious manner.
By elementary calculation and a Taylor expansion   we have
\begin{align*}
S_n(t)\begin{bmatrix}\hat{\mu}_{b_n}(t)-\mu(t)\\ b_n(\hat{\dot{\mu}}_{b_n}(t)-\dot{\mu}(t))\end{bmatrix}=\begin{bmatrix}
       \frac{1}{nb_n}\sum_{i=1}^ne_iK_{b_n}(t_i-t)+\frac{1}{2}\ddot{\mu}(t)\mu_2b_n^2\\
        \frac{1}{nb_n}\sum_{i=1}^ne_iK_{b_n}(t_i-t)(\frac{t_i-t}{b_n})
     \end{bmatrix}+O(b_n^3+b_n/n)
     \end{align*}
     uniformly with respect to $t\in \mathfrak{T}_n$.
Note that  $S_{n,0}(t)=1+O(\frac{1}{nb_n})$ and $S_{n,1}(t)=O(\frac{1}{nb_n})$, uniformly with respect to  $t\in \mathfrak{T}_n$, which yields
\begin{align*}
\sup_{t\in\mk{T}_n}\Big|\hat{\mu}_{b_n}(t)-\mu(t)-\frac{\mu_2\ddot{\mu}(t)}{2}b_n^2-\frac{1}{nb_n}\sum_{i=1}^ne_iK_{b_n}^{}(t_i-t)\Big|=O\big(b_n^3+\frac{b_n}{n}\big).
\end{align*}
Therefore the lemma follows from the definition of the estimate $\hat \mu_{b_n}$ in  (\ref{defhatmu}). $\hfill \Box$

\medskip

\begin{lemma}\label{boundhat}
    Assume that the conditions of Lemma \ref{mu} hold,
 then
\begin{align*}
\sup_{t\in \mk{T}_n'}
\Big|c(t)(\hat{\mu}_{b_n}(t)-\mu(t))-\frac{1}{nb_n}\sum_{i=1}^n\Big[\nu_{2,b_n}(t)-\nu_{1,b_n}(t)\Big(\frac{t_i-t}{b_n}\Big)\Big]e_iK_{b_n}(t_i-t)+\notag\\
\frac{b_n^2}{2}\ddot{\mu}(t)(\nu_{2,b_n}^2(t)-\nu_{1,b_n}(t)\nu_{3,b_n}(t))\Big|=O(b_n^3+\frac{b_n}{n}),
\end{align*}
where
$\mk{T}_n'=[0,b_n]\cup[1-b_n,1]$,  $\nu_{j,b_n}(t)=\int_{-t/b_n}^{(1-t)/b_n}x^jK(x)dx$   and $c(t)=\nu_{0,b_n}(t)\nu_{2,b_n}(t)-\nu^2_{1,b_n}(t)$.
\end{lemma} \

{\it Proof.} 
For any $t \in [0,b_n]\cup[1-b_n,1]$, using $(\ref{11})$, we obtain
\begin{align*}
S_n(t)\begin{bmatrix}\hat{\mu}_{b_n}(t)-\mu(t)\\ b_n(\hat{\dot{\mu}}_{b_n}(t)-\dot{\mu}(t))\end{bmatrix}=\begin{bmatrix}
       \frac{1}{nb_n}\sum_{i=1}^n  [Y_i-\mu(t)-\dot{\mu}(t) (t_i-t)]K_{b_n}(t_i-t)\\
        \frac{1}{nb_n}\sum_{i=1}^n[Y_i-\mu(t)-\dot{\mu}(t)(t_i-t)]K_{b_n}(t_i-t)(\frac{t_i-t}{b_n})
     \end{bmatrix},
     \end{align*}
and a Taylor expansion     yields
\begin{align}\label{new.15}
S_n(t)\begin{bmatrix}\hat{\mu}_{b_n}(t)-\mu(t)\\ b_n(\hat{\dot{\mu}}_{b_n}(t)-\dot{\mu}(t))\end{bmatrix}=\begin{bmatrix}
       \frac{1}{nb_n}\sum_{i=1}^ne_iK_{b_n}(t_i-t)+\frac{b_n^2}{2}\nu_{2,b_n}(t)\ddot{\mu}(t)\\
        \frac{1}{nb_n}\sum_{i=1}^ne_iK_{b_n}(t_i-t)(\frac{t_i-t}{b_n})+\frac{b_n^2}{2}\nu_{3,b_n}(t)\ddot{\mu}(t)
     \end{bmatrix}+O(b_n^3+b_n/n)
\end{align}
uniformly with respect to    $t\in [0,b_n]\cup[1-b_n,1]$.
On the other hand, uniformly with respect to $t\in[0,b_n]\cup[1-b_n,1]$, we have that
\begin{align}\label{new.16}
S_n(t)=\begin{bmatrix}
       \nu_{0,b_n}(t) &  \nu_{1,b_n}(t) \\
        \nu_{1,b_n}(t)  & \nu_{2,b_n}(t)
     \end{bmatrix}+O(\frac{1}{nb_n}).
\end{align}
Therefore, combining (\ref{new.15}) and (\ref{new.16}), it follows that
\begin{align*}
c(t)(\hat{\mu}_{b_n}(t)-\mu(t))=\frac{1}{nb_n}\sum_{i=1}^n\big[\nu_{2,b_n}(t)-\nu_{1,b_n}(t)\big(\frac{t_i-t}{b_n}\big)\big]e_iK_{b_n}(t_i-t)+\notag\\
\frac{b_n^2}{2}\ddot{\mu}(t)\big(\nu_{2,b_n}^2(t)-\nu_{1,b_n}(t)\nu_{3,b_n}(t)\big)+O\Big(b_n^3+\frac{b_n}{n}\Big)
\end{align*}
uniformly with respect to $t\in [0,b_n] \cup [1-b_n,1]$.
   $\hfill \Box$

\medskip

The next lemma concerns the order of deviations of $\hat{\mu}_{b_n}$ from $\mu$ in the $\| \cdot \|_4$-norm.
 \begin{lemma}\label{maxdevvar1}
Assume that assumptions (A1)-(A4) are satisfied and that $nb^3_n\rightarrow \infty$, $nb_n^6\rightarrow 0$, then
\begin{align}
&\sup_{t\in [0,1]}\|\hat{\mu}_{b_n}(t)-\mu(t)\|_4=O(b_n^2+(nb_n)^{-1/2}),\label{revision2-34}
\\&\Big\|\sup_{t\in [0,1]}|\hat{\mu}_{b_n}(t)-\mu(t)|\Big\|_4=O(b_n^2+(nb_n)^{-1/2}b_n^{-1/4}).\label{revision2-35}
\end{align}
\end{lemma}

{\it Proof.} Observing the stochastic expansion in Lemma \ref{mu} we first
 evaluate  $\|\sum_{i=1}^ne_iK_{b_n}(t_i-t)\|_4$ and $\|\frac{\partial}{\partial t}\sum_{i=1}^ne_iK_{b_n}(t_i-t)\|_4$.
 {Recalling the definition of $\mathcal{P}_i$  we note that \begin{align*}\Big \|\sum_{i=1}^ne_iK_{b_n}(t_i-t)\Big \|_4=\Big \|\sum_{i=1}^n\sum_{k=0}^{\infty}\pp_{i-k}e_iK_{b_n}(t_i-t)\Big \|_4\leq\sum_{k=0}^\infty \Big \|\sum_{i=1}^n\pp_{i-k}e_iK_{b_n}(t_i-t)\Big \|_4.
 \end{align*}
 Since for each $k$, $\pp_{i-k}e_iK_{b_n}(t_i-t)$, $1\leq i\leq n$ is a martingale difference sequence, it follows from Burkholder's inequality
 \begin{align*}
& \Big\|\sum_{i=1}^n\pp_{i-k}e_iK_{b_n}(t_i-t)\Big \|^2_4\leq C \Big\|
 \big( {\sum_{i=1}^n(\pp_{i-k}e_iK_{b_n}(t_i-t))^2} \big)^{1/2} \Big \|_4^2\notag
 \\&\leq C\sum_{i=1}^n\|(\pp_{i-k}e_iK_{b_n}(t_i-t))^2 \|_2\notag =C \sum_{i=1}^n\|(\pp_{i-k}e_iK_{b_n}(t_i-t))\|^2_4,
 \end{align*}
  and condition (A4)  implies
  $\|\sum_{i=1}^n\pp_{i-k}e_iK_{b_n}(t_i-t) \|_4=O(\sqrt{nb_n}\chi^k)$,
 uniformly with respect to $t \in [0,1]$. This yields
 \begin{align}\label{revision2-38}\sup_{t\in[0,1]}\Big\|\sum_{i=1}^ne_iK_{b_n}(t_i-t)\Big\|_4=O(\sqrt{nb_n}).\end{align}
 Similar arguments show $\sup_{t\in [0,1]}\|\frac{\partial}{\partial t}\sum_{i=1}^ne_iK_{b_n}(t_i-t)\|_4=O(\sqrt{nb_n}b_n^{-1})$, and
by Proposition \ref{prop:2.1} in Section \ref{proofstechnical4} it follows that
\begin{equation}\|\sup_{t\in [0,1]}|\sum_{i=1}^ne_iK_{b_n}(t_i-t)/(nb_n)|\|_4=O((nb_n)^{-1/2} b_n^{-1/4}),\label{revision2-38.1}\end{equation}
and by Lemma \ref{mu}, we obtain \begin{align*}
\Big \|\sup_{t\in \mathfrak{T}_n}\Big|(\hat{\mu}_{b_n}(t)-\mu(t))^2-\big( \frac{1}{nb_n}\sum_{i=1}^ne_iK_{b_n}(t_i-t)+\frac{\mu_2\ddot{\mu}(t)}{2}b_n^2 \big )^2\Big | \Big \|_2=O\Big(\frac{\chi_n}{\sqrt{nb_n}b_n^{1/4}}+\chi^2_n\Big),
\end{align*}
where $\chi_n=b_n^3+b_n/n$. Hence
  $\|\sup_{t\in \mathfrak{T}_n}(\hat{\mu}_{b_n}(t)-\mu(t))^2   \|_2=O (\frac{1}{nb_n^{3/2}}+b_n^4 ).$
 By similar argument and Lemma \ref{boundhat} it follows that
  $\|\sup_{t\in \mathfrak{T}'_n}(\hat{\mu}_{b_n}(t)-\mu(t))^2 \|_2=O (\frac{1}{nb_n^{3/2}}+b_n^4 )$,
and a combination of the last two estimates gives   (\ref{revision2-35}).
 On the other hand, Lemma \ref{mu} and (\ref{revision2-38}) also imply
 \begin{align*}
\sup_{t\in \mathfrak{T}_n}\Big \|(\hat{\mu}_{b_n}(t)-\mu(t))^2-\big( \frac{1}{nb_n}\sum_{i=1}^ne_iK_{b_n}(t_i-t)+\frac{\mu_2\ddot{\mu}(t)}{2}b_n^2 \big )^2\Big)^2 \Big \|_2=O\Big(\frac{\chi_n}{\sqrt{nb_n}}+\chi^2_n\Big),
\end{align*}which further yields
$\sup_{t\in \mathfrak{T}_n}  \|(\hat{\mu}_{b_n}(t)-\mu(t))^2 \|_2=O ((\frac{1}{\sqrt{nb_n}}+b_n^2)^2 )$.
 Similar arguments show the estimate
$\sup_{t\in \mathfrak{T}'_n}  \|(\hat{\mu}_{b_n}(t)-\mu(t))^2 \|_2=O((\frac{1}{\sqrt{nb_n}}+b_n^2)^2)$,
which proves the remaining estimate (\ref{revision2-34}).
\hfill $\Box$

\bigskip

The following results give a   uniform bound for the $p$-mean of $\hat{\sigma}^2_{}(t) - \sigma^2(t) $, where
$\hat{\sigma}^2 =  \hat \sigma^2_{c_n,b_n}$ is the
variance estimator  defined in \eqref{defhatmu1}.
We  begin with a uniform asymptotic stochastic expansion for the difference  $\hat{\sigma}^2_{}(t) - \sigma^2(t) $.

\begin{lemma}\label{hatsigma}
Suppose that Assumptions (A1)-(A4) are satisfied,   $c_n\rightarrow 0$, $nc_n\rightarrow \infty$, and the variance function $\sigma^2$ is strictly positive, twice differentiable with a Lipschitz continuous second derivative $\ddot{\sigma}^2$. Then the estimate $\hat \sigma^2= \hat \sigma_{c_n,b_n}^2$ defined in \eqref{defhatmu1} satisfies  \begin{align}\label{boundedsigma1}
\sup_{t\in \mk{T}_n}\Big|\hat{\sigma}^2(t)-\sigma^2(t)-\frac{\mu_2\ddot{\sigma}^2(t)c_n^2}{2}-\frac{1}{nc_n}\sum_{i=1}^n(\hat{e}_i^2-\E{e}_i^2)K_{c_n}(t_i-t)\Big|=O\Big(c_n^3+\frac{c_n}{n}\Big),
\end{align}
\begin{align}\label{boundedsigma2}
\sup_{t\in \mk{T}_n'}
\Big|c(t)(\hat{\sigma}^2(t)-\sigma^2(t)-\frac{1}{nc_n}\sum_{i=1}^n\Big[\nu_{2,c_n}(t)-\nu_{1,c_n}(t)\Big(\frac{t_i-t}{c_n}\Big)\Big][\hat{e}^2_i-\E(e^2_i)]K_{c_n}(t_i-t)+\notag\\
\frac{c_n^2}{2}\ddot{\sigma}^2(t)(\nu_{2,c_n}^2(t)-\nu_{1,c_n}(t)\nu_{3,c_n}(t))\Big|=O\Big(c_n^3+\frac{c_n}{n}\Big),
\end{align}
where   $c(t)$, $\nu_{j,c_n}(t)$ are defined in Lemma \ref{boundhat},  $\mk{T}_n=[c_n,1-c_n]$ and $\mk{T}_n'=[0,c_n] \cup [1-c_{n},1]$.
\end{lemma}
{\it Proof.}
 Following the argument given in the proof of Lemma \ref{mu}, we have that
\begin{align*}
S_n(t)\begin{bmatrix}(\hat{\sigma}^2_{}(t)-\sigma^2(t))\\c_n(\hat{\dot{\sigma}}^2_{}(t)-\dot{\sigma}^2(t))\end{bmatrix}=
\begin{bmatrix}\frac{1}{nc_n}\sum_{i=1}^n(\hat{e}_i^2-\sigma^2(t)-\dot{\sigma}^2(t)(t_i-t))K_{c_n}(t_i-t)\\
\frac{1}{nc_n}\sum_{i=1}^n(\hat{e}_i^2-\sigma^2(t)-\dot{\sigma}^2(t)(t_i-t))(\frac{t_i-t}{c_n})K_{c_n}(t_i-t)\end{bmatrix},
\end{align*} where $S_n(t)$ is defined in the proof of Lemma \ref{mu}.
The lemma now follows by the same arguments as given in the proof of Lemma \ref{mu} and Lemma \ref{boundhat}, which are omitted for the sake of brevity.\hfill $\Box$
\begin{corol}\label{corosigma1}
Suppose that the conditions of Lemma \ref{hatsigma} hold  with $\iota\geq 8$, then
\begin{align*}
\sup_{t\in [0,1]}\|\hat{\sigma}^2(t)-\sigma^2(t)\|_4=O\Big(c_n^2+\frac{1}{\sqrt{nc_n}}+b_n^2+\frac{1}{\sqrt{nb_n}} \Big).
\end{align*}
\end{corol}
{\it Proof.}
\ By (\ref{boundedsigma1}), we have   for some large constant $C$,
\begin{align*}
\sup_{t\in \mathcal{T}_n}\|\hat{\sigma}^2(t)-\sigma^2(t)\|_4\leq Cc_n^2+ {\sup_{t\in \mathcal{T}_n}} \Big\|\frac{1}{nc_n}\sum_{i=1}^n(\hat{e}_i^2-\E{e}_i^2)K_{c_n}(t_i-t)\Big\|_4 \leq   \notag\\
Cc_n^2+ {\sup_{t\in \mathcal{T}_n}}\Big\|\frac{1}{nc_n}\sum_{i=1}^n({e}_i^2-\E{e}_i^2)K_{c_n}(t_i-t)\Big\|_4+\sup_{t\in \mathcal{T}_n}\Big\|\frac{1}{nc_n}\sum_{i=1}^n({e}_i^2-\hat{e}_i^2)K_{c_n}(t_i-t)\Big\|_4.
\end{align*}
It is easy to verify that the first term satisfies \begin{align*}
 {\sup_{t\in \mathcal{T}_n}}\Big \|\frac{1}{nc_n}\sum_{i=1}^n({e}_i^2-\E{e}_i^2)K_{c_n}(t_i-t)\Big\|_4=O\Big(\frac{1}{\sqrt{nc_n}}\Big).
\end{align*}
By the proof of Lemma
 {\ref{maxdevvar1}}, we obtain (note that   $\iota\geq 8$)
\begin{align*}
\sup_{t\in [0,1]}\|\hat{\mu}_{b_n}(t)-\mu(t)\|_8=O\Big(b_n^2+\frac{1}{\sqrt{nb_n}}\Big),
\end{align*}
which yields (note that
$\hat{e}_i=e_i+\mu(t_i)-\hat{\mu}_{b_n}(t_i)$)
\begin{align*}
\sup_{t\in \mathcal{T}_n}\Big\|\frac{1}{nc_n}\sum_{i=1}^n({e}_i^2-\hat{e}_i^2)K_{c_n}(t_i-t)\Big\|_4=
O\Big(b_n^2+\frac{1}{\sqrt{nb_n}}\Big).
\end{align*}
Hence
$\sup_{t\in \mathcal{T}_n}\|\hat{\sigma}^2(t)-\sigma^2(t)\|_4=O (c_n^2+\frac{1}{\sqrt{nc_n}}+b_n^2+\frac{1}{\sqrt{nb_n}} )$.
 Similarly, using the estimate (\ref{boundedsigma2}) we obtain
$\sup_{t\in \mathcal{T}'_n}\|\hat{\sigma}^2(t)-\sigma^2(t)\|_4=O (c_n^2+\frac{1}{\sqrt{nc_n}}+b_n^2+\frac{1}{\sqrt{nb_n}} )$,
 which completes the proof. \hfill $\Box$

\begin{corol}\label{corosigma2}
Suppose the conditions of Lemma \ref{hatsigma} hold, with $\iota\geq 8$. Then
\begin{align*}
\Big\|\sup_{t\in (0,1)}|\hat{\sigma}^2(t)-\sigma^2(t)|\Big\|_4=O\Big(c_n^2+\Big(\frac{1}{\sqrt{nc_n}}+b_n^2+\frac{1}{\sqrt{nb_n}}\Big)c_n^{-1/4}\Big).
\end{align*}
\end{corol}
{\it Proof}: The lemma follows from   Proposition \ref{prop:2.1} in Section \ref{proofstechnical4}, the triangle inequality and simple calculations.
{Note that the first assumption of Proposition \ref{prop:2.1} is satisfied by the arguments in Corollary \ref{corosigma1}.
The second assumption regarding  the derivative can be shown by similar arguments as given in (\ref{revision2-38}) and (\ref{revision2-38.1})}.  
\hfill $\Box$

\begin{corol}\label{corosigmastar}
Suppose that the conditions of Corollary \ref{corosigma2} hold. Let $\alpha=1-4/\iota-\nu$ for some $\nu>0$, and $\iota$ is defined in condition (A3) and condition (A4). Then we have
\begin{align*}
\sup_{t\in [0,\tilde{t}_v-n^{-\alpha}]\cup [\tilde{t}_v+n^{-\alpha},1]}|\sigma_n^{2*}(t)-\sigma^2(t)|=O_p ( \rho_n )~,
\end{align*}
where
$ \rho_n=
n^{-\alpha}c_n^{-5/4}+c_n^2+\big(\tfrac{1}{\sqrt{nc_n}}+b_n^2+\frac{1}{\sqrt{nb_n}}\big)c_n^{-1/4}
$
\end{corol}
{\it Proof.} Define
\begin{align*}
\bar S_{n,l}^k=\frac{1}{nc_n}\sum_{i=1}^k\Big(\frac{t_i-t}{c_n}\Big)^lK_{c_n}(t_i-t),\ \bar R_{n,l}^k=\frac{1}{nc_n}\sum_{i=1}^k\Big(\frac{t_i-t}{c_n}\Big)^lK_{c_n}(t_i-t)\hat e_i^2,\\
\underline S_{n,l}^k=\frac{1}{nc_n}\sum_{i=k}^n\Big(\frac{t_i-t}{c_n}\Big)^lK_{c_n}(t_i-t),\ \underline R_{n,l}^k=\frac{1}{nc_n}\sum_{i=k}^n\Big(\frac{t_i-t}{c_n}\Big)^lK_{c_n}(t_i-t)\hat e_i^2.
\end{align*}
Recall the definition of $\sigma_n^{2*}(t)$ in \eqref{varest1}. With the notation $k^*=\lf nt_n^*\rf$ elementary calculations yield
\begin{align}\label{singletilde}
\tilde \sigma_1^2(t):=\tilde \sigma_{1,k^*}^2(t)=\big(\bar R_{n,0}^{k^*}(t)\bar S_{n,2}^{k^*}(t)-\bar R_{n,1}^{k^*}(t)\bar S_{n,1}^{k^*}(t)\big)/\big(\bar S_{n,0}^{k^*}(t)\bar S_{n,2}^{k^*}(t)-(\bar S_{n,1}^{k^*}(t))^2),\\ \notag
\tilde \sigma_2^2(t):=\tilde \sigma_{2,k^*+1}^2(t)=\big(\underline R_{n,0}^{k^*+1}(t)\underline S_{n,2}^{k^*+1}(t)-\underline R_{n,1}^{k^*+1}(t)\underline S_{n,1}^{k^*+1}(t)\big)/\big(\underline S_{n,0}^{k^*+1}(t)\underline S_{n,2}^{k^*+1}(t)-(\underline S_{n,1}^{k^*+1}(t))^2\big).\end{align}
 We will show below that
\begin{eqnarray*}
{ \sup_{|\tilde{t}_v-t_h|\leq n^{-\alpha}}}\sup_{t\in [0,\tilde{t}_v-n^{-\alpha}]}|\tilde \sigma_{1,h}^2(t)-\sigma_1^2(t)|=O_p( \rho_n ),\label{newth-reve2}\\
{ \sup_{|\tilde{t}_v-t_h|\leq n^{-\alpha}}}\sup_{t\in [\tilde{t}_v+n^{-\alpha},1]}|\tilde \sigma_{2,h}^2(t)-\sigma_2^2(t)|=O_p(\rho_n ).\label{newth-reve2-2}\end{eqnarray*}
The assertion of Corollary \ref{corosigmastar} then follows    since $t_n^*-\tilde{t}_v=O_p(n^{-\alpha})$.
 For a proof of the two remaining estimates we assume without loss of generality  $t_h>\tilde{t}_v$ and define
$$
\bar{ \mathcal{R}}_{n,l}^h(t)=\frac{1}{nc_n}\Big(\sum_{i=1}^{\lf n\tilde{t}_v\rf}\Big(\frac{t_i-t}{c_n}\Big)^lK_{c_n}(t_i-t)\hat e_i^2+\sum_{i= \lf n\tilde{t}_v+1\rf}^{h}\Big(\frac{t_i-t}{c_n}\Big)^lK_{c_n}(t_i-t) \tilde e_i^2\Big),
$$
where $\{\tilde e_i^2 | \lf n\tilde{t}_v\rf+1\leq i\leq t_h\}$ are independent of $\{\FF_i, i\in \mathbb{Z}\}$ with bounded $\iota_{th}$ moment,  $\E(\tilde e_i)=0$,  Var$(\tilde e_i){ :=}\tilde \sigma^2(t_i)$ such that
$\ddot{\tilde \sigma}^2(t_i)$ are Lipschitz continuous in $[\tilde{t}_v,t_h]$ and $\ddot{\tilde \sigma}^2(\tilde{t}_v^+)=\ddot{\tilde \sigma}^2(\tilde{t}_v)=\ddot{\sigma}^2(\tilde{t}_v^-)=\ddot{\sigma}^2_1(\tilde{t}_v-)$.
Define \begin{align}\label{doubletilde}
\tilde {\tilde \sigma}_{1,h}^2(t)=\big(\bar {\mathcal{R}}_{n,0}^{h}(t)\bar S_{n,2}^{h}(t)-\bar{\mathcal{ R}}_{n,1}^{h}(t)\bar S_{n,1}^{h}(t)\big)/\big(\bar S_{n,0}^{h}(t)\bar S_{n,2}^{h}(t)-\left(\bar S_{n,1}^{h}(t)\right)^2\big).
\end{align}
Using similar arguments as given in the proof of Corollary \ref{corosigma2} it follows that
\begin{align}\label{7.18}
\sup_{\lf n\tilde{t}_v\rf <h<\lf n\tilde{t}_v\rf+\lf n^{1-\alpha}\rf}\sup_{t\in [0,\tilde{t}_v-n^{-\alpha}]}|\tilde {\tilde \sigma}_{1,h}^2(t)-\sigma_1^2(t)|=O_p\Big(c_n^2+\Big(\frac{1}{\sqrt{nc_n}}+b_n^2+\frac{1}{\sqrt{nb_n}}\Big)c_n^{-1/4}\Big).
\end{align}
On the other hand, by   Proposition \ref{prop:2.1}  we can show that for $l=0,1$ 
\begin{align*}
\sup_{\lf n\tilde{t}_v\rf <h<\lf n\tilde{t}_v\rf+\lf n^{1-\alpha}\rf}\sup_{t\in [0,\tilde{t}_v-n^{-\alpha}]}|\bar{ \mathcal{R}}_{n,l}^h(t)-{ \bar{R}}_{n,l}^h(t)|=O_p(n^{-\alpha}c_n^{-5/4}).
\end{align*}
Combining \eqref{singletilde} and \eqref{doubletilde}, the last estimate implies that
\begin{align*}
\sup_{\lf n\tilde{t}_v\rf <h<\lf n\tilde{t}_v\rf+\lf n^{1-\alpha}\rf}\sup_{t\in [0,\tilde{t}_v-n^{-\alpha}]}|\tilde {\tilde \sigma}_{1,h}^2(t)-\tilde \sigma_1^2(t)|=O_p(n^{-\alpha}c_n^{-5/4}).
\end{align*}
By \eqref{7.18}  we have 
\begin{align}\label{7.21}
\sup_{\lf n\tilde{t}_v\rf <h<\lf n\tilde{t}_v\rf+\lf n^{1-\alpha}\rf}\sup_{t\in [0,\tilde{t}_v-n^{-\alpha}]}| {\tilde \sigma}_{1,h}^2(t)-\sigma_1^2(t)|=O_p\
 ( \rho_n ) ,
\end{align}
and it  is easy to see that
\begin{align} \nonumber
&\sup_{\lf n\tilde{t}_v\rf <h<\lf n\tilde{t}_v\rf+\lf n^{1-\alpha}\rf}\sup_{t\in [\tilde{t}_v+n^{-\alpha},1]}| {\tilde \sigma}_{2,h}^2(t)-\sigma_2^2(t)|=O_p
 ( \rho_n )  \\  \nonumber
&\sup_{\lf n\tilde{t}_v\rf-\lf n^{1-\alpha}\rf <h<\lf n\tilde{t}_v\rf\rf}\sup_{t\in [0,\tilde{t}_v-n^{-\alpha}]}|{\tilde \sigma}_{1,h}^2(t)-\sigma_1^2(t)|=O_p ( \rho_n ) ,\\
&\sup_{\lf n\tilde{t}_v\rf-\lf n^{1-\alpha}\rf <h<\lf n\tilde{t}_v\rf}\sup_{t\in [\tilde{t}_v+n^{-\alpha},1]}|{\tilde \sigma}_{2,h}^2(t)-\sigma_2^2(t)|=O_p\ ( \rho_n ).\label{7.24}
\end{align}
The assertion now follows from \eqref{7.21}--\eqref{7.24}, the definition of $\sigma^{2*}_n(t)$, and the fact that $t_n^*-\tilde{t}_v=o_p(n^{-\alpha})$.\hfill$\Box$
\subsection{Two additional technical results} \label{proofstechnical4}
\begin{proposition}\label{prop:2.1}
Let $\{\Upsilon_n(t)\}_{t \in[0,1]}$ be a sequence of stochastic processes with   differentiable paths.
Assume   that for some   $p\geq 1$ and any $t\in[0,1], \|\Upsilon_n(t)\|_p=O(m_n), \|\dot{\Upsilon}_n(t)\|_p=O(l_n), $ where $m_n, l_n $ are sequences of real numbers, $m_n=O(l_n)$, then
$$\Big\|\sup_{t\in[0,1]}|\Upsilon_n(t)|\Big\|_p=O\Big(m_n\Big(\frac{m_n}{l_n}\Big)^{-\frac{1}{p}}\Big).$$
 In particular, if $p=2$, we have   $\|\sup_{t\in[0,1]}|\Upsilon_n(t)|\|_2=O(\sqrt{m_nl_n})$.
\end{proposition}
\noindent{\it Proof}. For a sequence $b_n$ define $\tilde{b}_n=\lfloor {b_n}^{-1}\rfloor$ and let $\tau_i=ib_n$, $i=1,2,...,\tilde{b}_n$ and $\tau_i=1$ for $i= \tilde{b}_n+1$.   Then by the triangle inequality, we have
 \begin{align*}
\sup_{t\in{(0,1)}}|\Upsilon_n(t)|\leq  \max_{0\leq i \leq \tilde{b}_n+1}|\Upsilon_n(\tau_i)|+\max_{1\leq i\leq \tilde{b}_n+1}Z_{in},\end{align*}
where $Z_{in}=\sup_{\tau_i-b_n<t<\tau_i}|\Upsilon_n(t)-\Upsilon_n(\tau_i)|.$\
Observing the inequalities
  $$ \|Z_{in}\|_p \leq \Big\| \int^{\tau_i}_{\tau_i -b_n}|  \dot{\Upsilon} (t)|dt \Big\|_p \leq\int_{\tau_i-b_n}^{\tau_i}\|\dot{\Upsilon}_n(t)\|_p dt=O(b_n l_n)$$
and $ \max_{1\leq i\leq \tilde{b}_n+1}Z^p_{in}\leq \sum_{i=1}^{\tilde{b}_n+1}Z^p_{in}$,
we have $$ \Big\|\max_{1\leq i\leq\tilde{b}_n+1} Z_{in}\Big\|_p=O((l_n^pb_n^{(p-1)})^{1/p})=O(l_nb_n^{(p-1)/p}).$$
Similarly, we obtain the estimate
 $\|\max_{0\leq i \leq\tilde{b}_n+1}|\Upsilon_n(t_i)|\|_p=O_p(b_n^{-1/p}m_n)$,
and picking $b_n=m_n/l_n$ proves the assertion.
\hfill$\Box$

\bigskip

\begin{proposition}\label{prop01}
Suppose $A_n$ are sets such that $\p(A_n)\rightarrow 0$ as $n\rightarrow\infty$, and $X_nI(\bar{A}_n)=O_p(1)$. Then $X_n=O_p(1)$.
\end{proposition}
{\it Proof.} For any $\epsilon>0$,   let $N$ be a large constant such that $\p(A_n)\leq \epsilon/2$ for $n\geq N$, and $M$ be a large constant such that $\p(|X_n|I(\bar{A}_n)\geq M/2)\leq \epsilon/2$ for $n\geq N$. Then
\begin{align*}\p(|X_n|\geq M)\leq \p(|X_n|I(A_n)\geq M/2)+\p(|X_n|I(\bar{A}_n)\geq M/2)\notag
\\\leq \p(A_n)+\p(|X_n|I(\bar{A}_n)\geq M/2)\leq \epsilon\end{align*} for all $n\geq N$.
\hfill$\Box$

\end{document}